\documentclass[fleqn,usenatbib]{mnras}
\usepackage[T1]{fontenc}

\usepackage{graphicx}	
\usepackage{amsmath}	
\usepackage{amssymb}	

\usepackage{multirow}
\usepackage[table,xcdraw]{xcolor}
\usepackage{booktabs}
\usepackage{lscape}

\usepackage{color,soul}
\definecolor{lightblue}{rgb}{.70,.95,1}
\sethlcolor{lightblue} 

\usepackage{xspace} 
\renewcommand{\AA}{\normalfont\r{A}\xspace} 

\newcommand{\teff}{\ensuremath{T_{\mathrm{eff}}}\xspace}
\newcommand{\teffa}{\ensuremath{T_{\mathrm{eff,A}}}\xspace}
\newcommand{\kms}{\ensuremath{\rm{km}\,s^{-1}}\xspace}
\newcommand{\logg}{\ensuremath{\log g}\xspace}
\newcommand{\feh}{\rm{[Fe/H]}\xspace}
\newcommand{\cfe}{\rm{[C/Fe]}\xspace}
\newcommand{\nfe}{\rm{[N/Fe]}\xspace}
\newcommand{\ch}{\rm{[C/H]}\xspace}
\newcommand{\ac}{\rm{A(C)}\xspace}

\newcommand{\Gaia}{\textit{Gaia}\xspace}
\newcommand{\CaHK}{\emph{CaHK}\xspace}
\newcommand{\Pristine}{\emph{Pristine}\xspace}

\defcitealias{arentsen20b}{A20b}
\defcitealias{placco14}{P14}
\defcitealias{aguado19}{AY19}

\usepackage{newtxtext,newtxmath}

\title[On the inconsistency of carbon abundances]{On the inconsistency of [C/Fe] abundances and the fractions of carbon-enhanced metal-poor stars among various stellar surveys}
 
\author[A. Arentsen et al.]{Anke Arentsen,$^{1}$\thanks{E-mail: anke.arentsen@astro.unistra.fr},
Vinicius M. Placco,$^{2}$
Young Sun Lee,$^{3}$
David S. Aguado,$^{4,5}$
Nicolas F. Martin,$^{1,6}$
\newauthor
Else Starkenburg,$^{7}$ 
Jinmi Yoon$^{8,9}$
\\
$^{1}$Universit\'e de Strasbourg, CNRS, Observatoire astronomique de Strasbourg, UMR 7550, F-67000 Strasbourg, France\\
$^{2}$NSF's NOIRLab, 950 N. Cherry Ave., Tucson, AZ 85719, USA \\
$^{3}$Department of Astronomy and Space Science, Chungnam National University, Daejeon 34134, Republic of Korea\\
$^{4}$Dipartimento di Fisica e Astrofisica, Univerisità degli Studi di Firenze, via G. Sansone 1, I-50019 Sesto Fiorentino, Italy \\
$^{5}$INAF/Osservatorio Astrofisico di Arcetri, Largo E. Fermi 5, I-50125 Firenze, Italy\\
$^{6}$Max-Planck-Institut f\"ur Astronomie, K\"onigstuhl 17, D-69117 Heidelberg, Germany\\
$^{7}$Kapteyn Astronomical Institute, University of Groningen, Postbus 800, 9700 AV, Groningen, the Netherlands\\
$^{8}$Space Telescope Science Institute, 3700 San Martin Dr., Baltimore, MD 21218, USA\\
$^{9}$Joint Institute for Nuclear Astrophysics -- Center for the Evolution of the Elements (JINA-CEE), USA
}

\date{Accepted 2022 July 20. Received 2022 July 19, in original form 2022 June 8}

\pubyear{2022}

\begin{document}
\label{firstpage}
\pagerange{\pageref{firstpage}--\pageref{lastpage}}
\maketitle

\begin{abstract}

Carbon-enhanced metal-poor (CEMP) stars are a unique resource for Galactic archaeology because they 
probe the properties of the First Stars, early chemical evolution and
binary interactions at very low metallicity. Comparing the fractions and properties of CEMP stars in different Galactic environments can provide us with unique insights into the formation and evolution of the Milky Way halo and its building blocks.  
In this work, we investigate whether directly comparing fractions of CEMP stars from different literature samples of very metal-poor ($\feh < -2.0$) stars is valid.  
We compiled published CEMP fractions and samples of Galactic halo stars from the past 25 years, and find that they are not all consistent with each other. Focusing on giant stars, we find significant differences between various surveys when comparing their trends of \feh versus \cfe and their distributions of CEMP stars. 
To test the role of the analysis pipelines for low-resolution spectroscopic samples, we re-analysed giant stars from various surveys with the SSPP and FERRE pipelines. We found systematic differences in \cfe of $\sim0.1-0.4$~dex, partly independent of degeneracies with the stellar atmospheric parameters. These systematics are likely due to the different pipeline approaches, different assumptions in the employed synthetic grids, and/or the comparison of different evolutionary phases.
We conclude that current biases in (the analysis of) very metal-poor samples limit the conclusions one can draw from comparing different surveys. We provide some recommendations and suggestions that will hopefully aid the community to unlock the full potential of CEMP stars for Galactic archaeology.

\end{abstract}

\begin{keywords}
stars: chemically peculiar -- stars: Population II -- stars: abundances -- techniques: spectroscopic -- methods: data analysis
\end{keywords}


\section{Introduction}

Very metal-poor stars are useful probes for Galactic archaeology, since they are old and contain key information about the early Universe. However, they are rare and often dedicated or very large surveys are needed to find them. A surprise in the early HK prism survey of metal-poor stars \citep[][hereafter BPS]{beers92} was that many very metal-poor stars appeared to have exceptionally strong carbon features. \citet{norris97} calculate the fraction and find that $\sim 10$\% of the BPS stars have `stronger than normal G-band strengths' -- broken down by metallicity these fractions are 14\%, 4\% and 4\% for \feh\footnote{[X/Y] $ = \log(N_\mathrm{X}/N_\mathrm{Y})_* - \log(N_\mathrm{X}/N_\mathrm{Y})_{\odot}$, where the asterisk subscript refers to the considered star, and N is the number density.} $< -2.5$, $-2.5 < \feh < -1.5$ and $-1.5 < \feh < -0.5$, respectively. They also point out that \citet{luck91} find 1\% of carbon-rich stars (CH stars) among metal-poor disk stars (with average $\feh = -0.4$), and conclude that the fraction of carbon-rich stars increases with decreasing metallicity. These stars were dubbed carbon-enhanced metal-poor (CEMP) stars by \citet{beerschristlieb05}, who set the criterion for a star to be CEMP at $\cfe > +1.0$ and $\feh < -1.0$. Since then, many CEMP stars were found in (metal-poor) surveys such as the Hamburg/ESO (HES) survey \citep{christlieb08} and the Sloan Digital Sky Survey \cite[SDSS,][]{sdss} \citep[e.g.][]{carollo12, lee13}. Reported CEMP fractions were between $10-30\%$ for very metal-poor (VMP, $\feh < -2.0$) stars \citep[e.g.][]{lucatello06, lee13, placco14}, and typically increase with decreasing metallicity. 

Many CEMP stars have been studied in detail in recent years. The two most common types of CEMP stars are CEMP-s stars, which have an excess of slow neutron-capture (s-process) elements as well as a high carbon abundance, and CEMP-no stars, which do not show the s-process enhanced signature but do have a high carbon abundance. There are also CEMP-r, CEMP-r/s stars and CEMP-i stars, enhanced with rapid and/or intermediate neutron-capture processes, which are more rare and will not be considered in this work. The CEMP-s stars are common at $\feh > -3.0$ and are thought to form through binary interaction with a former asymptotic giant branch (AGB) companion. This is supported by their chemical abundance patterns and a high binary fraction \citep[e.g.][]{lucatello05,bisterzo10,abate15,hansen16b}. The CEMP-no stars, on the other hand, are dominant at $\feh < -3.0$ and are thought to have been born with their high carbon abundance, from gas polluted by the first generations of stars \citep[for a review on extremely metal-poor stars including CEMP-no stars, see][]{frebelnorris15}. Their binary fraction is much lower than that of CEMP-s stars \citep{norris13b,starkenburg14,hansen16a}, although it appears to be higher than that of normal metal-poor stars \citep{arentsen19_cemp}. There are several mechanisms related to the First Stars that could cause high yields of carbon, such as their explosion as mixing-and-fallback (faint) supernovae \citep[e.g.][]{umedanomoto03,nomoto13,tominaga14} and/or their exceptionally high rotation rates \citep[e.g.][]{chiappini06,meynet06,hegerwoosley10}. 

Knowing the fraction of CEMP stars relative to carbon-normal stars, especially for CEMP-s and CEMP-no stars separately, is important for our interpretation of populations of metal-poor stars. The fraction of CEMP-s stars teaches us about the binary fraction of low-metallicity stars in a population, about the properties of those binary systems and about the way they interact with their (former) AGB companions. The fraction of CEMP-no stars can shed light on the importance of peculiar processes in the First Stars and on the rate of chemical evolution in the birth environments of extremely metal-poor stars in the early Universe. Differences in the fractions of CEMP stars between Galactic environments can provide us with valuable clues about the different conditions in structures in the early Universe.  

Almost all samples of low-metallicity stars contain high numbers of CEMP stars, but the discussion is still ongoing what exactly the fraction of CEMP stars is as a function of metallicity, how it changes for the different classes of CEMP stars and different stellar evolutionary phases \citep{beerschristlieb05,cohen05,frebel06,lucatello06,aoki07,carollo12,lee13,yong13,lee14,placco14,beers17,yoon18,placco18,placco19,limberg21,whitten21,shank21}, and how it may be different in different Galactic environments, such as different regions of the halo \citep{frebel06,carollo12,carollo14,lee17,beers17,lee19,yoon18}, dwarf galaxies \citep[e.g.][]{starkenburg13,salvadori15,chiti18,yoon19, ji20}, globular clusters \citep{dorazi10}, stellar streams \citep{aguado21} and the Galactic bulge \citep{howes16,arentsen21}. There are many uncertainties in determining the CEMP fraction, e.g. due to selection effects in samples of metal-poor stars, challenges in measuring stellar parameters and carbon abundances, and difficulties with the classification of different types of CEMP stars. Additionally, comparisons between literature fractions are sometimes hampered by the adoption of different CEMP star definitions and whether or not corrections for the evolutionary stage of the stars have been taken into account. 

Ideally, when comparing different VMP/CEMP samples, stars of similar evolutionary phases should be used because they have experienced similar evolutionary effects. 
Additionally, their spectral analysis will suffer from similar systematics due to e.g. the assumption of local thermodynamic equilibrium (LTE) and calculations in one dimension (1D) for the computation of synthetic model spectra. 
In environments other than the Galactic halo, such as the Galactic bulge, dwarf galaxies, globular clusters or stellar streams, VMP samples typically consist of giant stars because these are bright. It is therefore important to study the CEMP fraction among giant stars, if we want to interpret the CEMP fraction in different environments. This is the focus of this paper. 

Giants are cool ($\teff < 5700$~K) and have clear carbon features, which can easily be detected in low-resolution spectroscopy. Their CEMP fraction, however, can be challenging to interpret. Giants can suffer from evolutionary effects that change their surface \cfe, which needs to be corrected for, and samples could be a mix of red giant branch (RGB) and early AGB stars, which have experienced different evolutionary effects. It is also not trivial to analyse spectra of cool, carbon-rich stars, which are dominated by molecular bands \citep[e.g.][]{beers99, rossi05, Goswami06, yoon20}, and finally, their photometry is also affected due to the large carbon features, leading to sample selection effects.

We review the various reported CEMP fractions in the literature in Section~\ref{sec:lit}, and discuss issues in comparing them with each other. In Section~\ref{sec:spec}, we compare two pipelines for low-resolution spectroscopic analysis of VMP giants and discuss implications for some previously derived CEMP fractions. We summarise our results in Section~\ref{sec:discussion}, concluding that there is still more work to be done before we can safely compare and interpret CEMP fractions of giant stars.


\section{Literature CEMP fractions}\label{sec:lit}

\begin{figure}
\centering
\includegraphics[width=1.0\hsize,trim={0.0cm 1.0cm 0.0cm 0.0cm}]{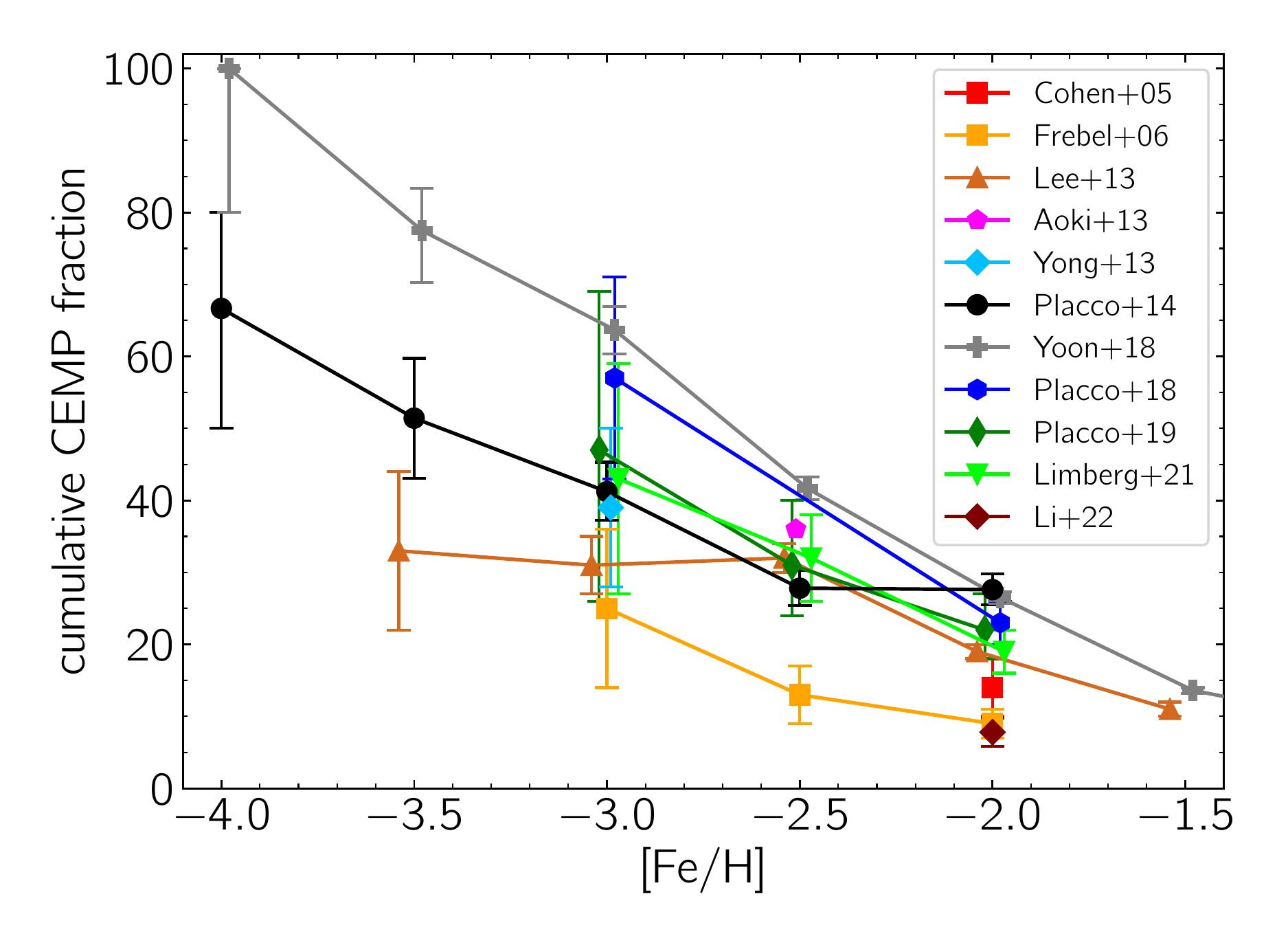}
\caption{Cumulative fraction of CEMP stars as function of metallicity for samples of giant stars (see Table~\ref{tab:cempfrac}), including all sub-types of CEMP stars. Follow-up samples from narrow-band photometric surveys are not included. The $\cfe > +0.7$ CEMP definition was used in all samples, except those of \citet{cohen05} and \citet{frebel06}, who used $\cfe > +1.0$. These are plotted with square symbols. Evolutionary corrections (following either \citealt{aoki07} or \citealt{placco14}) have been taken into account in all cases, except for \citet{cohen05}, \citet{frebel06} and \citet{lee13}. Small offsets in \feh have been applied to avoid overlap between points.} 
    \label{fig:allfrac}
\end{figure}

There are many reported values of CEMP fractions among metal-poor halo stars in the literature. In this section, we discuss various reports throughout the years, separated into low ($R \lesssim 2500$) and high ($R \gtrsim 15\,000$) resolution spectroscopic samples. An overview is presented in Table~\ref{tab:cempfrac} (not all of those will be mentioned in the text, and the compilation may also not be exhaustive). For giant stars, the different cumulative CEMP fractions are shown in Figure~\ref{fig:allfrac}, summing all stars below a given metallicity. We also discuss the relative occurrence of CEMP-s and CEMP-no stars in various samples. We end this section with a summary and a discussion of the literature comparisons. 

\subsection{Low/medium resolution samples}\label{sec:lr_lit}

\subsubsection{Early estimates}

The first large spectroscopic samples of very metal-poor (VMP, $\feh < -2.0$) stars were built by following up exciting, metal-poor looking candidates from objective prism efforts such as the HK and HES surveys. In the review by \citet{beerschristlieb05}, a carbon-enhanced ($\cfe > +1.0$) fraction of $\sim 20\%$ among VMP stars is quoted, determined from `moderate-resolution follow-up spectra' of objective prism surveys (but no further details are provided). Another early number which is regularly cited is $25\%$ by \citet{marsteller05}, but again no details are given as to how this number was derived. 

\citet{cohen05} analyse a sample of HES stars with high-resolution spectroscopy and discuss issues with the \citet{beers99} metallicity and carbon abundance scale for cool stars used in previous HES and HK medium-resolution analyses -- if they correct for those effects they find a slightly lower fraction of CEMP stars of $14\% \pm 4\%$. \citet{frebel06} use an updated calibration for the derivation of \cfe from medium-resolution spectra \citep{rossi05}, and find a CEMP fraction of $9\% \pm 2\%$ in a HES medium-resolution follow-up VMP sample of giants.
They comment that this is similar to the fraction by \citet{cohen05}, although slightly lower, possibly due to limitations in the analysis for stars with very strong carbon bands. They also found that the CEMP fraction rises to 13\% $\pm$ 4\% and 25\% $\pm$ 11\% for $\feh < -2.5$ and $< -3.0$, respectively, and that the CEMP fraction appears to be larger for stars further away from the Galactic disc. 

\subsubsection{Change in definitions}
We note that \citet{aoki07} proposed a new definition of CEMP stars, lowering the CEMP threshold to $\cfe = +0.7$ instead of $+1.0$ because that appeared to be a more natural division between carbon-normal and carbon-rich stars. They also added a dependence of the threshold on the evolutionary status/luminosity of a star, which is important for giants since the surface carbon abundance of evolved stars decreases during their ascent on the RGB. This was further studied by \citet{placco14}, who derived \cfe corrections based on the evolutionary stage of a star and its metallicity and measured carbon abundance. Some have argued that the $\cfe = +0.7$ threshold is not optimal since it may lead to spurious classifications \citep{bonifacio15}, but it has been widely adopted. All fractions described above considered stars with $\cfe > +1.0$ to be CEMP and did not take into account evolutionary effects. The following CEMP fractions in the remainder of this section assume a CEMP definition of $\cfe > +0.7$, unless otherwise specified. They are inhomogeneous in whether or not they took into account any evolutionary corrections. 

\subsubsection{SDSS}
Other than employing dedicated metal-poor surveys such as HK and HES to determine CEMP fractions, one can use large surveys that do not directly target very metal-poor stars. These have the advantage that they are expected to be less biased in their targeting (apart from e.g. cuts in brightness and colour), whereas the HK and HES samples for example could be biased by human choices in the selection for follow-up. 

\citet{carollo12} analyse $\sim 30\,000$ calibration stars in SDSS/SEGUE (a mixture of dwarf, turn-off and subgiant/giant stars) using the SEGUE Stellar Parameter Pipeline (SSPP, \citealt{lee08a, lee08b}) and report a CEMP fraction of $12\%$ among VMP stars (with $\feh < -2.0$). The fraction decreases to $8\%$ for $\feh < -1.5$, and increases to $20\%$ for stars with $\feh < -2.5$. They also find trends between the CEMP fraction and the height above the disc, as in 
\citet{frebel06}. 

\citet{lee13} developed a method for determining \cfe and implemented it into the SSPP, and derive carbon abundances for a much larger sample of several hundred thousand SDSS stars (of all metallicities, not just VMP stars). They combine their sample with high-resolution literature samples for the extremely metal-poor stars ($\feh < -3.0$), and find CEMP fractions consistent with \citet{carollo12} to within $1\%$. If they adopt a CEMP definition of $\cfe > +1.0$, their fractions are consistent with those from \citet{frebel06} as well. \citeauthor{lee13} also report CEMP fractions for extremely metal-poor stars, which are further increasing to $28\% \pm 3\%$ for $\feh < -3.0$ and $43\% \pm 11\%$ for $\feh < -3.5$. The authors find that there is a large discrepancy between CEMP fractions determined from giants and turn-off stars (they also test dwarfs, but most VMP stars are in the other categories) -- the cumulative fractions for $\feh < -2.0, -2.5$ and $-3.0$ are $\sim 2 - 3$ times higher among giants compared to the hotter turn-off stars (the giant fractions are $19\%, 32\%$ and $31\%$, respectively). They interpret this as a result of the difficulty to measure carbon abundances for hot stars (e.g., turn-off stars) -- at higher temperatures, the CH G-band can only be detected for stars with very high carbon abundances, and stars with moderate carbon-enhancement would be missed. There is also a systematic effect with metallicity, as it is more difficult to detect the G-band for more metal-poor stars. No evolutionary corrections were applied in either \citet{carollo12} or \citet{lee13}. 

\subsubsection{Recent estimates}
In recent years, CEMP fractions were determined for many more large samples of metal-poor stars with low-resolution spectroscopy. All stellar parameter and carbon abundance determinations for these samples were performed using an adapted version of the SSPP for non-SDSS/SEGUE spectra (called n-SSPP), except for samples from the \Pristine survey (which will be discussed later). For example \citet{beers17}, using on a mixed sample of dwarfs, turn-off stars and giants selected from HES, found CEMP fractions largely consistent with previous estimates from SDSS.  Other examples are \citet{yoon18} using stars from the AAOmega Evolution of Galactic Structure (AEGIS) survey (a follow-up survey based on SkyMapper data, \citealt{wolf18}), \citet{placco18} for metal-poor stars selected from the RAdial Velocity Experiment (RAVE, \citealt{ravedr1}), and \citet{placco19} and \citet{limberg21} for stars selected from the Best \& Brightest (B\&B, \citealt{schlaufman14}) survey. All of these followed up mostly (sub)giant stars (due to selection biases in the original surveys), and typically find CEMP fractions higher than any other reported fractions in the literature based on low-resolution spectroscopy. For example, \citet{yoon18} find CEMP fractions of $27\%, 42\%$ and $64\%$ for $\feh < -2.0, -2.5$ and $-3.0$ in the AEGIS survey.
The higher fraction in these recent samples might be because they do not contain hot stars and therefore they miss fewer moderately carbon-rich objects, and also because the \citet{placco14} corrections for evolutionary effects on \cfe were applied in all these works. However, we will show in Section~\ref{sec:spec} that there might be another reason for the high CEMP fractions in these samples. 

A lot of work has also been done to identify and specifically search for CEMP stars, for example using colours, prism-spectroscopy or infrared spectroscopy \citep[e.g.][]{beers92, christlieb01, placco10, placco11, kielty17}. Such samples cannot be used to determine the fraction of CEMP stars, but are still valuable sources for high-resolution spectroscopic follow-up.

\subsubsection{CEMP-no vs. CEMP-s}
Low-resolution samples typically cannot distinguish between CEMP-no stars and CEMP-s stars based on their neutron-capture abundances, because of the intrinsic weakness of the absorption features. However, fortunately CEMP stars of different types not only differ in their neutron-capture abundances, but they also have different distributions in their metallicity and carbon abundances \citep[e.g.][]{spite13,bonifacio15,yoon16}. CEMP-s stars tend to have $\feh > -3.5$ and absolute carbon abundance\footnote{$A$(C) $= \log{\epsilon (C)} = \log(N_C/N_H) + 12$} $\ac > 7.0$ (called Group I stars by \citealt{yoon16}), while CEMP-no stars occupy the regions with lower \feh and/or \ac (\citealt{yoon16} identify two different regions, which they call Group II and III). \citet{yoon18} present separate CEMP frequencies for Group I (or CEMP-s) and Group II and III (CEMP-no) for their AEGIS low-resolution spectroscopic sample. For $\feh < -2.0$, they find Group II+III and Group I fractions of $17\%$ and $10\%$, respectively. The fraction of Group II+III CEMP stars among stars with $\feh < -2.5, -3.0$ and $-3.5$ rises to $35\%$, $54\%$ and $68\%$, respectively, while the Group I fraction stays relatively constant with $6\%$, $10\%$ and $10\%$, respectively. 


\subsection{High resolution samples}\label{sec:hr_lit}

\subsubsection{\citet{lucatello06}}
After the first confirmations of very metal-poor stars with low/medium-resolution spectroscopy, extensive follow-up campaigns with high-resolution spectroscopy started. Abundances from high-resolution spectroscopy are more precise and extensive, a clear advantage over using low-resolution samples for the determination of CEMP fractions, although the sample sizes are typically smaller. From high-resolution spectroscopic follow-up of 270 very metal-poor stars from the HES R-process enhanced star (HERES, \citealt{barklem05}) survey, \citet{lucatello06} derive a CEMP fraction ($\cfe > +1.0$) of 21\% $\pm$ 2\% for $\feh < -2.0$. They find that the fraction does not change when considering only unevolved stars. The CEMP fraction in their sample increases slightly, to 24\% $\pm$ 3\%, when adopting the \citet{aoki07} definition. They do not report fractions for different metallicity ranges, although the CEMP fraction for stars with $\feh < -3.0$ appears much smaller than the fraction for $\feh < -2.0$ from their Figure~1. 

\citet{lucatello06} discuss that the criteria to select HERES stars from the HES medium-resolution spectroscopy sample were specifically set to minimise biases in, for example, carbon abundance, and therefore assumed that their derived CEMP fraction was not biased. However, it was later recognised that some other high-resolution HES samples did have biases towards C-rich stars \citep[see e.g.][]{cohen13}. This was due to systematics in the early analysis of the medium-resolution spectra, producing biased metallicities and carbon abundances.

\subsubsection{\citet{yong13}}
\citet{yong13} report a CEMP fraction focusing especially on the extremely metal-poor regime ($\feh < -3.0$). They re-analyse 152 stars with high-resolution spectroscopy from the literature and combine it with a HES follow-up sample of 38 stars from their earlier work \citep{norris13a}. Most of the literature stars originate from follow-up of the HK and HES surveys, and there is also a small number from other sources such as high proper motion surveys. Detailed studies of the HES metallicity distribution function (MDF) completeness were presented in \citet{schorck09} and \citet{li10}, who concluded that the HES selection is complete for $\feh < -3.0$. \citet{yong13} carefully propagate this information (for both HK and HES stars) to their sample, but little is known about the selection function of the remainder of the literature stars. 

From the 69 stars in their sample with $\feh < -3.0$, they report a CEMP fraction of $32\% \pm 8\%$ (adopting the \citealt{aoki07} definition). They find that the CEMP fraction appears to be higher for dwarfs, which they interpret as the result of the difficulty to measure carbon abundances in these hotter stars -- several have upper limits on the carbon abundance above $\cfe = +0.7$, and these are not taken into account in the fraction computation. \citet{yong13} do not report CEMP fractions for $\feh > -3.0$ as they were worried that there may possibly be a bias towards CEMP stars for these metallicities in the sample. For example such a bias could be due to the \citet{norris13a} part of the sample for which the strategy included observing `objects with prominent G bands in their medium-resolution spectra' for the more metal-rich candidates. 

\subsubsection{\citet{placco14}}
The CEMP fraction published by \citet[][hereafter P14]{placco14} is based on a compilation from the high-resolution literature as well, combining stars from the SAGA database \citep{suda08} and the \citet{frebel10} compilation. The final sample consists of 505 stars with $\feh < -2.0$ and a measurement of \cfe (no stars with only upper limits were included, which could introduce some biases). A significant difference with previous works is the use of individual evolutionary carbon corrections, based on stellar evolution models and using the \logg, \feh and measured \cfe of each star. This allows for fairer CEMP fraction estimates from giant stars. \citetalias{placco14} assume that their literature sample is not biased towards or against CEMP stars. The contributions to the literature come from many different sources, which all have different selection effects that may cancel each other out, but this is not necessarily true. 

The overall CEMP fractions that they find are $33\%$, $34\%$, $48\%$ and $60\%$ for $\feh < -2.0, -2.5, -3.0$ and $-3.5$, respectively (using the CEMP definition of $\cfe > + 0.7$). This is higher than all previous estimates (although similar to some of the more recent low-resolution estimates). It is expected to be somewhat higher due to the correction for evolutionary effects, but there may also be certain selection effects which are particularly troublesome for warmer stars (as discussed in multiple previous works). Re-calculating the numbers for only the giants ($\teff < 5700$~K and $\logg < 3.8$) in the \citetalias{placco14} sample, the fractions go down to $28\%$, $29\%$, $41\%$ and $51\%$ for $\feh < -2.0, -2.5, -3.0$ and $-3.5$, respectively. This brings them somewhat closer to previous CEMP fractions. \citetalias{placco14} also compute separate fractions for CEMP-no and CEMP-s stars. The fraction of CEMP-s stars decreases with decreasing \feh from $13\%$ to $10\%$ to $5\%$ to $0\%$ for $\feh < -2.0, -2.5, -3.0, -3.5$, respectively, whereas the fraction of CEMP-no stars increases with decreasing metallicity: $20\%, 24\%, 43\%$ and $60\%$. 

\subsubsection{SkyMapper}

In recent years, many new metal-poor stars are being discovered in dedicated narrow-band photometric surveys. The SkyMapper survey in the Southern hemisphere employs the narrow-band $v$ filter which includes the Ca H\&K lines, and is therefore metallicity-sensitive \citep{keller07}. The CEMP fraction in the high-resolution follow-up SkyMapper sample of \citet{yong21} is 67$\%$ (6/9) for $\feh < -3.5$, 29$\%$ (19/66) for $\feh < -3.0$ and 20$\%$ (19/93) for $\feh < -2.5$, after \citetalias{placco14} carbon-corrections, assuming a CEMP definition of $\cfe > +0.7$ and excluding stars with upper limits on \cfe. For $\feh < -2.5$ and $-3.0$ this is lower than the $34\%$ and $48\%$ reported by \citetalias{placco14} for $\feh < -3.0$ and $< -3.5$, respectively. The CEMP stars that are present in the sample all have relatively low $\cfe$ ($\lesssim 1.5$), and almost none of them have $\feh > -3.0$. These lower frequencies of CEMP stars (and the lack of very carbon-rich CEMP stars) are interpreted as the result of photometric selection effects. The flux in the metallicity-sensitive $v$ filter is affected by large molecular features for carbon-rich stars, making them look more metal-rich, hence creating a bias against selecting such stars for VMP follow-up \citep[see e.g.][]{dacosta19, chiti20}. 

The Extremely Metal-poor BuLge stars with AAOmega (EMBLA) survey \citep{howes14,howes15,howes16} is the follow-up of SkyMapper in the inner regions of the Milky Way. They selected targets for high-resolution spectroscopic follow-up from a large sample of low-resolution spectroscopic SkyMapper follow-up observations with the Anglo Australian Telescope (AAT). Out of 33 very metal-poor stars in their high-resolution sample, they found only one to be clearly carbon-enhanced (with $\cfe > +1.0$). After correcting for evolutionary effects, there are five additional CEMP stars with $+0.7 < \cfe < +1.0$, and the derived CEMP fractions are $44\%$, $18\%$ and $20\%$ for $\feh < -3.0, -2.5$ and $-2.0$, respectively \citep{arentsen21}. \citet{howes16} discuss the possibility that the low frequency of CEMP stars in the bulge could hint at differences in the early chemical evolution between the inner Galaxy and the more distant halo, but also recognise that selection effects could be playing a role in the photometric SkyMapper selection.

\subsubsection{LAMOST: \citet{li22}}

Recently, a large high-resolution spectroscopic follow-up sample of VMP stars in LAMOST\footnote{\url{http://www.lamost.org/public/?locale=en}} (Large sky Area Multi-Object fiber Spectroscopic Telescope) has been published by \citet{li22}, with carbon abundances for 281 stars. They find very different CEMP fractions for giants and turn-off stars in their sample, 7.8\% and 31\%, respectively, adopting $\cfe > +0.7$ for CEMP stars. The turn-off CEMP fraction decreases to 22\% when counting all stars with \cfe upper limits as carbon-normal, and to 11\% when adopting $\cfe > +1.0$ as CEMP definition. The difference in CEMP fractions between giants and turn-off stars is most striking for the CEMP-s stars, but also visible for CEMP-no stars. The authors suggest that this may be related to non-effective mixing during the turn-off phase, resulting in higher carbon abundances in turn-off stars than in giants. 

The CEMP fraction in the \citet{li22} sample is much lower for VMP giants than in other literature samples, whereas for turn-off stars it is similar to previous estimates. The target selection was based on promising very metal-poor candidates from the analysis of the low-resolution LAMOST spectra \citep{li18}, with a mixed strategy: mostly focussing on extremely low-metallicity candidates, and partly randomly extending the sample up to $\feh = -2.0$ for bright stars \citep{aoki22}. The authors do not comment on possible selection effects towards or against carbon-rich stars in this process. 

\begin{figure*}
\centering
\includegraphics[width=0.35\hsize,trim={0.0cm 0.0cm 0.5cm 0.0cm}]{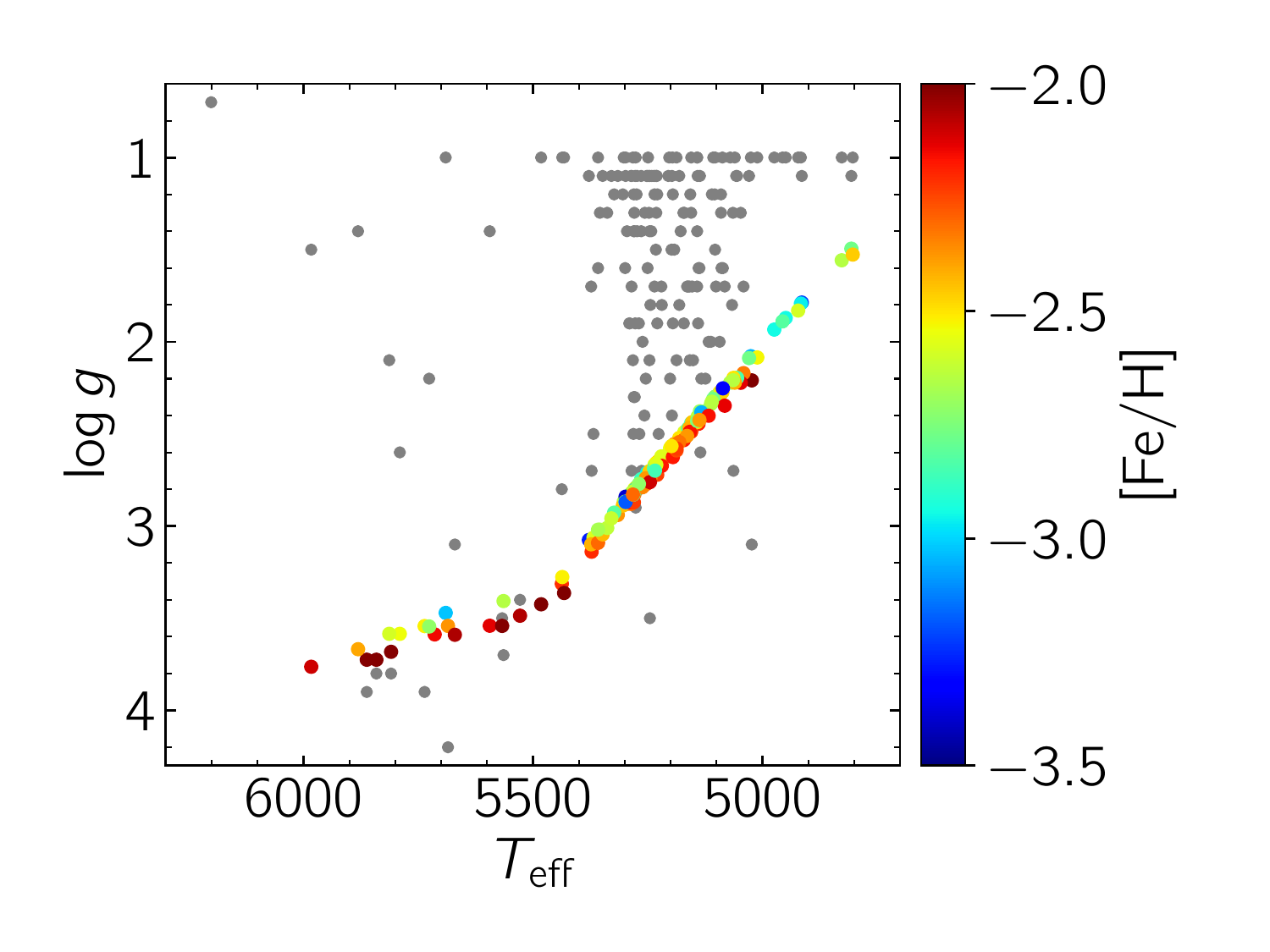}
\includegraphics[width=0.62\hsize,trim={0.5cm 0.0cm 0.0cm 0.0cm}]{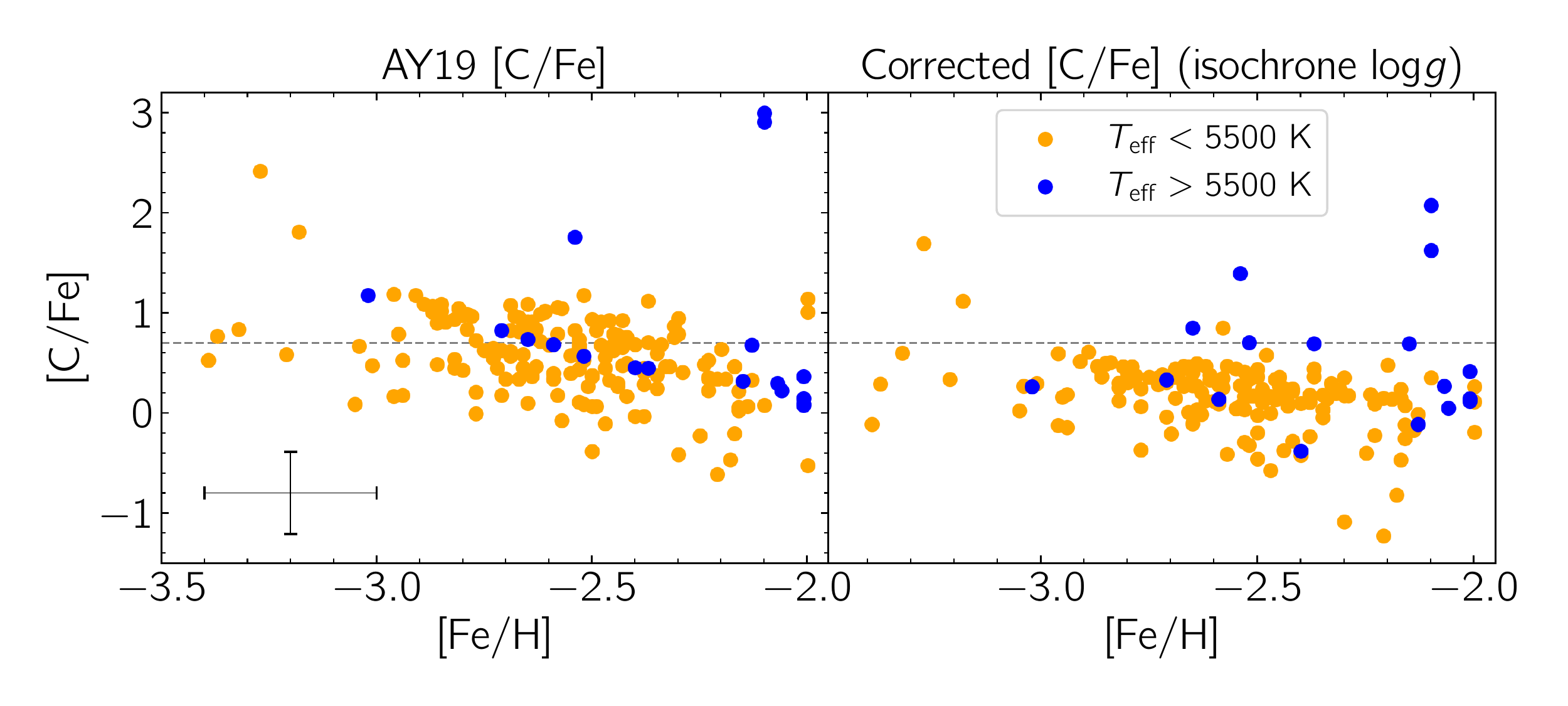}
\caption{Determination of the \cfe correction for the sample of VMP stars from \citetalias{aguado19} with reliable \cfe. Left-hand panel: \teff -- \logg diagram, with the FERRE \logg's in grey and the isochrone \logg's coloured by \feh from FERRE. 
Right-hand panels: \feh versus $\cfe_\mathrm{AY19}$ (left) or $\cfe_\mathrm{iso}$ (right), for cool stars with $\teff < 5500$~K in orange and hotter stars in blue. The dashed line is placed at $\cfe = +0.7$. The median uncertainties for \feh and \cfe are shown on the bottom left of the AY19 \cfe panel. }
    \label{fig:halo-iso}
\end{figure*}

\subsection{CEMP fractions in the \Pristine survey}\label{sec:pris}

The \Pristine survey uses narrow-band photometry combined with broad-band photometry to select metal-poor stars \citep{starkenburg17b}. The \CaHK filter employed by the \Pristine survey is narrower than the $v$ filter of the SkyMapper survey, and is expected to be less affected by molecular features in carbon-rich stars. However, \Pristine also uses broad-band photometry, which can also be affected by carbon. The main \Pristine survey, targeting the Galactic halo, has done extensive follow-up over the last years. We will discuss the CEMP fractions in various \Pristine halo samples next. 

\subsubsection{The CEMP fraction in \citet{aguado19}} 

A sample of 1007 very metal-poor candidates followed up with medium-resolution was published in \citet[][hereafter AY19]{aguado19}, analysed with the FERRE code\footnote{FERRE \citep{allende06} is available from \url{http://github.com/callendeprieto/ferre}}. Due to the relatively low signal-to-noise and the fact that many stars had relatively high temperatures, \cfe could only be reliable derived for a sample of 173 stars with $\feh < -2.0$. It was also not possible to derive good \logg values and many stars had \logg values close to the limit of the grid ($\logg = 1$ for giants). A bad \logg is not expected to strongly affect the \teff and \feh values, but \citetalias{aguado19} showed that there is a strong correlation between \logg and \cfe in the FERRE analysis (also discussed in \citealt{arentsen21}). No corrections for this effect were applied in \citetalias{aguado19}, the uncertainties on \logg were simply inflated accordingly. The authors concluded that the fraction of CEMP stars was consistent with the literature, with $41 \pm 4 \%$ for VMP stars with $-3.0 < \feh < -2.0$ and $58 \pm 14 \%$ for EMP stars with $\feh < -3.0$. 

Here we apply empirical corrections to the \cfe values from \citetalias{aguado19}, based on the mean behaviour between \logg and \cfe found in the FERRE analyses of \citetalias{aguado19} and \citet{arentsen21}, and quantified in \citet{arentsen21} as $\Delta \cfe = -0.37 \Delta \logg$. We determine the difference between the FERRE \logg values and \logg values adopted from Yonsei-Yale isochrones \citep{demarque04} with an age of 12 Gyr in a grid from $-3.5 < \feh < -2.0$ in steps of 0.1 dex, and use this to estimate corrections to \cfe. The results are shown in Figure~\ref{fig:halo-iso}, where the left-hand panel shows a Kiel diagram with the old (grey) and new (coloured) \logg values. Most stars in the sample have $\logg_\mathrm{iso} > 2.0$, therefore the \cfe should not yet be affected significantly by evolutionary effects along the RGB. The right-hand panels show the difference between the \cfe from \citetalias{aguado19} and those corrected using the isochrone \logg values. The total number of CEMP stars is reduced from 64 to 13 -- these are the stars above the $\cfe = +0.7$ lines in the right-hand panels. Additionally, the dispersion in \cfe reduces as well. 

How does this \cfe correction affect the CEMP fraction in the \citetalias{aguado19} sample? It varies with temperature and metallicity, as visible in the right-hand panels of Figure~\ref{fig:halo-iso}. For warm VMP stars with $\teff > 5500$~K the CEMP fraction is $7/17 = 41.2 \%$, which is comparable to what was reported in \citetalias{aguado19}. There is only one EMP star in this temperature range, and it is not carbon-rich. For cooler EMP stars with $5000$~K~$< \teff < 5500$~K the fraction is $3/8 = 37.5 \%$, consistent with \citetalias{aguado19}. For $\feh < -2.0$ in this temperature range, however, there is a strikingly low fraction of CEMP stars: only $3/135 = 2.2 \%$. In fact none of these three has $\cfe > +1.0$. This is in severe disagreement with previously derived CEMP fractions. 

\subsubsection{The CEMP fraction in other Pristine samples}

In the \Pristine-selected sample followed up with FORS2 medium-resolution spectroscopy by \citet{caffau20}, the authors noted a lower CEMP fraction compared to \citetalias{aguado19}, with $3/28 = 11\%$ for $-3.0 < \feh < -2.0$ and $4/13 = 31\%$ for $\feh < -3.0$. If one only considers the stars in their sample with $\teff < 5500$~K, actually none out of the 63 stars with $\feh < -2.0$ are considered CEMP, whereas for hotter stars the fraction is 32\% (8/25). 

Similar effects with temperature are seen in the high-resolution \Pristine ESPaDOnS sample, by \citet{lucchesi22}. The fraction of CEMP stars appears to be consistent with that of \citetalias{placco14} for dwarfs/unevolved giants. For cooler giants, however, they find a CEMP fraction of only 11\%, with none of the CEMP giants having $\cfe > +1.0$. 

In the \Pristine Inner Galaxy Survey \citep[PIGS,][]{arentsen21}, which contains only giant stars, the fractions of CEMP stars are found to be 5.7\% for $\feh < -2.0$ and 16.4\% for $\feh < -2.5$, both much lower than in \citetalias{placco14}. For $\feh < -3.0$, the CEMP fraction is consistent with \citetalias{placco14}. The authors interpret the discrepancy to be a combination of selection effects and a real difference in CEMP stars between the inner Galaxy and the rest of the halo (see the discussion in \citealt{arentsen21}). 

The general trend appears to be that the CEMP fractions in \Pristine are largely consistent with previous estimates for extremely metal-poor stars and for warmer stars (e.g. $\teff > 5500$~K), but not for cooler giants with $\feh > -3.0$. This is likely related to photometric selection effects, as for example in the SkyMapper survey \citep{dacosta19}, which will be discussed in a forthcoming paper.

\subsection{Relative CEMP-no/-s fractions}\label{sec:relnos}

Above, we mainly discussed the fraction of CEMP stars with respect to carbon-normal stars. In several cases, separate fractions for CEMP-s and CEMP-no stars were reported as well. Since these CEMP types have very different progenitors, their relative fraction teaches us about the importance of the different physical processes with metallicity. Here we will compare a number of CEMP samples and the relative frequency of CEMP-no vs. CEMP-s stars. 

A compilation of more than 300 CEMP stars ($\cfe > + 0.7$) which have high-resolution spectroscopy was published by \citet{yoon16}. This sample does not contain carbon-normal stars and can therefore not be used to estimate the CEMP fraction, but it is an interesting sample to investigate the relative fraction of CEMP-s and CEMP-no stars. Given the inhomogeneity of the literature, it is not possible to say whether or not the sample is unbiased with respect to the different types of CEMP stars. Different selection biases may have canceled each other out, but it is not certain. For practical purposes, for now we assume that there is no (significant) bias towards a certain type of CEMP star at a given metallicity. There could conceivably be an overall bias towards more metal-poor stars in the sample, given previous follow-up strategies. 

\begin{figure}
\centering
\includegraphics[width=0.95\hsize,trim={0.0cm 0.9cm 0.0cm 0.0cm}]{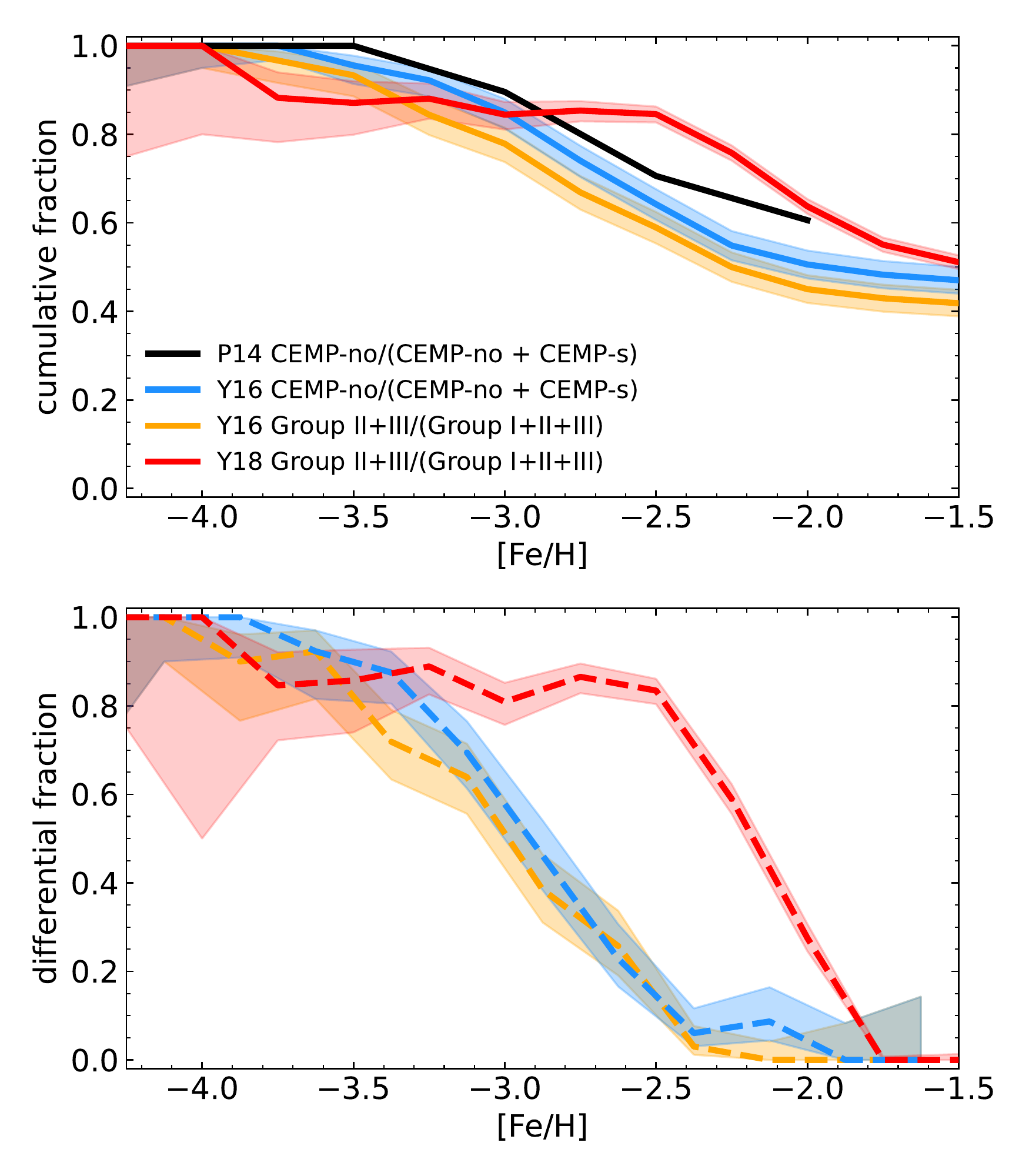}
\caption{Fraction of CEMP-no (blue) or CEMP Group II+III (orange) stars among all classified CEMP stars in the \citet{yoon16} sample. The CEMP-no fraction from \citetalias{placco14} is shown in black, the Group II+III fraction from \citet{yoon18} in red.  The top and bottom panel show the cumulative fraction (for all stars below a certain \feh) and the differential fraction (for stars in a small range of \feh), respectively.} 
    \label{fig:yoon16_frac}
\end{figure}

We determine the fraction of CEMP-no stars among all the CEMP stars in the \citet{yoon16} compilation, and also the fraction of Group II+III stars (these have been selected using \ac instead of using [Ba/Fe]), which are expected to be similar \citep{yoon16}. The results for cumulative and differential fractions as a function of metallicity are shown in the top and bottom panels of Figure~\ref{fig:yoon16_frac}, respectively. Indeed they are similar for CEMP-no and Group II+III. The cumulative fraction is sensitive to the underlying metallicity distribution -- the fractions for $\feh > -3.0$ are strongly affected by what happens at lower metallicities. This is especially true if a sample has relatively many extremely metal-poor stars compared to very metal-poor stars, which is often the case for high-resolution spectroscopic samples. The differential fractions are more relevant, they describe the relative importance of each type of CEMP star for a given metallicity bin. They clearly show that the CEMP-s/Group I stars dominate for $\feh > -3.0$ and the CEMP-no/Group II+III stars for $\feh < -3.0$. 

The cumulative fractions of the high-resolution \citetalias{placco14} compilation are also shown in black in the top panel of Figure~\ref{fig:yoon16_frac}. It follows the shape of the cumulative distribution in \citet{yoon16} -- comparing the black and blue curves. This is not surprising, given that they are both based on similar literature samples. The \citetalias{placco14} curve lies slightly higher, meaning larger fractions of CEMP-no stars, which may also be unsurprising given that they mainly set out to determine the CEMP-no fraction and therefore may have included fewer CEMP-s stars in the compilation.

Finally, the cumulative and differential fractions of Group~II+III stars in the low-resolution \citet{yoon18} AEGIS sample are shown in red. They have a very different behaviour compared to the high-resolution samples -- comparing to the orange curves for \citet{yoon16} Group~II+III stars. AEGIS has both a higher cumulative and a higher differential fraction of Group~II+III stars for $\feh > -3.0$, while it is slightly lower for extremely metal-poor stars. The difference is particularly striking for the differential fraction, which is the fairer comparison because it is not sensitive to the underlying metallicity distribution of the sample. The reasons for the difference in CEMP-no fraction between the samples is unclear. It could be related to selection effects in one or both samples -- for example, AEGIS is a follow-up survey based on SkyMapper photometry, which could possibly have introduced a bias against the more carbon-rich Group~I stars. The fraction could also be affected by the precision of derived \cfe and \feh estimates from low-resolution spectroscopy. This makes the division into the different CEMP groups using \ac less precise. Having larger uncertainties also means that more stars may scatter above the $\cfe > +0.7$ limit, into the region of Group~II+III CEMP stars. This will be a topic of discussion in Section~\ref{sec:spec} as well. 

\subsection{Discussion on selection effects and biases}\label{sec:lit_disc}

The goal of this section on the literature is to show how different various samples used for the determination of CEMP fractions are, how they may be plagued by different systematic issues, and how it can be difficult to compare them with each other. Although the general trend of an increasing CEMP fraction with decreasing metallicity is clear, there still exist discrepancies between the actual derived fractions that we do not yet understand. 

The difficulty with high-resolution (and small) samples is that they are often selected for the most metal-poor stars (e.g. HK, HES), and they are not complete at higher metallicities ($\feh > -3.0$). Additionally, there are other human selection factors involved in the building of those samples as well, e.g., preference for or against carbon-rich stars, and these are hard to retrace for individual samples, let alone for literature compilations. The advantage of high-resolution spectroscopy is that the carbon abundances are (likely) more precisely determined. 

The difficulty with low-resolution (often the larger, likely less biased) samples is that the stellar parameters and \cfe are more challenging to determine. Some biases were for example visible in the early \cfe estimates from \citet{beers98, beers99} mentioned above, or in the \Pristine \citet{aguado19} analysis, discussed above as well. More on this topic will be discussed in Section~\ref{sec:spec}. Low-resolution samples also have larger uncertainties associated with their \feh and \cfe determinations, leading to more stars randomly scattering into the CEMP regime. 

Finally, the difficulty with photometrically selected VMP samples (e.g. SkyMapper, \Pristine) are the biases coming from colour selections. The photometry in different bands can be affected by large molecular carbon features, especially for the cooler carbon-rich objects. 

All CEMP fractions discussed in this section are for samples of halo stars (except for PIGS). However, not all halo samples are the same. For example, they have different metallicity distributions, which strongly affects the derivation of \textit{cumulative} CEMP fractions. If samples are large enough, \textit{differential} fractions should be determined instead, making comparisons between samples more straightforward and less dependent on the underlying MDF. Additionally, different samples of halo stars can probe very different regions of the halo, for example because they select a different magnitude range, or use different tracers such as turn-off stars or giants. There may be real differences between CEMP fractions and the relative number of CEMP-s/-no stars in different parts of the Galaxy (as suggested by e.g. \citealt{frebel06, carollo14, lee17, lee19, yoon18, arentsen21}), but part of the observed differences may also be due to systematics caused by the use of different tracers, differences in spectral analysis pipelines and/or different assumptions for the synthetic grids. This makes comparisons between different samples even more challenging.

There is a huge caveat in \textit{all} CEMP fraction determinations that we have not yet discussed. The iron and carbon abundances that go into these fractions are usually determined using synthetic spectra calculated in one dimension and in local thermodynamic equilibrium (LTE), but for very metal-poor stars the 3D and/or non-LTE effects on \feh and/or \cfe can be large (e.g. \citealt{asplund05, gallagher17, amarsi19}). \citet{norrisyong19} applied 3D and non-LTE corrections (based on computations in the literature for a small number of stars) to the CEMP sample of \citet{yoon16} and find that the number of stars that can be classified as CEMP drops by 70\% after the corrections. For the sample of extremely metal-poor stars in \citet{yong13}, the CEMP-no fraction drops from 24\% to 8\%. The impact of 3D/non-LTE on the CEMP fractions is significant, and, if correct, completely changes the picture. Preferentially, one would have 3D/non-LTE calculations for each star, but these are extremely expensive and this will not be possible in the near future. 

While this means that we may not (yet) know the \textit{absolute} number of CEMP stars, the hope is that it is still possible to compare the CEMP fraction between different samples and/or Galactic environments. Since 3D/non-LTE corrections depend on the evolutionary stage of a star, ideally, samples with stars in similar stages would be used for such comparisons. This is another reason why we mostly focus on giants in this work.


\section{Spectral analysis comparison of low-resolution VMP samples}\label{sec:spec}

It is clear from the previous section that low-resolution spectroscopic samples play a key role in understanding populations of CEMP stars. In this section, we compare various samples and analyses of low-resolution ($\mathrm{R} \sim 1000-2000$) spectroscopic VMP stars to test whether their carbon abundances are on the same scale, and hence whether their CEMP fractions can be compared. We focus on giants in the metallicity range of $-3.0 < \feh < -2.0$, where both CEMP-no and CEMP-s stars are expected to be important and where there is still unclarity regarding the CEMP fraction for the two classes.

\subsection{Carbon trends in various literature samples}\label{sec:trends}

We first compile a number of low-resolution VMP samples from the literature and study their \cfe trends to see if there are systematic differences between them. The main sample comes from the SDSS, analysed with the SSPP as described in \citet{lee13}. Second, we use the VMP LAMOST DR3 sample \citep{yuan20}, whose spectra are quite similar to those from SDSS, analysed with the n-SSPP \citep{beers14, beers17}. Furthermore, we include the low-resolution optical follow-up from RAVE \citep{placco18} and the Best \& Brightest (B\&B) survey (\citealt{placco19, limberg21}, recently superseded by \citealt{shank21}), analysed with the n-SSPP. Then, we include two samples analysed with FERRE \citep{allendeprieto14}, from the \Pristine halo survey (\citetalias{aguado19}, \cfe corrected as described in the previous section) and from the \Pristine Inner Galaxy Survey \citep[PIGS,][]{arentsen21}. Finally, we compare with the \citetalias{placco14} and \citet{li22} high-resolution samples. 

We focus on cool giants ($\teff < 5700$~K, $\logg < 3.8$) for the reasons described in the introduction, and mainly because the sample of warm stars is biased towards stars with higher carbon abundances due to difficulty in measuring carbon in warmer stars. We further limit our analysis to $-3.0 < \feh < -2.0$ because this is the metallicity range where the CEMP fraction is less well-constrained and the various low-resolution samples have most of their stars. We use carbon abundances with \citetalias{placco14} evolutionary corrections where available (for the RAVE, B\&B, PIGS and \citetalias{placco14} samples, and we computed them for the \citealt{li22} sample as well), and if they are not available (for the SDSS, LAMOST and Pristine samples) we limit ourselves to giants with $\logg > 2.3$, for which evolutionary effects are not yet important.

\begin{figure*}
\centering
\includegraphics[width=0.75\hsize,trim={0.0cm 0.0cm 0.0cm 0.0cm}]{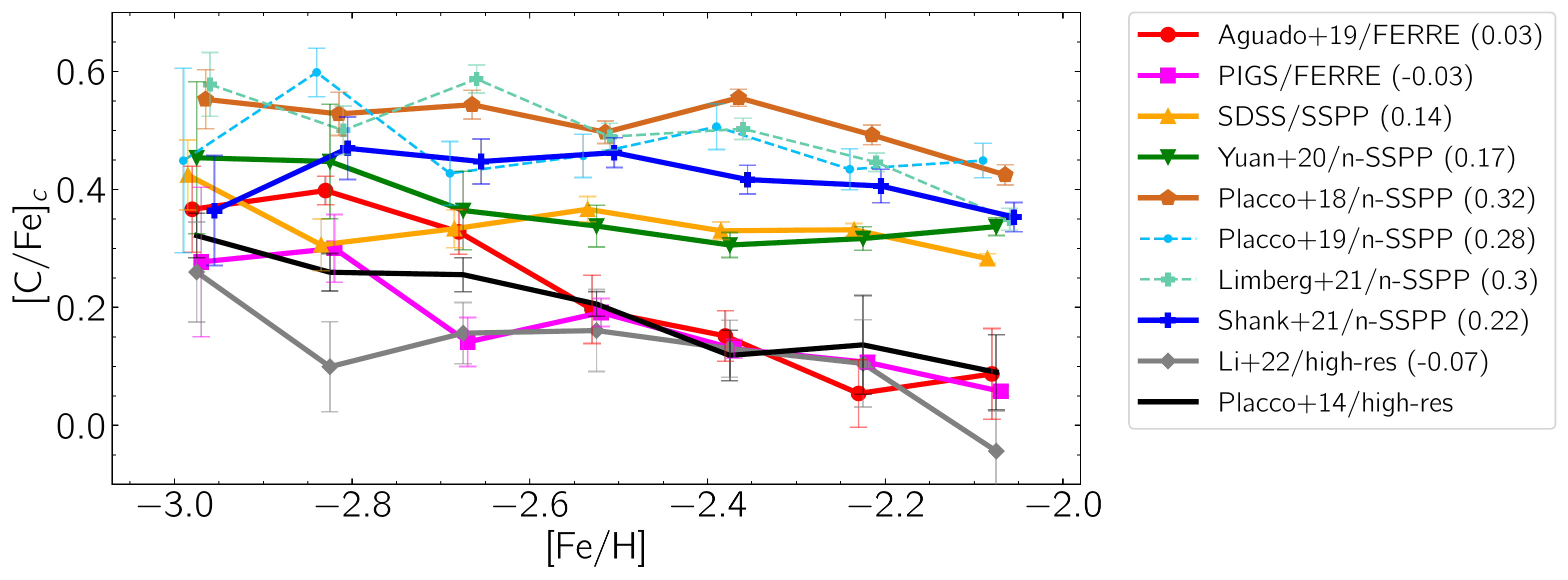}
\includegraphics[width=0.88\hsize,trim={0.0cm 0.0cm 0.0cm 0.0cm}]{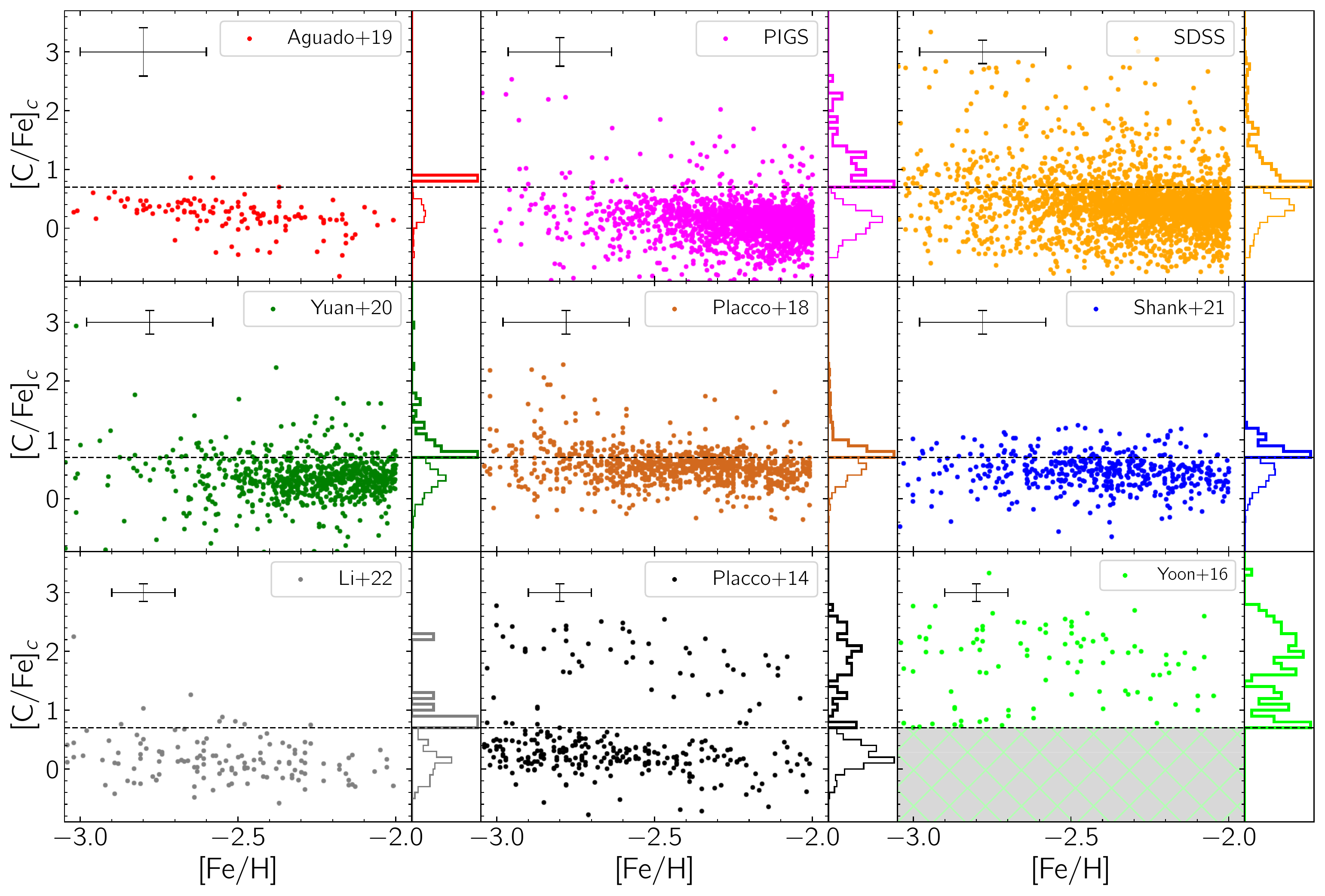}
\caption{Top: the average $\cfe_c$ with metallicity for relatively carbon-normal stars ($-0.5 < \cfe_c < +1.0$) for various giant samples and methods (see text for details), where the error bars represent $\sigma / \sqrt{N}$. All included stars have been corrected for evolutionary effects, or have $\logg > 2.3$. Small offsets in \feh have been applied for visibility of the error bars. The \citet{placco19} and \citet{limberg21} analyses were superseded by \citet{shank21}, they are shown with dashed lines. In the legend, the average $\cfe_c$ difference between the \citet{placco14} high-resolution trend and each sample is indicated in parenthesis. Bottom: distributions in \feh and \cfe for the same samples, and an additional panel for the CEMP compilation by \citet{yoon16}. The \cfe histograms are shown on the right-hand side of each panel, with the carbon-normal stars represented by thin lines and the CEMP stars by thick lines (separated at $\cfe = +0.7$). They are normalised separately to highlight their distributions. Representative error bars are indicated in the top left of each panel -- for the \citetalias{aguado19} and PIGS samples this is the median of the provided individual uncertainties, for the (n-)SSPP samples this is 0.2~dex for both \feh and \cfe \citep{lee08a, lee13, beers14, beers17}, and for the high-resolution samples they are 0.1~dex for \feh and 0.15~dex for \cfe. 
} 
    \label{fig:compcfe}
\end{figure*}

In the top panel of Figure~\ref{fig:compcfe}, we show the mean carbon abundance with metallicity for the various samples. We only include stars not strongly enhanced or depleted in carbon ($ -0.5 < \cfe < +1.0$), to get a sense of the general trends. Although the trend is similar in each sample (increasing \cfe with decreasing \feh), there are large offsets between them, especially for \feh higher than $-2.5$. The FERRE samples more or less follow the \citet{placco14} high-resolution $\cfe$, whereas the SDSS and LAMOST samples have a higher average \cfe by $0.14-0.17$~dex, and the n-SSPP samples a higher average \cfe by $0.22-0.32$~dex. It is notable that samples analysed with similar methods have similar systematic differences compared to the \citet{placco14} high-resolution trend. 

In the bottom panels of Figure~\ref{fig:compcfe}, we show the distributions of the VMP stars in each sample, this time including the CEMP stars. We also include the high-resolution CEMP compilation from \citet{yoon16}, making the same cool giant cuts and adopting the carbon abundances with \citetalias{placco14} evolutionary corrections. In both the \citetalias{placco14} and \citet{yoon16} samples, almost all CEMP stars have $\cfe > +1.5$, whereas in most of the low-resolution samples the majority of the CEMP stars is actually between $+0.7 < \cfe < +1.0$ (and almost none with $\cfe > +1.5$). In most low-resolution giant samples, especially those analysed with a version of the SSPP, $\cfe > +0.7$ does not appear to be a natural division of carbon-normal and CEMP stars. 

The CEMP distributions of VMP giants in low- and high-resolution samples are very different. This could be a reflection of the high-resolution follow-up efforts over the years -- the interest was mostly in extremely metal-poor stars and very carbon-rich stars, and they may have a relative lack of more metal-rich, less carbon-rich stars. The \citet{li22} high-resolution spectroscopic follow-up LAMOST sample also does not have many very carbon-rich stars. This sample was chosen ``randomly'' from the VMP candidates sample in LAMOST \citep{aoki22}. However, even randomly selected samples can have selection effects, e.g. due to stars needing to be observable at the high-resolution facility at a given time. Additionally, the VMP candidates sample itself could already have biases against carbon-rich stars due to the analysis of the low-resolution spectra.

With the exception of the PIGS sample (which is in the inner Galaxy), all samples are supposed to be ``typical'' halo samples and there is therefore no expectation that the differences in the $\cfe$ trend or distribution of CEMP stars would have a physical origin. Sample selection effects and/or differences in spectral analysis systematics must play an important role, and they strongly affect the derived CEMP fraction. Next, we compare the analyses from two different low-resolution spectral analysis tools to asses whether some of the systematic differences we have seen here can be clarified. 

\subsection{The SSPP and FERRE}\label{sec:pipelines}

Some of the differences in \cfe between various analyses/samples may come from systematic differences in the main stellar parameters (\teff, \logg, \feh), due to degeneracies of \cfe with all of them. The other two parameters kept the same, the strength of the CH G-band decreases with increasing \teff, with decreasing \logg or with decreasing \feh. But there may also be more fundamental differences in the determination of \cfe, even if all other parameters agree. We test this in this section. 

Two commonly used codes to analyse low-resolution spectra of VMP stars are the SSPP and FERRE. In FERRE\footnote{FERRE is available from \url{http://github.com/callendeprieto/ferre}} \citep{allende06}, the four parameters \teff, \logg, \feh and \cfe are typically determined simultaneously. The application of FERRE to low-resolution spectra to derive these four parameters for VMP stars was first described in \citet{aguado17}. 
Within the SSPP, the main stellar parameters \teff, \logg and \feh are determined first \citep{lee08a,lee08b,lee13}. They are derived by computing many different spectroscopic and/or photometric estimates for each parameter and combing them to an adopted value. In a second step, \teff and \logg are fixed to the previously determined values, and only \feh and \cfe can vary in a comparison with synthetic spectra in a limited wavelength range around the G-band (between $4290-4318$~\AA). The adopted \cfe comes from this second step. 

Both methods use grids of synthetic spectra to derive stellar parameters and carbon abundances. Carbon enhancement strongly affects the structure of stellar atmospheres, resulting in differences in the synthesised stellar spectra. It is therefore important to use carbon-enhanced atmosphere models, which has been done for both the SSPP and FERRE grids, although they have been computed using different models: for SSPP the MARCS models \citep{Gustafsson08} were used and for FERRE the Kurucz models \citep{meszaros12}. The spectra have been also been synthesised using different codes, namely TURBOSPECTRUM \citep{alvarez98,plez12} for SSPP and ASSET \citep{koesterke08} for FERRE. Different line lists were used as well, where most relevant here is what has been adopted for molecular carbon features -- for SSPP those come from \citet{masseron14}, while for FERRE the Kurucz\footnote{\url{http://kurucz.harvard.edu/linelists.html}} line list was adopted. Additionally, different assumptions were made for the micro-turbulence, with a fixed value of $\xi_\mathrm{t}=2$~\kms for the FERRE grid (representative for giant stars) and adopting the relation $\xi_\mathrm{t} \, [\kms] =-0.345 \cdot \logg +2.225$ for the SSPP grid (derived from high-resolution spectra of SDSS/SEGUE stars used to calibrate the SSPP). Finally, both grids assume [$\alpha$/Fe] $=+0.4$ for very metal-poor stars, but different assumptions were made for the nitrogen abundances: in the FERRE grid [N/Fe] = 0.0 and in the SSPP grid nitrogen follows carbon, meaning that $\cfe = \nfe$ (or [C/N] = 0.0). For stars with large carbon over-abundances, this makes a significant difference. Furthermore, these assumptions do not affect the analysis of all CEMP types in the same way, since the CNO abundances are different for CEMP-s and CEMP-no stars due to differences in the nucleosynthetic processes in AGB stars and the (supernovae of) the First Stars. 

The different adopted synthetic grids may result in systematic differences in derived carbon abundances. We performed a preliminary analysis to test the magnitude of this effect, fitting spectra from the SSPP grid with FERRE and spectra from the n-SSPP grid with FERRE. Details are given in Appendix~A. In summary, we find that for stars with $\cfe < +1.5$ there can be systematics in \cfe of $\sim 0.1-0.2$~dex, depending on the \teff and the adopted pipeline. 

For n-SSPP analyses, additional empirical corrections based on a comparison with high-resolution spectroscopy are usually applied to each of the four parameters \citep{beers14}. These corrections have been determined for a mixture of stars of different temperatures, evolutionary phases and metallicities, and are only a function of the parameter itself (no cross-terms with other parameters). Conceivably, these corrections could be quite different for stars of different types. This should be tested in the future. For \cfe, the parameter of most interest in this paper, the empirical correction is the following: $\cfe_\mathrm{B14} = \cfe - (-0.068 \times \cfe + 0.273)$. It is therefore of the order of $-0.2$ to $-0.3$ dex for carbon-normal stars ($-0.5 < \cfe < +1.0$).

\subsection{Direct comparisons of SSPP and FERRE analyses}

The analysis of low-resolution spectra of cool, very carbon-rich stars is a particularly challenging task (see e.g. discussions in \citealt{beers99, rossi05, Goswami06, yoon20}). It is beyond the scope of this paper to investigate in detail the differences for very carbon-rich stars. For example, it would be important to further analyse the underlying assumptions in the synthetic spectroscopy grids used (especially the adopted CNO abundances) and/or whether photometry is included or not in the parameter determination. Instead, we focus on carbon-normal stars and stars are not very carbon-enhanced ([C/Fe] $< +1.0$) in this section, where we directly compare results from the (n-)SSPP and FERRE. 

\subsubsection{Re-analysis of SDSS, LAMOST and RAVE follow-up with FERRE}

Using the FERRE code, we re-analyse stars from three different halo samples previously analysed with the (n-)SSPP. We select random cool, carbon-normal ($\cfe < +1.0$) VMP giants in the same $\teff-\logg$ range as in Section~\ref{sec:trends} from the SDSS (700 stars, $R \sim 2000$, $3850 - 9200$~\AA), analysed with the SSPP, and the LAMOST VMP sample of \citet{yuan20} (400 stars, $R \sim 1800$, $3700-9000$~\AA), analysed with the n-SSPP. We apply a signal-to-noise (S/N) cut of S/N $> 20$. 
We additionally select all VMP giants from the optical RAVE follow-up from \citet{placco18} observed with KPNO/Mayall-RC (80 stars, $R~\sim 1500$, $3500-6000$~\AA) and ESO/NTT (160 stars, $R~\sim 1200$, $3300-5100$~\AA), analysed with the n-SSPP. The spectra in these samples typically have very high S/N, because they are the follow-up of bright stars. 
We do not apply \citetalias{placco14} evolutionary carbon corrections to any of the sample \cfe values, because here we are only interested in the comparison between direct results from the spectroscopic analyses. For the RAVE samples, \citet{placco18} applied the \citet{beers14} empirical n-SSPP corrections to the n-SSPP output.  

The synthetic spectral grid used in the FERRE analysis is an extension of the grid published in \citet{aguado17} to lower temperatures and higher metallicities, and has free parameters \teff, \logg, \feh and \cfe (first used in \citealt{arentsen20b}, see details there). We use the cubic B\'ezier interpolation in the models, and search for the best fit using the Nelder-Mead algorithm. The synthetic spectra are smoothed to the respective observed spectral resolutions, and the observed and synthetic spectra are normalised using a running mean of 30 pixels. We limit the fitted wavelength range to $3700-5500$~\AA, or the available part thereof. 

The comparison between the parameters from our FERRE analysis and the published values from the (n-)SSPP analyses is shown in Figure~\ref{fig:ferresspp} (some cross-comparisons can be found in Figure~C1 in Appendix~C), with the SDSS and \citet{yuan20}/LAMOST samples in the left-hand column in green and orange, respectively, and the two RAVE/\citet{placco18} samples in the middle column. In this work we will not investigate the origins of all the stellar parameter differences between various methods and samples in detail, but we will focus on a few key observations regarding \cfe. 

For the SDSS and RAVE/NTT samples, the agreement between the stellar parameters (\teff, \logg, \feh) from FERRE and those from the original (n-)SSPP analyses is in general excellent (except for the lower temperatures for SDSS, with $\teff \mathrm{(orig)} < 5000$~K -- this is likely a bias in the FERRE analysis, see Appendix~B.The \cfe values for these two samples are, however, systematically offset compared to the FERRE analysis, despite the main stellar parameters typically being in good agreement. For the other two samples, from LAMOST and RAVE/Mayall, the agreement for the stellar parameters is less good between the original and the FERRE analysis -- there are systematic offsets. There is also a difference in the \cfe values.

\begin{figure*}
\centering

\includegraphics[width=0.3\hsize,trim={0.0cm 0.0cm 0.0cm 0.0cm}]{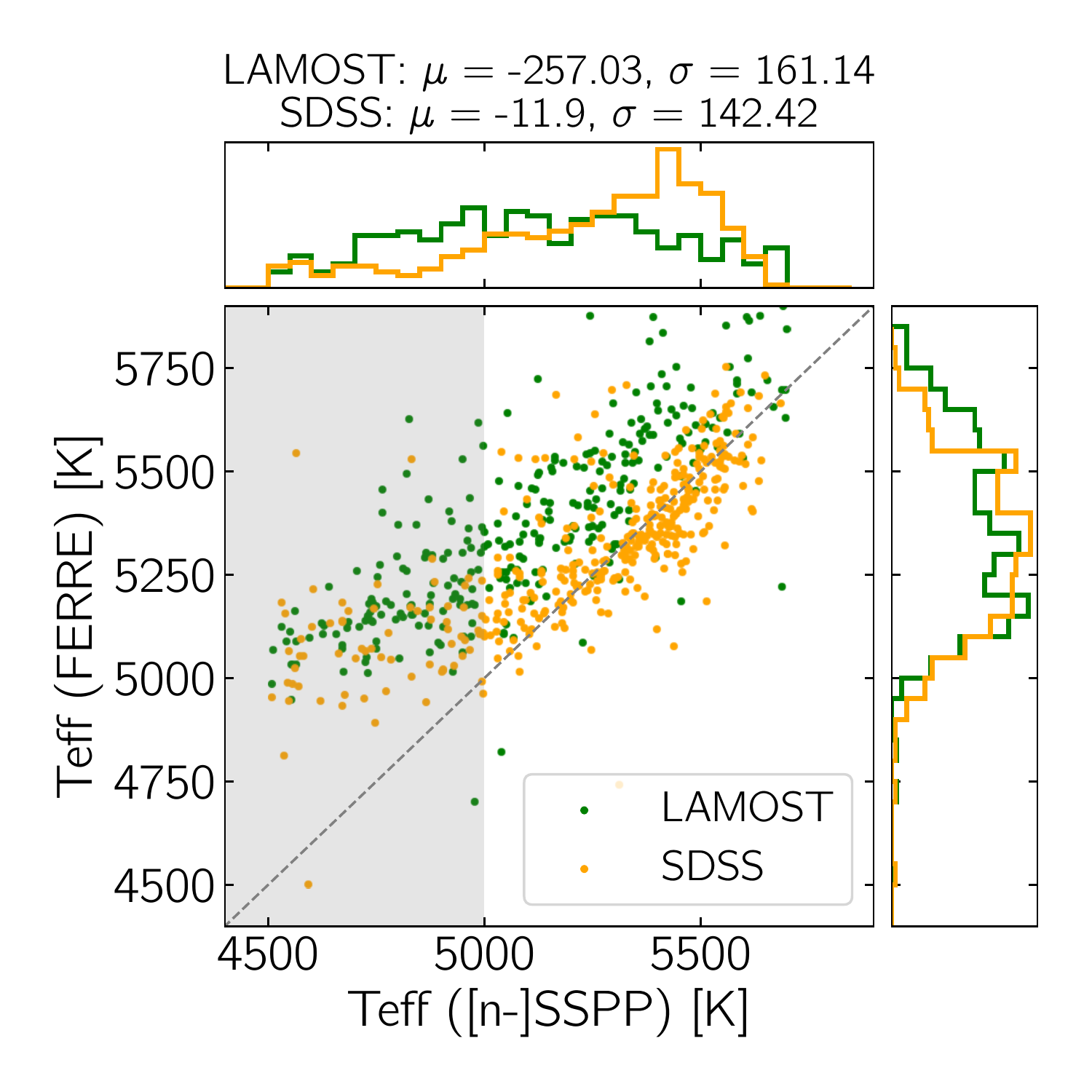}
\includegraphics[width=0.3\hsize,trim={0.0cm 0.0cm 0.0cm 0.0cm}]{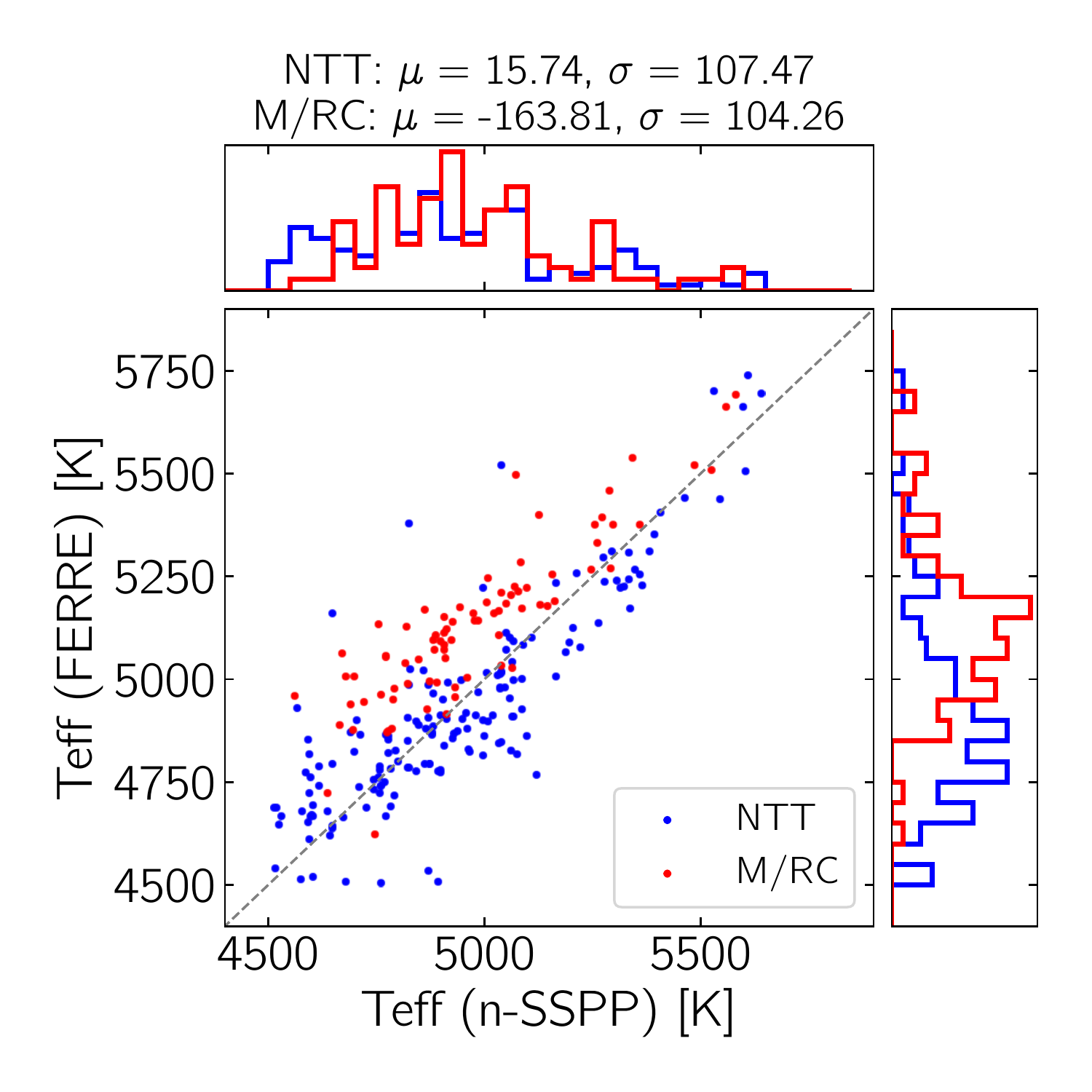}
\includegraphics[width=0.3\hsize,trim={0.0cm 0.0cm 0.0cm 0.0cm}]{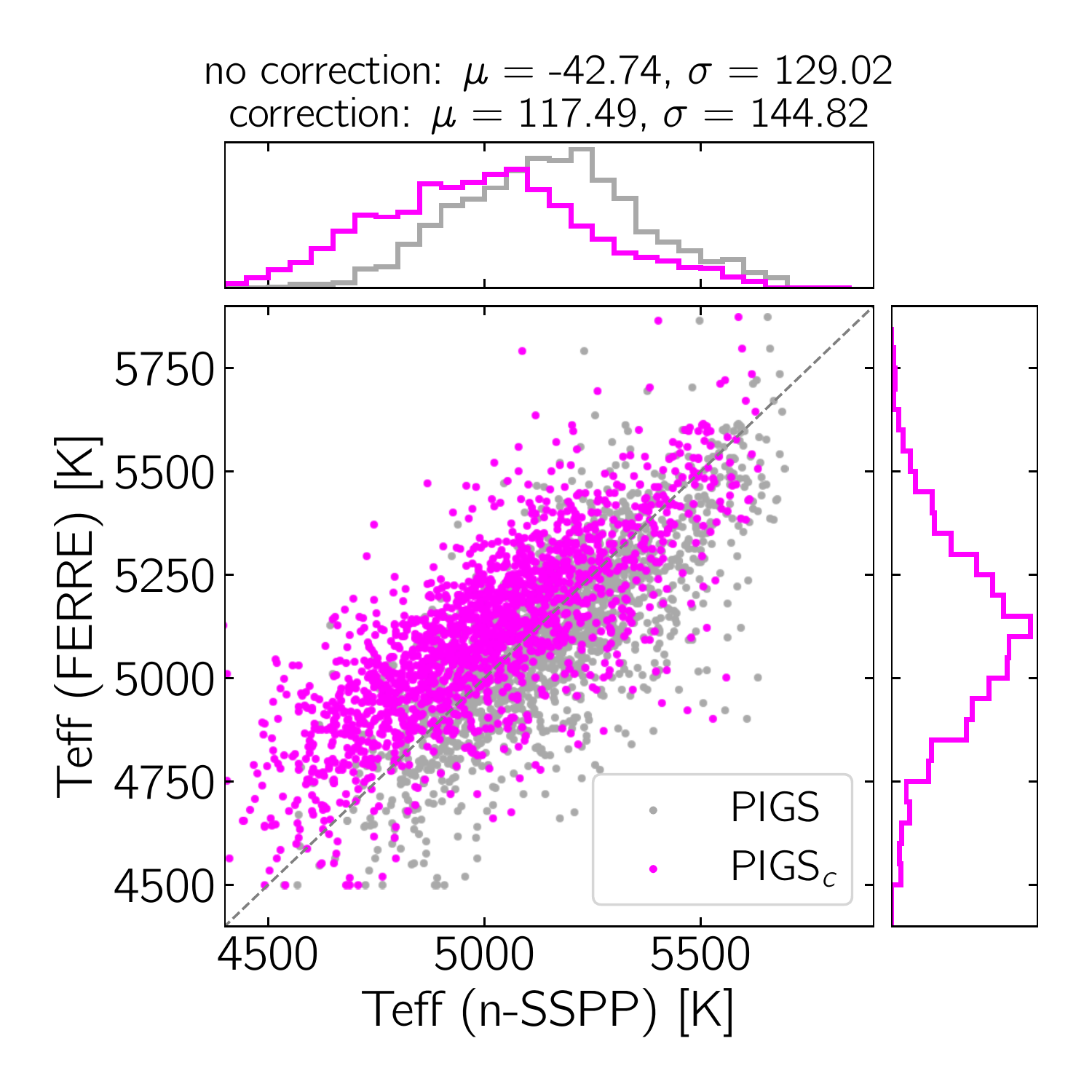}

\includegraphics[width=0.3\hsize,trim={0.0cm 0.0cm 0.0cm 0.0cm}]{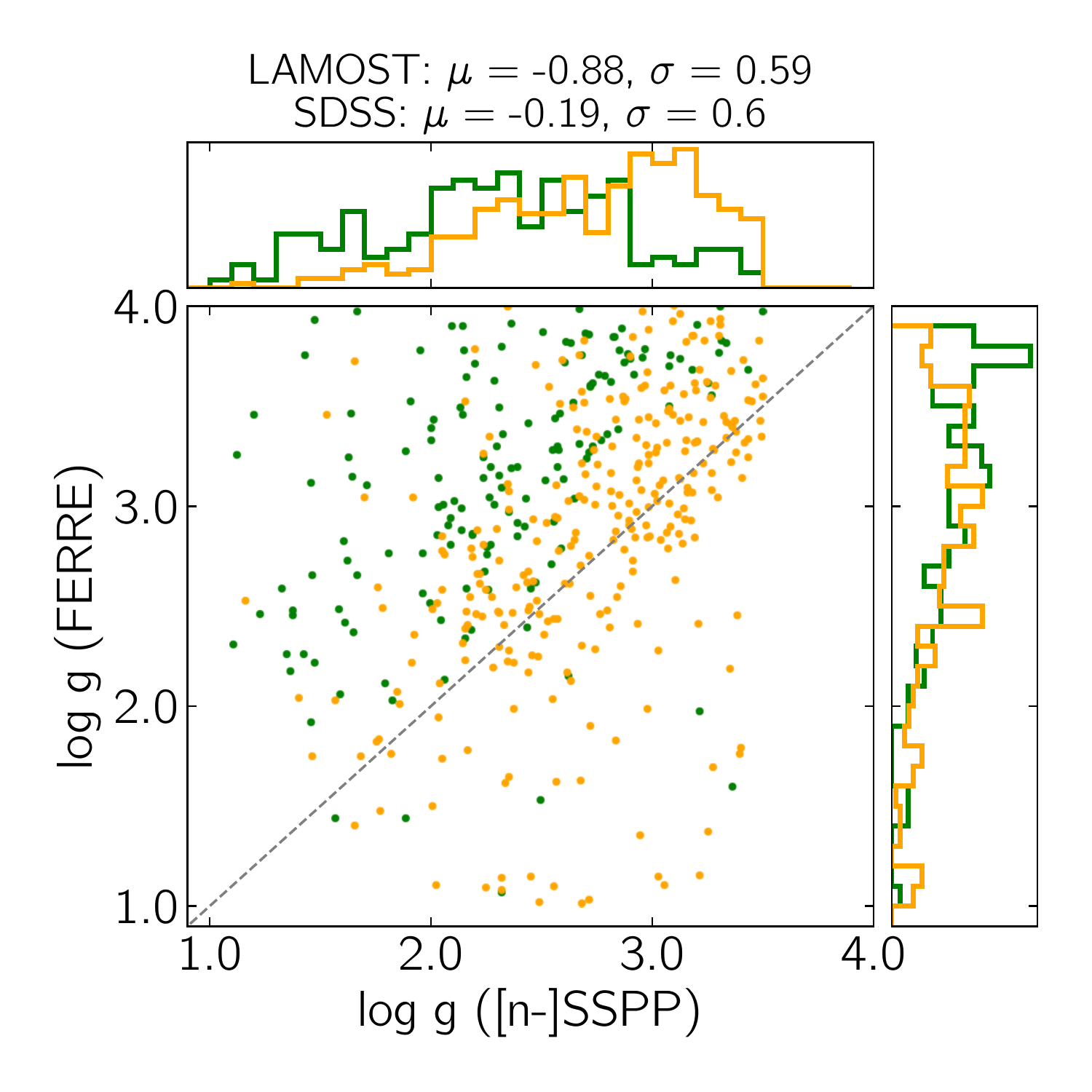}
\includegraphics[width=0.3\hsize,trim={0.0cm 0.0cm 0.0cm 0.0cm}]{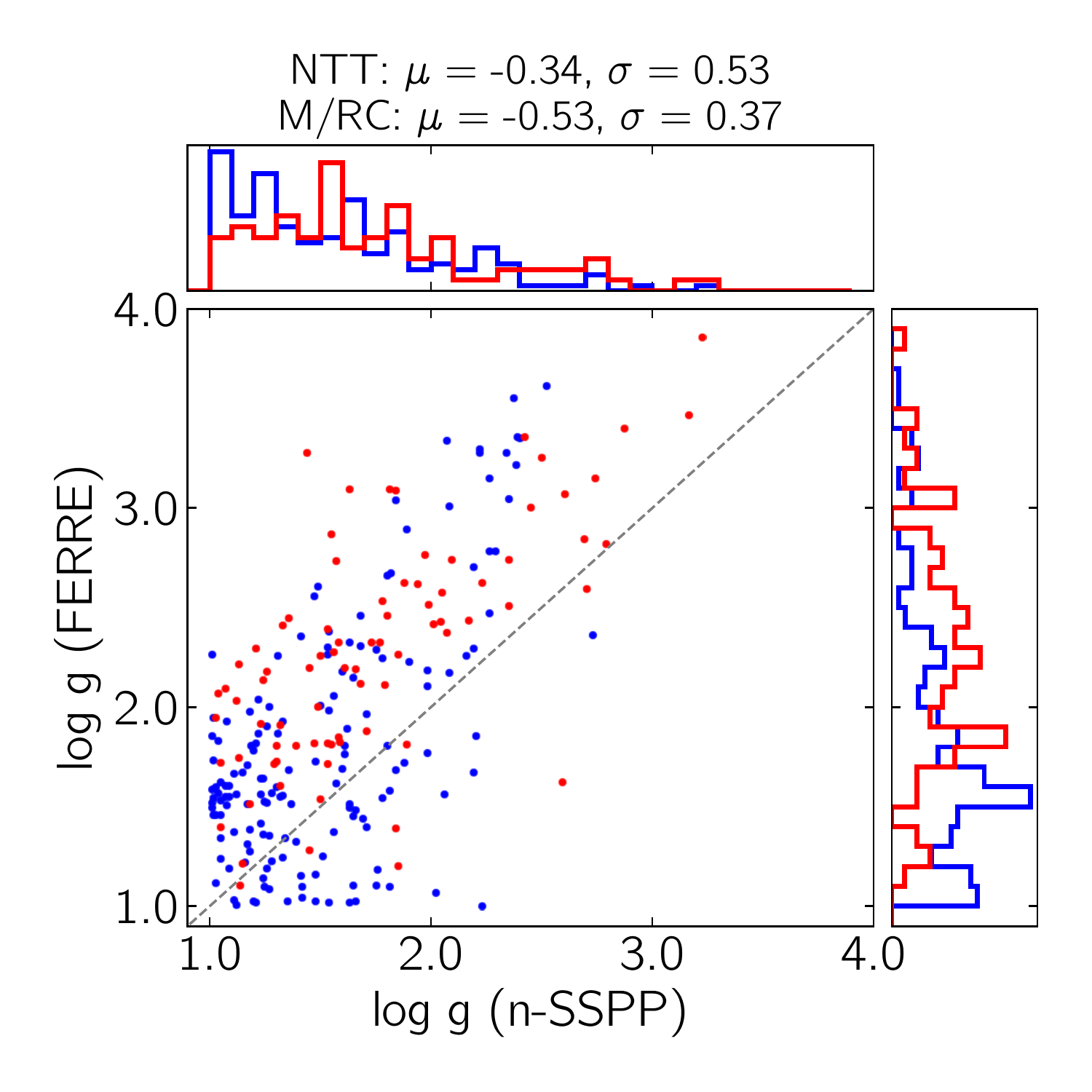}
\includegraphics[width=0.3\hsize,trim={0.0cm 0.0cm 0.0cm 0.0cm}]{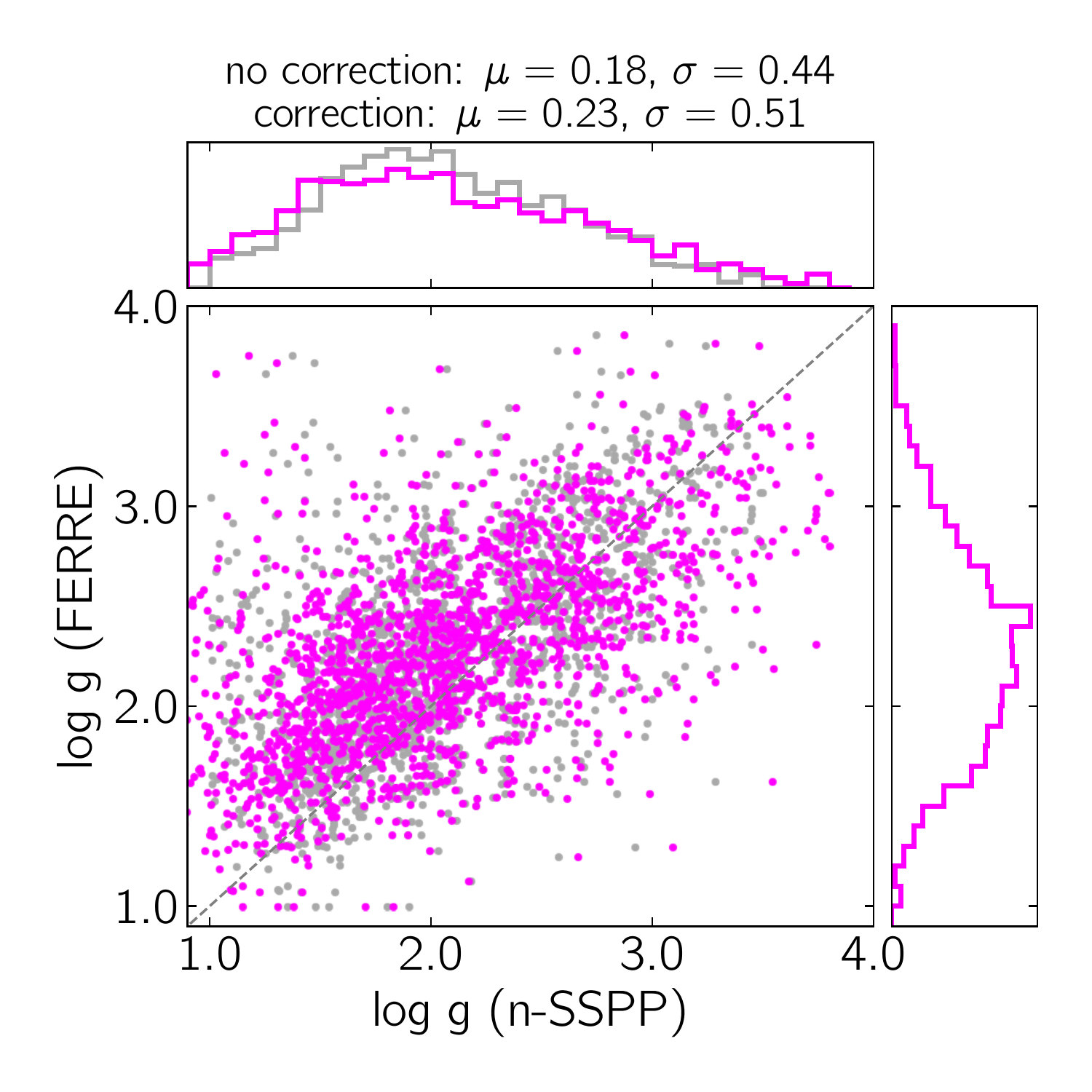}

\includegraphics[width=0.3\hsize,trim={0.0cm 0.0cm 0.0cm 0.0cm}]{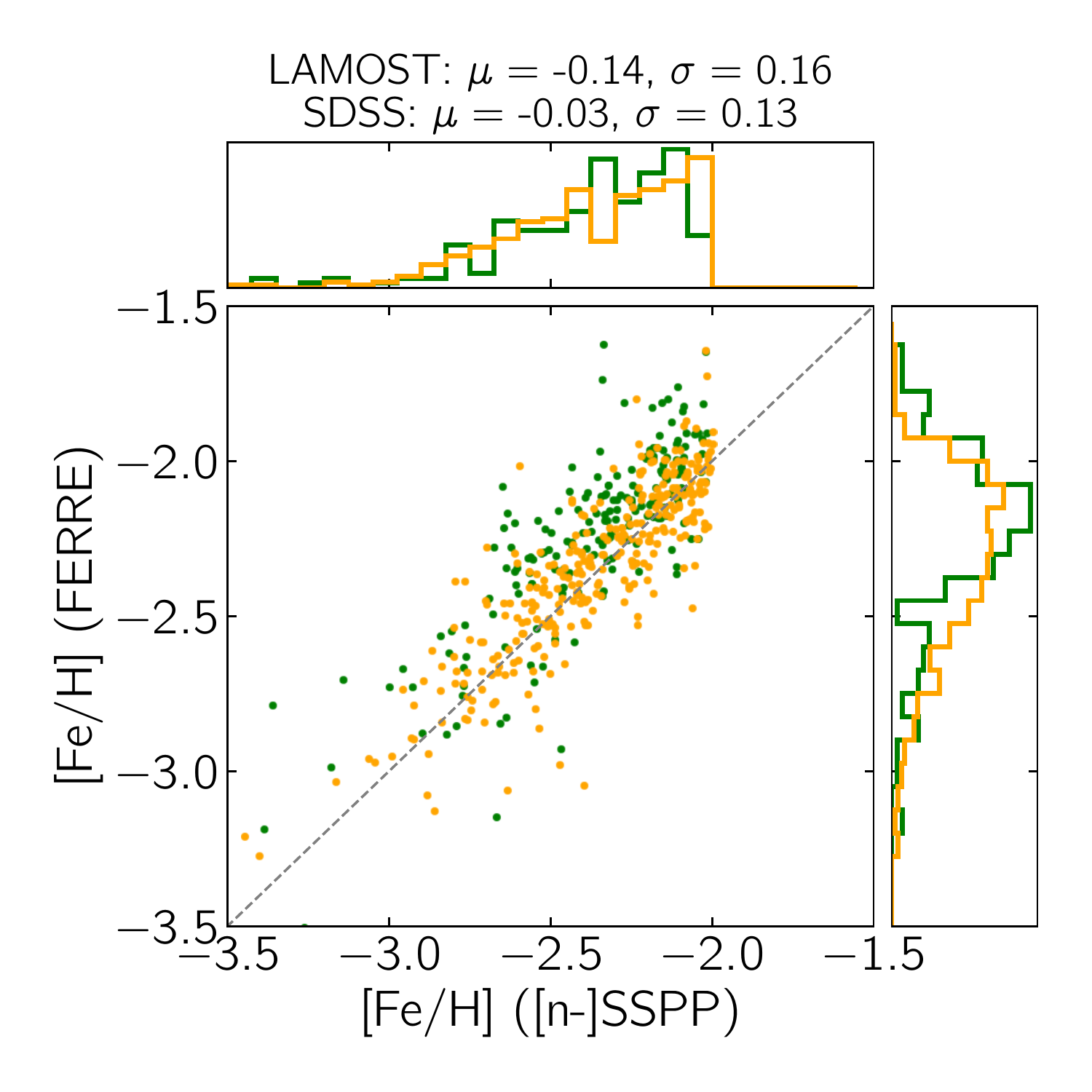}
\includegraphics[width=0.3\hsize,trim={0.0cm 0.0cm 0.0cm 0.0cm}]{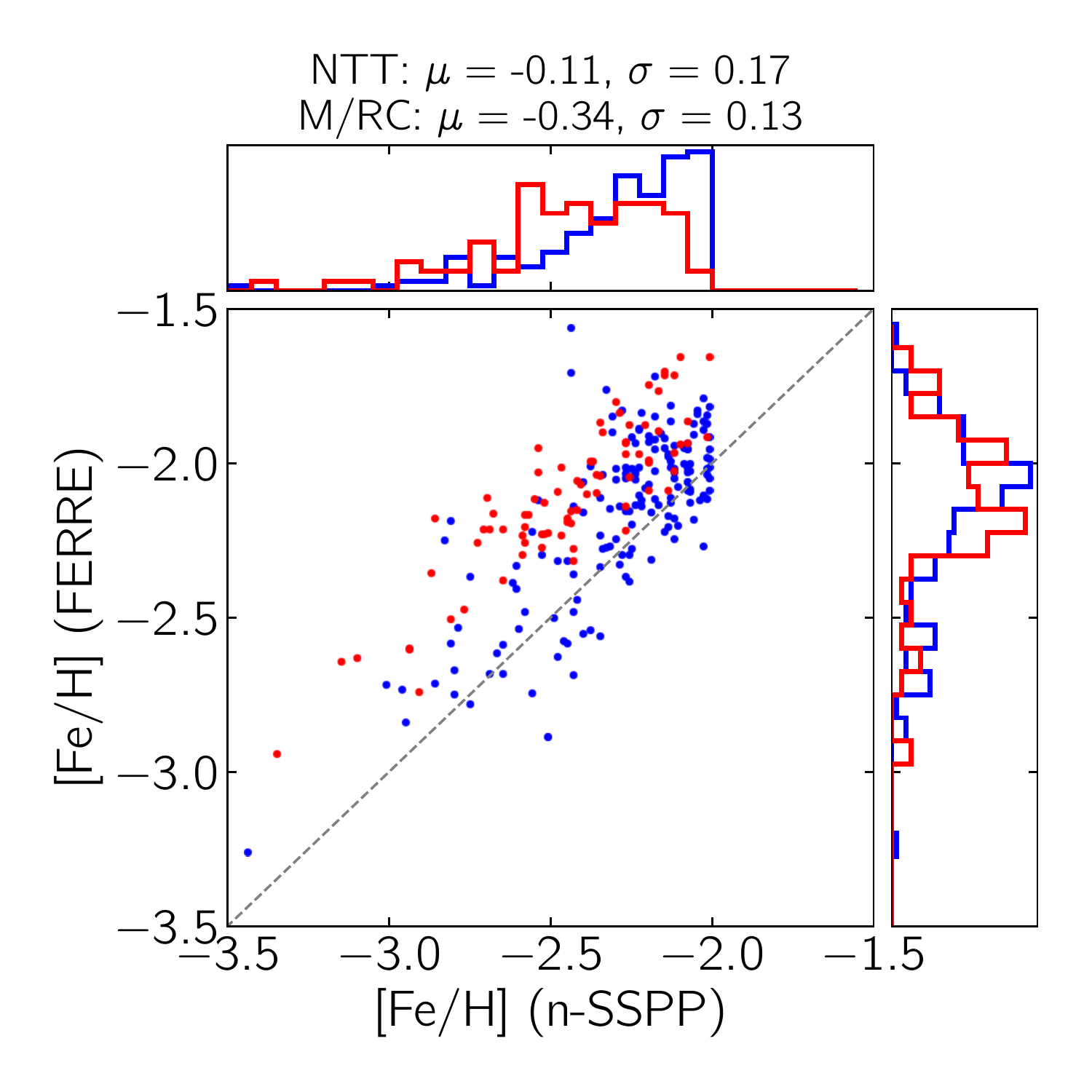}
\includegraphics[width=0.3\hsize,trim={0.0cm 0.0cm 0.0cm 0.0cm}]{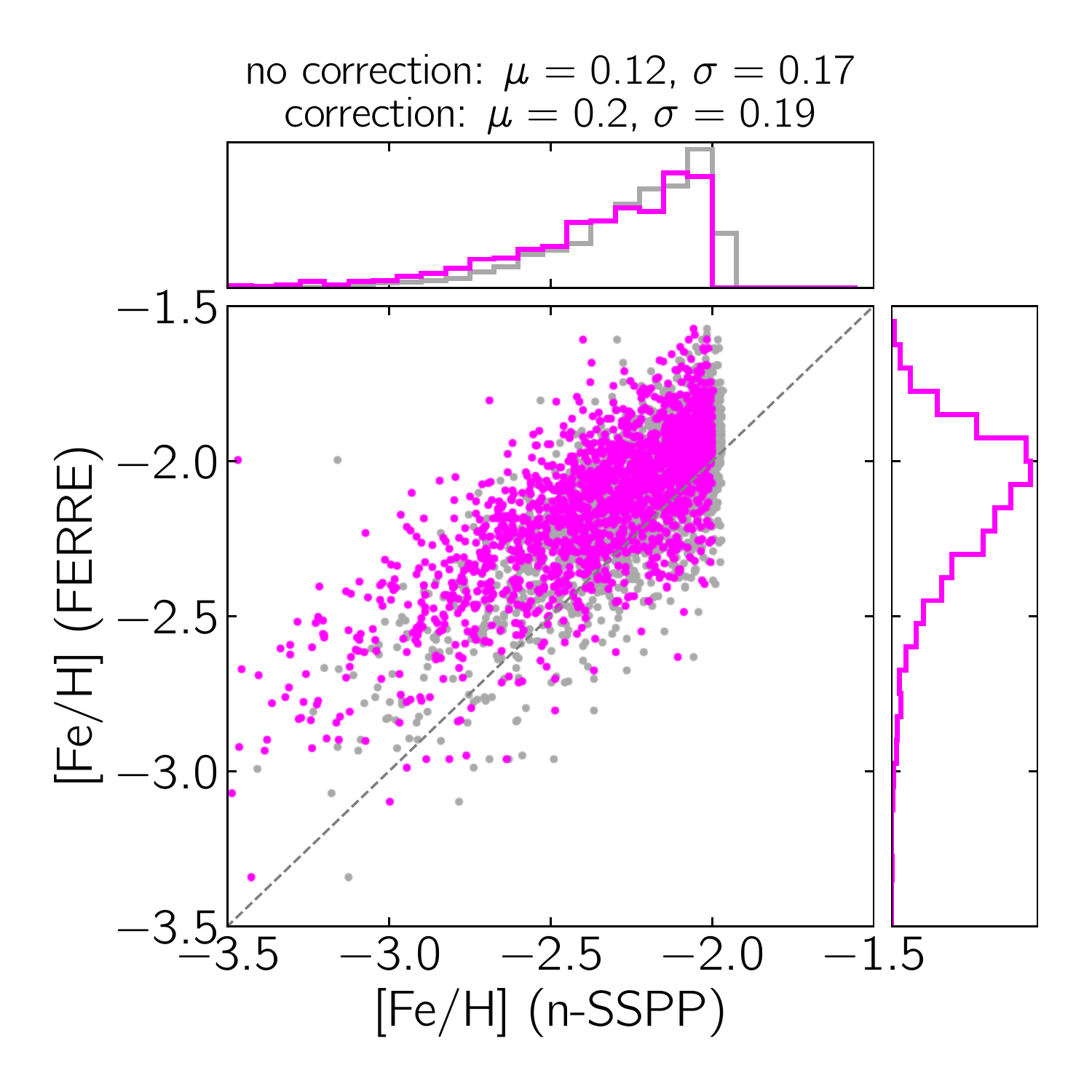}

\includegraphics[width=0.3\hsize,trim={0.0cm 0.0cm 0.0cm 0.0cm}]{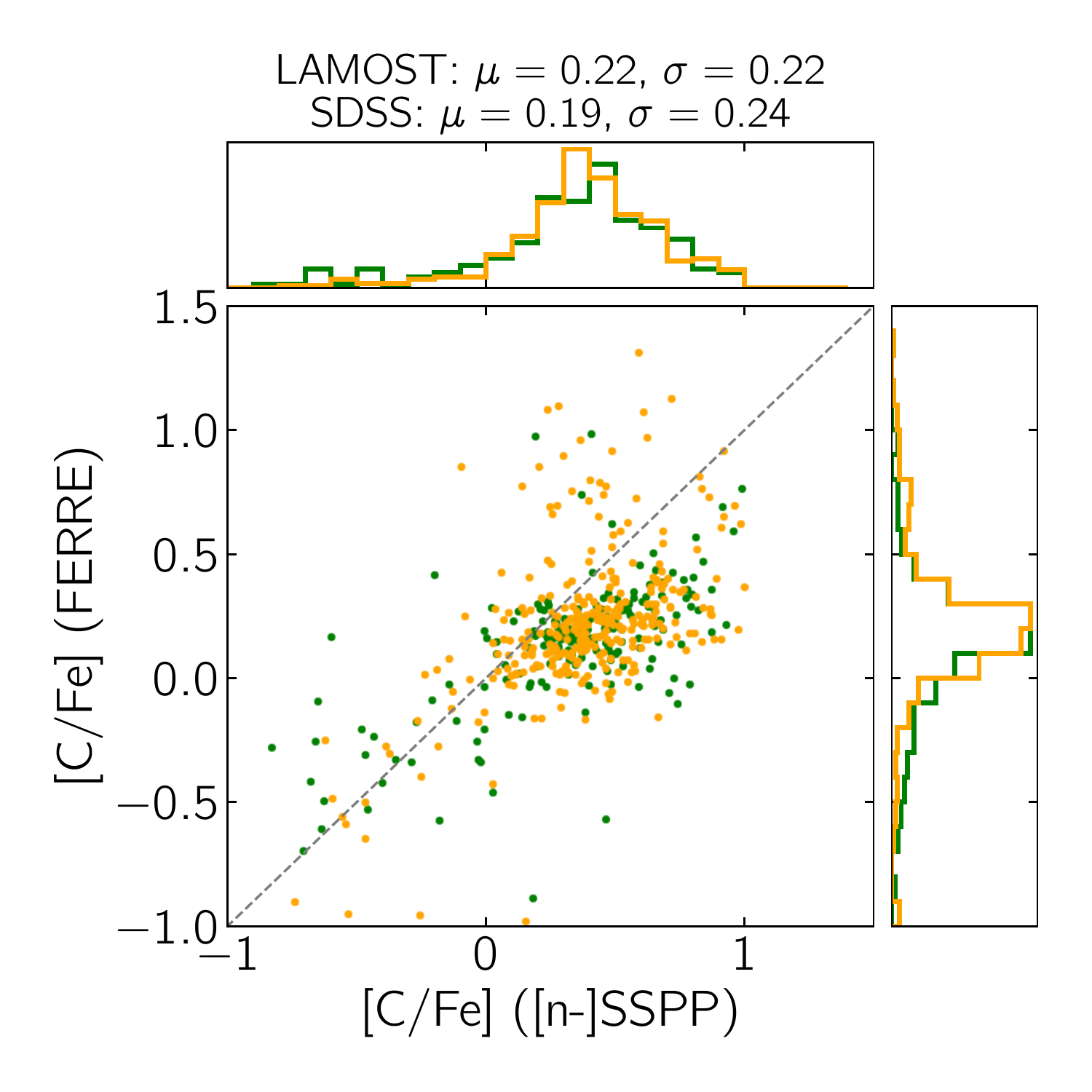}
\includegraphics[width=0.3\hsize,trim={0.0cm 0.0cm 0.0cm 0.0cm}]{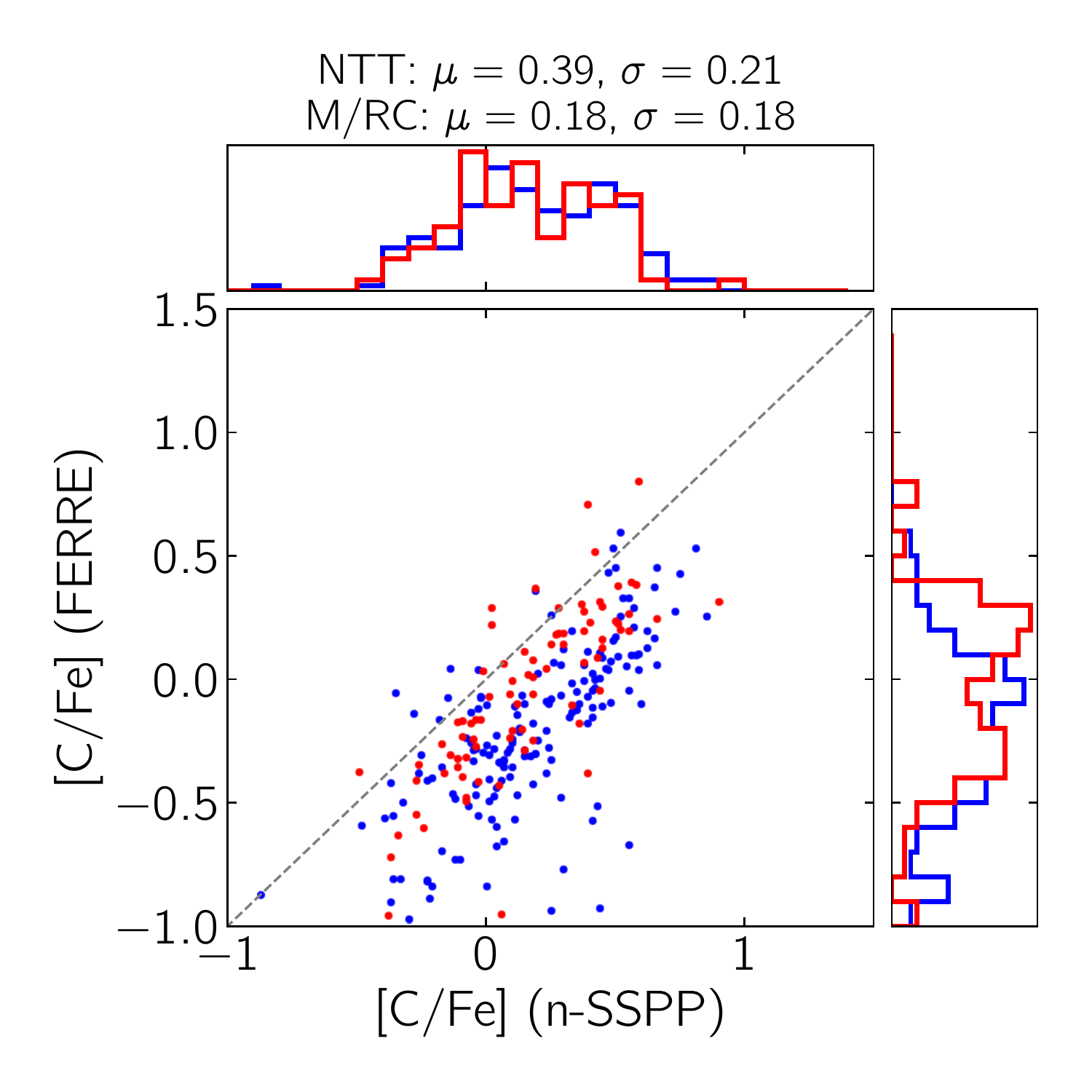}
\includegraphics[width=0.3\hsize,trim={0.0cm 0.0cm 0.0cm 0.0cm}]{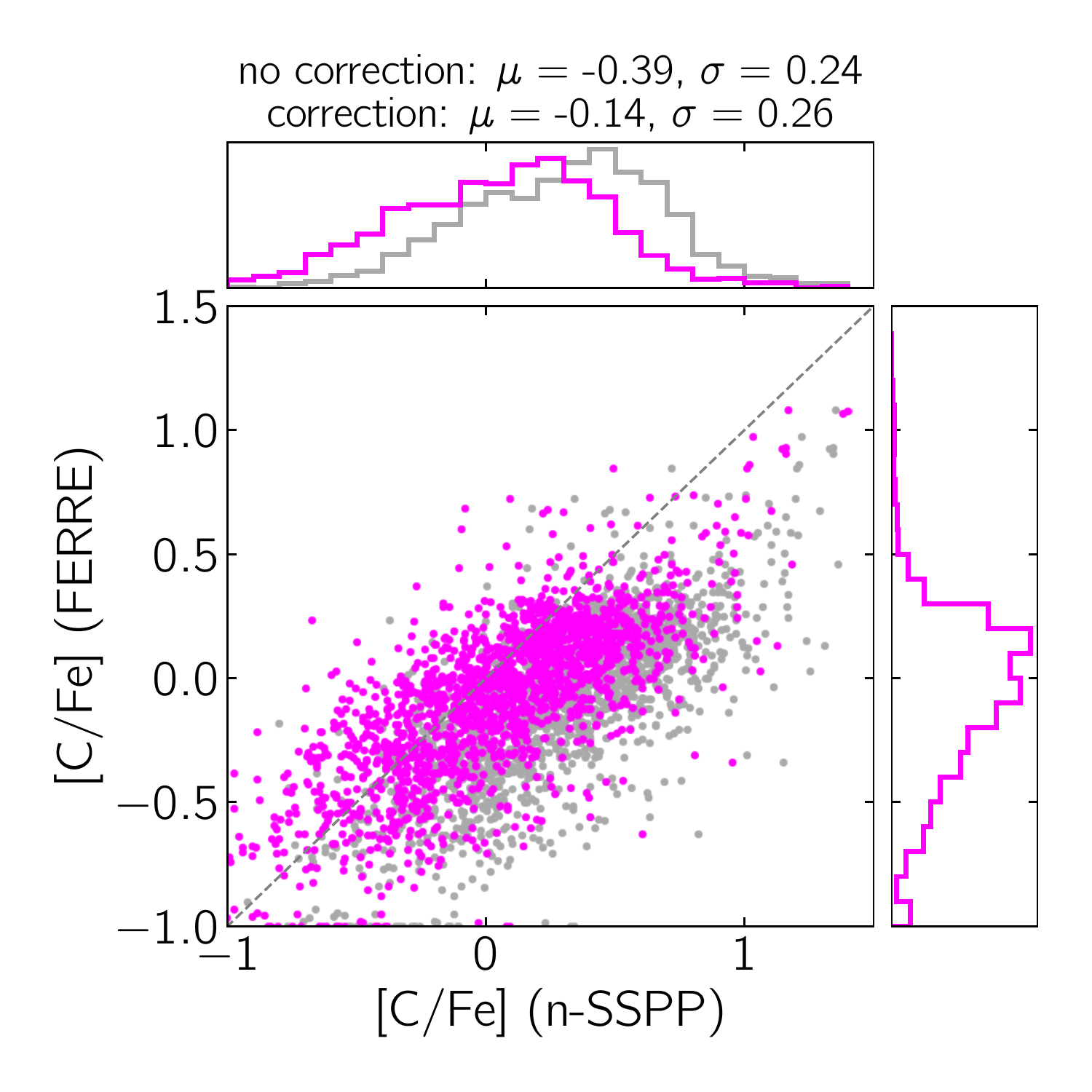}

\caption{Comparison of (n-)SSPP versus FERRE parameters, for samples from SDSS and LAMOST (left-hand column, orange and green, respectively), the RAVE follow-up by \citet{placco18} for two different instruments (middle column) and PIGS (right-hand column, magenta and grey for n-SSPP results with or without \citet{beers14} corrections applied, respectively). The median and standard deviation of $\Delta$par = FERRE $-$ (n-)SSPP are indicated in the title of each panel. For the SDSS/LAMOST panels, the stars with (n-)SSPP $ \teff < 5000$~K (the shaded region in the \teff panel) have not been included in the \logg, \feh and \cfe panels. }
    \label{fig:ferresspp}
\end{figure*}


\begin{figure}
\centering
\includegraphics[width=0.8\hsize,trim={0.0cm 0.7cm 0.0cm 0.0cm}]{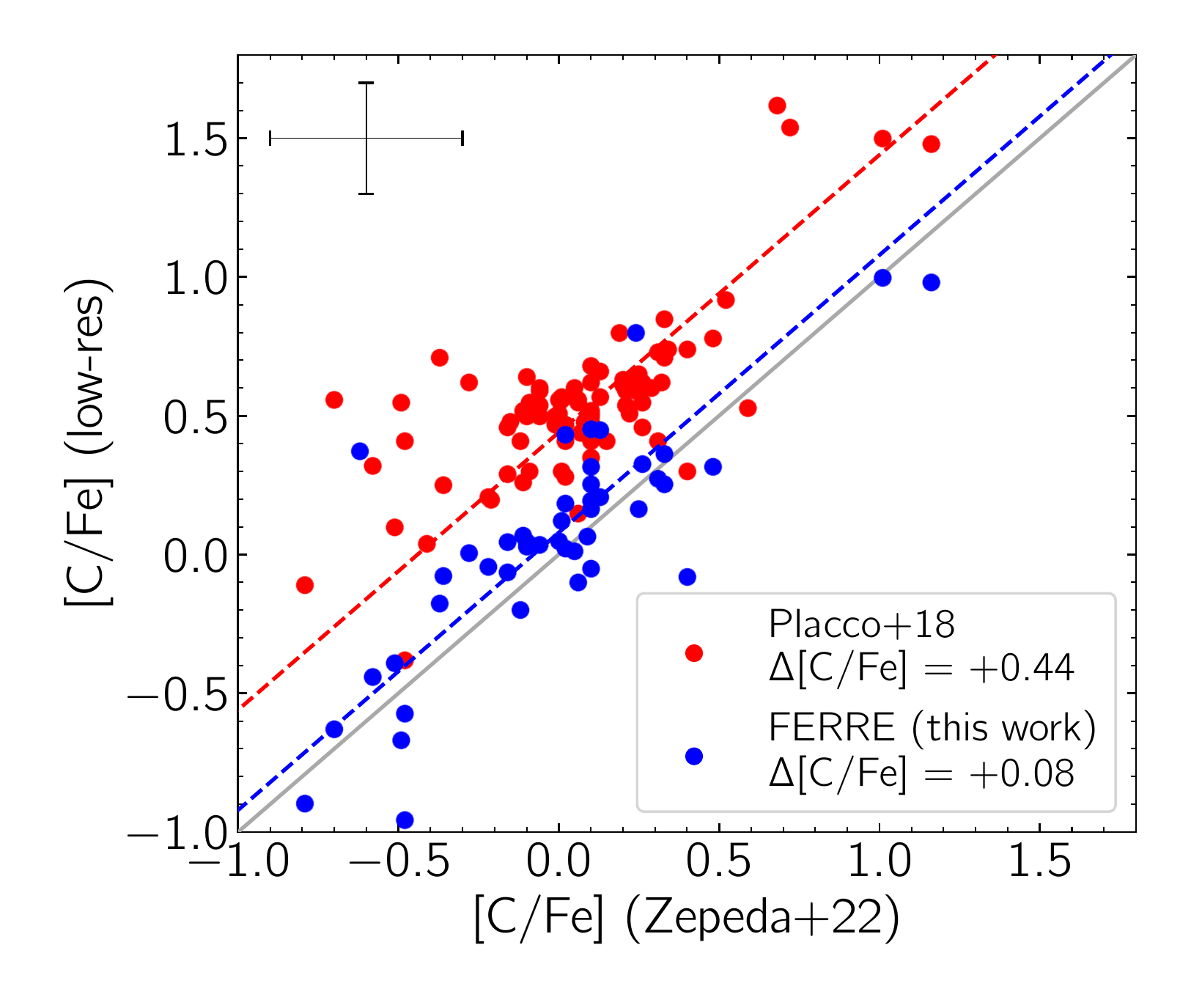}
\caption{Comparison of high-resolution \cfe from \citet{zepeda22} and low-resolution \cfe from \citet{placco18} in red and FERRE (this work) in blue. Median differences between the high and low-resolution abundances have been indicated in the legend and are represented with coloured dashed lines.
}  
    \label{fig:ravecomp}
\end{figure}

The mean $\Delta \cfe$ (original $-$ FERRE) for the SDSS, LAMOST, RAVE/Mayall and RAVE/NTT samples is $+0.19, +0.22, +0.18$ and $+0.39$, respectively. For the SDSS and LAMOST analyses the stars with $\teff$ (original) $< 5000$~K have been excluded in this comparison, because they show a clear \teff systematic in the comparison with FERRE. The differences between the (n-)SSPP and FERRE carbon abundances are of similar magnitudes as the differences seen in the \cfe trends in Figure~\ref{fig:compcfe} between the (n-)SSPP samples and the high-resolution trend. 

From Figure~\ref{fig:ferresspp} it can be seen that the difference in \cfe seems to be larger for relatively carbon-rich ($\cfe_\mathrm{(n-)SSPP} > +0.4$) stars, especially for the SDSS and LAMOST samples. Limiting the comparison to these stars, $\Delta \cfe$ (original $-$ FERRE) further increases to $+0.33$ and $+0.31$ for SDSS and LAMOST, respectively.

Some of the \cfe differences between samples may be due to stellar parameter differences. To quantify the differences in \cfe that are \emph{not} caused by the other stellar parameters, we make a comparison of stars with excellent agreement with the FERRE parameters only. For the best stars ($\Delta \teff < 50$~K, $\Delta \feh < 0.05$~dex and $\Delta \logg < 0.5$~dex), we find that there is still a systematic difference in the \cfe values in the SDSS sample. For the 55 stars that remain, we find $\Delta \cfe \mathrm{(original - FERRE)} = +0.22$. 
This goes up to $+0.35$ (based on 24 stars) for moderately carbon-rich stars ($\cfe_\mathrm{SSPP} > +0.4$). The other samples do not have enough stars with the strict comparison cuts. Increasing $\Delta \teff$ to $100$~K and $\Delta \feh$ to $0.1$~dex, there are 18 stars in the NTT sample, for which $\Delta \cfe \mathrm{(original - FERRE)} = +0.27$. 

Finally, for the RAVE follow-up sample, we compare the low-resolution [C/Fe] values directly with those from high-resolution spectroscopy from \citet{zepeda22}, see Figure~\ref{fig:ravecomp}. Most stars are from the NTT sample. There is a similarly large offset between the \citet{placco18} and high-resolution carbon abundances as before, whereas the FERRE carbon abundances more closely resemble the high-resolution results.

\subsubsection{Re-analysis of PIGS with n-SSPP}

We perform a similar exercise as above with the \Pristine Inner Galaxy Survey (PIGS) data, which have previously been analysed with FERRE \citep{arentsen20b} and which we now re-analyse with the n-SSPP. The PIGS data consist of two arms: one optical low-resolution arm ($R \sim 1300$, $3700-5500$~\AA) and one infrared medium-resolution arm ($R \sim 11\,000$, $8400-8800$~\AA). Both were used in the FERRE analysis, but here we use only the blue arm for the n-SSPP analysis, combined with 2MASS J and K photometry \citep{2mass}. Since the n-SSPP uses photometry, the results could be affected by incorrect corrections for the high extinction in the bulge region. However, since the used photometry is in the infrared, we expect the effect to be minimal. 

For each of the stellar parameters \teff, \logg and \feh, two versions are included in the n-SSPP output, namely A (adopted) and B (a bi-weight estimate). We use the adopted values, and use the difference between $\feh_A$ and $\feh_B$ as an extra quality cut. The carbon routine keeps \teff and \logg fixed to the adopted values and derives $\cfe_{car}$ and $\feh_{car}$, which together provide $\ch$. The adopted carbon abundance is $\cfe_A = \ch - \feh_A$. Finally, we apply the \citet{beers14} n-SSPP parameter corrections to the adopted parameters. 

We only consider stars passing all FERRE quality criteria \citep[see][]{arentsen20b}, and we additionally cut stars with $|(\feh_A - \feh_B)|$ $>0.2$~dex, uncertainty in J and K $> 0.15$~mag, number of \teff estimates going into $\teffa \leq 9$ or number of \feh estimates going into $\feh_A \leq 3$. These cuts are relatively strict, but leave us with a cleaner sample for the comparison. Limiting the sample to the same stellar parameter space as in the previous section, we are left with 1600 cool VMP giants. 

The comparison between the FERRE and n-SSPP analyses is shown in magenta in the right-hand column of Figure~\ref{fig:ferresspp}. We also include the parameters before applying the \citet{beers14} correction in grey. We make a few observations. First, for \cfe, the corrections appear to improve the agreement between FERRE and the n-SSPP, but they make the agreement slightly worse for \feh and significantly worse for \teff. This might be a hint that the n-SSPP corrections are not entirely applicable for giant VMP stars. It could also mean that the FERRE temperatures are off. Second, after the corrections, there are systematic differences in all parameters, and for \cfe the difference is $+0.13$. The \cfe discrepancy becomes more severe for more carbon-rich stars with $\cfe_\mathrm{SSPP} > 0.0$ and $+0.4$, with $\Delta \cfe$ going up to $+0.23$ and $+0.36$, respectively. 
Restricting the sample to stars which are in excellent agreement ($\Delta \teff < 50$~K, $\Delta \feh < 0.05$~dex and $\Delta \logg < 0.5$~dex), we find \cfe to be relatively consistent ($\Delta \cfe$ (n-SSPP $-$ FERRE) $= -0.03$, based on 60 stars), and $\Delta \cfe = +0.04$ for stars with $\cfe_\mathrm{SSPP} > 0.0$, based on 33 stars. The discrepancy in n-SSPP vs. FERRE \cfe for PIGS appears to be largely driven by the differences in the other stellar parameters, mostly the difference in metallicity. 

Since the average \cfe is higher in the n-SSPP than in the original FERRE analysis, especially for stars with n-SSPP $\cfe > +0.4$, the n-SSPP CEMP fraction in PIGS would increase compared to the published fraction in \citet{arentsen21}. However, most of those new CEMP stars would have $+0.7 < \cfe < +1.0$ and would not be clearly separate from the distribution of carbon-normal stars, whereas the CEMP stars identified in the original FERRE analysis~are.

\subsubsection{The $\cfe$ distribution of carbon-normal stars}

It can be seen in Figure~\ref{fig:ferresspp} that there are almost no stars in the different FERRE analyses with $\cfe_\mathrm{FERRE} > +0.3/+0.4$ -- there appears to be a limit in \cfe for carbon-normal stars. In the sample of VMP stars with high-resolution spectroscopy by \citetalias{placco14} there is also an upper bound to \cfe for carbon-normal stars, which is around $+0.4$ for stars with $-2.6 < \feh < -2.0$ and around $+0.7$ for stars down to $\feh = -3.0$ (see Figure~\ref{fig:compcfe}). 
In the SSPP-type analyses, the distribution is smoother and there does not appear to be a clear separation between carbon-normal and carbon-rich stars. This smoothness can also be seen in some of the (n-)SSPP samples shown in the bottom panels of Figure~\ref{fig:compcfe}. 

The smoother distributions for the (n-)SSPP samples could point to larger uncertainties in the abundances from those analyses compared to the FERRE analyses. Although the pipelines use the same spectra to derive \cfe, the (n-)SSPP only uses a very small wavelength range centred on the G-band, whereas FERRE uses a much larger wavelength range. An alternative possibility is that there could be a limitation in the FERRE code and/or synthetic grid leading to a spurious upper \cfe limit for carbon-normal stars. However, in the synthetic grid cross-analysis in Appendix~B we find no obvious discontinuities in the derivation of \cfe from FERRE for stars with $\cfe < +1.5$.

\subsection{Summary: $\cfe$ in low-resolution giant samples}

We compared the trend of \cfe as a function of \feh for carbon-normal giant stars in various low-resolution spectroscopic VMP samples, and found that there are systematic differences in \cfe between different samples analysed with different methods. Additionally, none of the samples have a CEMP star distribution similar to the CEMP stars in the high-resolution spectroscopic compilation of \citetalias{placco14} or \citet{yoon16} -- they all have a relatively low number of CEMP stars with $\cfe > +1.5$.

We re-analysed low-resolution spectra of various samples of VMP giants to be able to directly compare the parameters from different methods, namely (versions of) the SSPP and FERRE. We focussed on stars with low/intermediate carbon abundances ($\cfe < +1.0$). There are systematic differences in the determination of \cfe between FERRE and (versions of) the SSPP, of the order of $0.1-0.4$~dex (with the SSPP analyses finding systematically higher \cfe). The \cfe systematics become worse for stars with higher $\cfe_\mathrm{SSPP}$ (starting at $\cfe_\mathrm{SSPP}  > 0.0$, and worse for $> +0.4$). 

These differences are partly the result of systematic differences in the stellar parameters, because \cfe is very sensitive to \teff, \logg and \feh, but also partly \emph{independent} of degeneracies with the other stellar parameters. This is possibly the result of other differences in the analysis pipelines, such as the fitting methods and/or the synthetic spectra, as discussed briefly in Appendix~A. 

The quoted uncertainties on \cfe estimates from low-resolution spectroscopy are typically of the order of $\sim 0.2-0.3$~dex \citep[e.g.][]{lee13, arentsen21}. The offsets we have determined are smaller than or similar to these uncertainties. However, they cannot be ignored -- a systematically larger \cfe will lead to larger fractions of CEMP stars. Many of the CEMP stars in low-resolution samples have $+0.7 < \cfe < +1.0$ (see the bottom panels of Figure~\ref{fig:compcfe}), hence the effect of a small offset can be large, especially when the uncertainties on \feh and \cfe are large as well. The CEMP fraction for low-resolution samples is more sensitive to the exact CEMP definition, $\cfe > +0.7$ or the more conservative $\cfe > +1.0$, than the CEMP fraction for high-resolution samples. 


\section{Conclusions}\label{sec:discussion}

The frequencies of (different sub-classes of) CEMP stars in astrophysical different environments probe the properties of the First Stars, the early chemical evolution, and the number of interacting binary stars in those environments. In this work, we investigate whether directly comparing the CEMP fractions for various samples of giant stars from different surveys is a valid approach. 
To do so, we compile and compare the overall behaviours of different published CEMP fractions and distributions. Further, we make a direct comparison of carbon abundances from two different analysis methods, FERRE versus versions of the SSPP, for several low-resolution spectroscopic samples. Our results can be summarised as follows: 

\begin{itemize}

	\item There are various challenges when determining CEMP fractions, affecting each sample differently, such as selection effects and/or assumptions, uncertainties and biases in the spectral analysis. In particular, it is important to compare samples of stars of similar evolutionary stages to minimise such biases. 

	\item The published CEMP frequencies for different spectroscopic samples of halo giant stars do not always agree with one another, as summarised in Figure~\ref{fig:allfrac} and Table~\ref{tab:cempfrac}, although all frequencies are generally increasing with decreasing metallicity. The relative frequencies of CEMP-s and CEMP-no (or Group I and Group II+III, respectively) stars between low- and high-resolution spectroscopic samples also do not agree with each other (Figure~\ref{fig:yoon16_frac} and the bottom panels of Figure~\ref{fig:compcfe}). 
	
	\item We suspect that high-resolution spectroscopic compilations of very metal-poor stars might be biased towards (very) carbon-rich objects at the ``high'' metallicity end ($\feh > -3.0$ and especially $\feh > -2.5$) due to their follow-up strategies. This leads to an over-estimate of the CEMP fraction in this metallicity range.
	
	\item We found that there are systematic differences in \feh versus \cfe trends among various very metal-poor halo giant samples. We tested the role of the adopted pipeline for low-resolution spectroscopic samples by re-analysing stars from various surveys with the SSPP or FERRE, to compare the resulting stellar parameters and \cfe for carbon-normal stars ($\cfe < +1.0$). The \cfe from the SSPP is typically $\sim 0.15-0.2$~dex higher than the \cfe from FERRE, even after taking into account degeneracies between \cfe and other stellar parameters. Some of this may be due to the use of different synthetic grids in the two pipelines. 
	
	\item Systematic differences in \cfe can severely affect derived CEMP fractions, especially in low-resolution spectroscopic samples, which appear to have most of their CEMP stars in the range $+0.7 < \cfe < +1.0$ and have large \feh and \cfe uncertainties. This complicates the comparisons of CEMP fractions between different samples.
	
	\item Finally, we find that the frequency of CEMP stars in the low-resolution spectroscopic follow-up of the \Pristine survey is lower than previously reported in \citet{aguado19} due to a systematic effect in \logg, with a complete lack of CEMP stars with $\cfe > +1.0$ for $-3.0 < \feh < -2.0$. The low CEMP fraction and lack of very carbon-rich stars may be the result of photometric selection effects. 
	
\end{itemize}

Due to the inconsistent results in CEMP fractions from different methods and surveys, we provide the following recommendations for publishing CEMP fractions in future work:

\begin{itemize}

	\item It is more insightful to derive differential fractions (in bins of \feh) instead of cumulative fractions (all stars below a given \feh), because cumulative fractions depend on the underlying metallicity distribution function, which is not the same between different samples. 
	
	\item Apart from employing the typical CEMP definition of $\cfe > +0.7$, it would be useful to also derive the fraction of stars with very high carbon abundances (e.g. $\cfe > +1.5$), to probe differences in the carbon distributions between samples. For example, in this work we highlighted that most low-resolution spectroscopic samples lack very carbon-rich stars compared to high-resolution samples. Additionally, these very carbon-rich stars are less likely to become non-CEMP after e.g. 3D/NLTE corrections. 
	
\end{itemize}

We have already learned a lot from previous work on CEMP stars. However, due to the systematics between different large low-resolution spectroscopy samples of VMP and CEMP stars in the literature, it is still difficult to compare them directly and draw conclusions about differences or similarities between the properties of different Galactic environments. However, there are several ways to make progress in this area in the near future. These include:

\begin{itemize}
	
	\item performing more homogeneous analyses of different samples to allow for their direct comparisons (although there will always be some differences in the data, e.g. wavelength coverage and/or spectral resolution). This should likely also include a more sophisticated treatment of the uncertainties on \feh and \cfe, e.g. using Bayesian methods to derive statistical CEMP frequencies.
	
	\item deriving good estimates of selection biases in large previous, ongoing and upcoming spectroscopic surveys. This requires the selection function of surveys to be well-characterised, for which it is necessary to know how carbon-rich stars behave in various photometric bands. It would be useful to have access to carbon-enhanced stellar isochrones, for example. 
	
	\item using samples without strong selection effects due to limited follow-up, such as purely photometric surveys and/or the Gaia DR3 spectrophotometry (although these samples could still be biased in other ways). 
	
	\item making more direct comparisons of different pipelines and synthetic spectra to understand systematic differences, and improve the synthetic spectra and the assumptions that go into them. This likely should also include taking into account 3D/non-LTE effects on \feh and \cfe in the future, although this is presently computationally very expensive. 
	
\end{itemize}

Thanks to ongoing and upcoming large spectroscopic surveys such as WEAVE \citep{weave}, 4MOST \citep{4most} and DESI \citep{desi}, there will soon be much larger samples of very metal-poor stars than before, and with higher spectral resolution. These surveys will be a treasure trove for Galactic Archaeology with CEMP stars.

\section*{Acknowledgements}

	We thank Kim Venn, Lyudmila Mashonkina, Zhen Yuan, Martin Montelius, and other members of the Pristine collaboration for valuable discussions or insightful comments on a draft of this work. We thank Tadafumi Matsuno for sharing the \citet{li22} data before their final publication. 
		
	AA and NFM gratefully acknowledge funding from the European Research Council (ERC) under the European Unions Horizon 2020 research and innovation programme (grant agreement No. 834148).
	The work of VMP is supported by NOIRLab, which is managed by the Association of Universities for Research in Astronomy (AURA) under a cooperative agreement with the National Science Foundation.
	YSL acknowledges support from the National Research Foundation (NRF) of Korea grant funded by the Ministry of Science and ICT (NRF-2021R1A2C1008679).
	DSA acknowledges support from ERC Starting Grant NEFERTITI H2020/808240.
	NFM gratefully acknowledge support from the French National Research Agency (ANR) funded project ``Pristine'' (ANR-18-CE31-0017). 
	ES acknowledges funding through VIDI grant ``Pushing Galactic Archaeology to its limits'' (with project number VI.Vidi.193.093) which is funded by the Dutch Research Council (NWO).
	
	AA, DSA, NFM and ES thank the International Space Science Institute, Bern, Switzerland for providing financial support and meeting facilities to the ``Pristine'' and ``The early Milky Way'' international teams. 
	
	Based on data obtained at Siding Spring Observatory (via programs S/2017B/01, A/2018A/01, OPTICON 2018B/029 and OPTICON 2019A/045, PI: A. Arentsen and A/2020A/11, PI: D. B. Zucker). We thank the Australian Astronomical Observatory, which have made the PIGS observations possible. We acknowledge the traditional owners of the land on which the AAT stands, the Gamilaraay people, and pay our respects to elders past and present. 
	
	Horizon 2020: This project has received funding from the European Union's Horizon 2020 research and innovation programme under grant agreement No 730890. This material reflects only the authors views and the Commission is not liable for any use that may be made of the information contained therein.
	
	Based on observations at Kitt Peak National Observatory at NSF’s NOIRLab (NOIRLab Prop. ID 14A-0323, 14B-0231, 15A-0071, 15B-0071, 17A-0295; PI: V. Placco), which is managed by the Association of Universities for Research in Astronomy (AURA) under a cooperative agreement with the National Science Foundation. The authors are honored to be permitted to conduct astronomical research on Iolkam Du’ag (Kitt Peak), a mountain with particular significance to the Tohono O’odham.
	
	Based on observations collected at the European Organisation for Astronomical Research in the southern hemisphere under ESO programme(s) 095.D-0202(A), 096.D-0018(A), 097.D0196(A), 098.D-0434(A), and 099.D-0428(A).
	
	Funding for the SDSS and SDSS-II has been provided by the Alfred P. Sloan Foundation, the Participating Institutions, the National Science Foundation, the U.S. Department of Energy, the National Aeronautics and Space Administration, the Japanese Monbukagakusho, the Max Planck Society, and the Higher Education Funding Council for England. The SDSS Web Site is \url{http://www.sdss.org/}. The SDSS is managed by the Astrophysical Research Consortium for the Participating Institutions. The Participating Institutions are the American Museum of Natural History, Astrophysical Institute Potsdam, University of Basel, University of Cambridge, Case Western Reserve University, University of Chicago, Drexel University, Fermilab, the Institute for Advanced Study, the Japan Participation Group, Johns Hopkins University, the Joint Institute for Nuclear Astrophysics, the Kavli Institute for Particle Astrophysics and Cosmology, the Korean Scientist Group, the Chinese Academy of Sciences (LAMOST), Los Alamos National Laboratory, the Max-Planck-Institute for Astronomy (MPIA), the Max-Planck-Institute for Astrophysics (MPA), New Mexico State University, Ohio State University, University of Pittsburgh, University of Portsmouth, Princeton University, the United States Naval Observatory, and the University of Washington.
	
	Guoshoujing Telescope (the Large Sky Area Multi-Object Fiber Spectroscopic Telescope LAMOST) is a National Major Scientific Project built by the Chinese Academy of Sciences. Funding for the project has been provided by the National Development and Reform Commission. LAMOST is operated and managed by the National Astronomical Observatories, Chinese Academy of Sciences. 
	
	This work has made use of data from the European Space Agency (ESA) mission {\it Gaia} (\url{https://www.cosmos.esa.int/gaia}), processed by the {\it Gaia} Data Processing and Analysis Consortium (DPAC, \url{https://www.cosmos.esa.int/web/gaia/dpac/consortium}). Funding for the DPAC has been provided by national institutions, in particular the institutions participating in the {\it Gaia} Multilateral Agreement.
	
	This research has made use of the \textsc{matplotlib} \citep{hunter07}, \textsc{numpy} \citep{numpy}, \textsc{pandas} \citep{pandas}, \textsc{dustmaps} \citep{dustmaps} and \textsc{astropy} \citep{astropy13,astropy18} Python packages and of \textsc{Topcat} \citep{topcat}. We also used the spectroscopic analysis pipelines n-SSPP \citep{beers14} and FERRE \citep{allendeprieto14}.

\section*{Data Availability}

The SSPP analysis of the SDSS spectra with \cfe values and the \citet{placco18} spectra will be shared on reasonable request to Y. S. Lee and V. M. Placco, respectively. All other data underlying this article that is not public will be shared on reasonable request to A. Arentsen.


\bibliographystyle{mnras}
\bibliography{carbonbias.bib}   

\begin{landscape}
\begin{table}
\caption {Compilation of CEMP fractions for halo samples from the literature (samples for dwarf galaxies, globular clusters and stellar streams are not included)} \label{tab:cempfrac} 
\tabcolsep=0.11cm
\renewcommand{\arraystretch}{1.3}
\begin{tabular}{@{}| llllll | ccccc | lcccccl | l |@{}}
\toprule
 &  &  &  &  &  & \multicolumn{5}{c}{{[}C/Fe{]} \textgreater +1.0 \citep{beerschristlieb05}} &  & \multicolumn{5}{c}{{[}C/Fe{]} \textgreater +0.7 \citep{aoki07}} &  &  \\ \cmidrule(lr){6-18}
\multirow{-2}{*}{Reference} & \multirow{-2}{*}{Sample} & \multirow{-2}{*}{Type} & \multirow{-2}{*}{LR analysis} & \multirow{-2}{*}{Stellar type} 
&  & \textless $-$3.5 & \textless $-$3.0 & \textless $-$2.5 & \textless $-$2.0 & \textless $-$1.5 
&  & \textless $-$3.5 & \textless $-$3.0 & \textless $-$2.5 & \textless $-$2.0 & \textless $-$1.5 &  & \multirow{-2}{*}{evol. cor. ?} \\ \cmidrule(r){1-19} 

\citet{norris97} & HK & LR & by eye & mix &  &  &  & 14 &  &  &  &  &  &  &  &  &  & n \\
\citet{cohen05} & HES & LR/HR &  & giants &  &  &  &  & 14 $\pm$ 4 &  &  &  &  &  &  &  &  & n \\
\citet{frebel06} & HES & LR & Rossi+05 & giants &  &  & 25 $\pm$ 11 & 13 $\pm$ 4 & 9 $\pm$ 2 &  &  &  &  &  &  &  &  & n \\
\citet{lucatello06} & HES & HR &  & mix &  &  &  &  & 21 $\pm$ 2 &  &  &  &  &  & 24 $\pm$ 3 &  &  & n/y (a)  \\
 &  &  &  & unevolved &  &  &  &  & 21 $\pm$ 5 &  &  &  &  &  &  &  &  & - \\
\citet{carollo12} & SDSS & LR & SSPP & mix &  &  &  &  &  &  &  &  &  & 20 & 12 & 8 &  & n (b) \\
\citet{Aoki13} & SDSS & HR &  & giants &  &  &  &  &  &  &  &  &  & 36 &  &  &  & y (c) \\
 & SDSS & HR &  & turn-off &  &  &  &  &  &  &  &  &  & \textgreater{}9 &  &  &  & - \\
\citet{yong13} & HES+lit & HR &  & mix &  &  & 23 $\pm$ 6 &  &  &  &  &  & 32 $\pm$ 8 &  &  &  &  & y (c) \\
 &  &  &  & dwarfs &  &  &  &  &  &  &  &  & 50 $\pm$ 31 &  &  &  &  & - \\
 &  &  &  & giants &  &  &  &  &  &  &  &  & 39 $\pm$ 11 &  &  &  &  & y (c) \\
\citet{lee13} & SDSS & LR & SSPP & all &  & 41 $\pm$ 10 & 22 $\pm$ 3 & 15 $\pm$ 1 & 8 $\pm$ 1 & 4 $\pm$ 1 &  & 43 $\pm$ 11 & 28 $\pm$ 3 & 21 $\pm$ 1 & 12 $\pm$ 1 & 8 $\pm$ 1 &  & n \\
&  &  &  & giants &  & & & & & &  & 33 $\pm$ 11 & 31 $\pm$ 4 & 32 $\pm$ 2 & 19 $\pm$ 1 & 11 $\pm$ 1 &  & n \\
&  &  &  & dwarfs &  & & & & & &  & 100 $\pm$ 50 & 75 $\pm$ 22 & 15 $\pm$ 1 & 3 $\pm$ 1 & 1 $\pm$ 1 &  & - \\
&  &  &  & turn-off &  & & & & & &  & 50 $\pm$ 29 & 17 $\pm$ 4 & 10 $\pm$ 1 & 10 $\pm$ 1 & 8 $\pm$ 1 &  & - \\
\citet{placco14} & Lit & HR &  & mix &  & 51 & 38 & 28 & 27 &  &  & 60 & 48 & 34 & 33 &  &  & y (d) \\
\citet{beers17} & HES & LR & n-SSPP & mix &  &  &  &  &  &  &  &  & 39 $\pm$ 15 & 24 $\pm$ 6 & 19 $\pm$ 4 &  &  & n (e) \\
\citet{yoon18} & AEGIS & LR & n-SSPP & (sub)giants &  &  &  &  &  &  &  & 78$_{-7}^{+6}$ & 64 $\pm$ 3 & 42 $\pm$ 2 & 26.5 $\pm$ 0.8 & 13.6 $\pm$ 0.4 &  & y (d) \\
\citet{placco18} & RAVE & LR & n-SSPP & giants &  &  &  &  &  &  &  &  & 57$^{+13}_{-14}$ & 37 & 23 $\pm$ 3 &  &  & y (d) \\
\citet{placco19} & RPA/B\&B & LR & n-SSPP & giants &  &  &  &  &  &  &  &  & 47$^{+22}_{-21}$ & 31$^{+9}_{-7}$ & 22$^{+5}_{-4}$ &  &  & y (d) \\
\citet{limberg21} & B\&B & LR & n-SSPP & giants &  &  &  &  &  &  &  &  & 43$^{+16}_{-15}$ & 32 $\pm$ 6 & 19 $\pm$ 3 &  &  & y (d) \\
\citet{whitten21} (f) & S-PLUS & phot & SSPP-trained & K-dwarfs &  &  &  &  &  &  &  &  & 60 & 45 & 25 & 20 &  & - \\
\citet{li22} & LAMOST & HR &  & giants &  &  &  &  &  &  &  &  & &  & 7.8$^{+2.2}_{-1.8}$ & &  & y (c) \\
& & &  & turn-off &  &  &  &  & 11 &  &  &  & & & 31 $\pm$ 4 (g) & &  & - \\
\cmidrule(r){1-19} 

\multicolumn{19}{l}{Follow-up of VMP candidates from narrow-band photometric surveys}   \\

\cmidrule(r){1-19} 
\citet{howes16} (h) & SkyMapper & HR &  & giants (bulge) &  &  &  &  &  &  &  &  & 44$^{+16}_{-15}$ & 18$^{+10}_{-7}$ & 20$^{+8}_{-6}$  &  &  & y (d) \\
\citet{aguado19} (i) & Pristine & LR & FERRE & turn-off &  &  &  &  &  &  &  &  &  & &  41$^{+12}_{-11}$ & &  & - \\
 &  &  &  & giants &  &  &  &  &  &  &  &  &  38$^{+18}_{-15}$ & & 2$^{+2}_{-1}$ & &  & n (b) \\
\citet{caffau20} & Pristine & LR & MyGIsFOS  & mix &  &  &  &  &  &  &  &  & 31$^{+14}_{-11}$  & & 17$^{+7}_{-5}$  &  &  & n \\
\citet{arentsen21} & PIGS & LR & FERRE  & giants (bulge) &  & & 33.3$^{+14.4}_{-11.9}$ & 8.8$^{+2.4}_{-1.9}$ & 2.1$^{+0.4}_{-0.3}$ &  &  &  & 41.7$^{+14.3}_{-13.0}$ & 16.4$^{+3.0}_{-2.6}$ & 5.7$^{+0.6}_{-0.5}$  &  &  & y (d) \\
\citet{lucchesi22} & Pristine & HR &  & unevolved &  &  &  &  &  &  &  &  &  & & 60$^{+10}_{-11}$ &  &  & - \\
 &  &  &  & giants &  &  &  &  &  &  &  &  &  & & 11$^{+10}_{-5}$  &  &  & n \\
\citet{yong21} (j) & SkyMapper & HR &  & mix &  &  &  &  &  &  &  & 67$^{+13}_{-17}$ & 29$^{+6}_{-5}$ & 20$^{+4}_{-4}$ & &  &  & y (d) \\

\bottomrule

\end{tabular}

\vspace{0.1cm}

\noindent 
In general, if the absolute numbers of stars were given in a paper, we estimated uncertainties using binomial statistics. \\
\textbf{(a)} no for 1.0, yes for 0.7 using the \citet{aoki07} luminosity definition
\textbf{(b)} but few evolved stars
\textbf{(c)} using the \citet{aoki07} luminosity definition
\textbf{(d)} using the \citetalias{placco14} corrections
\textbf{(e)} if \citetalias{placco14} corrections are applied the number of CEMP stars goes up from 52 to 57
\textbf{(f)} estimated from their Figure~13
\textbf{(g)} the lower limit is 22 $\pm$ 3 \%, assuming all stars without detected CH bands are not carbon-enhanced
\textbf{(h)} as updated in \citet{arentsen21}
\textbf{(i)} as updated in this work, see Section~\ref{sec:pris}
\textbf{(j)} computed from their published table, only taking into account stars without upper limits on \cfe \\ 

\end{table}
\end{landscape}


\appendix

\section{Cross-analysis of the FERRE and SSPP synthetic grids}\label{sec:ap_cross}

The systematic differences in \cfe between the (n-)SSPP and FERRE analyses (even when the other stellar parameters are consistent) could come from different approaches to deriving carbon abundances in these codes, and/or from the use of different synthetic grids. FERRE fits the full spectrum and uses a running mean normalisation, whereas the (n)-SSPP only fits a small region in the G-band (between $4290-4318$~\AA) and normalises using a pseudo-continuum determined over the $4000-4650$~\AA range. The synthetic grids have different assumptions (see Section~3.2), those that are most likely to cause discrepancies in the derived \cfe are the adopted line lists for carbon features, the adopted nitrogen abundances, and possibly the difference in adopted micro-turbulence velocities. 

A preliminary look at the line lists shows that there do appear to be differences in the strengths of the CH lines between the Kurucz and Masseron line lists, with the Kurucz lines being stronger (M. Montelius, private communication). Regarding the assumptions on nitrogen abundances, they do not only have a direct effect on the CN features, but also indirectly affect e.g. the G-band and other CH/C$_2$ features due to the contribution of nitrogen to the overall metallicity of the stellar atmosphere. It is beyond the scope of this work, which is focused on the observations, to investigate detailed differences in the synthetic grids. Instead, we investigate the effects of adopting different grids and/or analysis methods for the derivation of \cfe. 

We performed a cross-analysis of the two synthetic grids: we analysed spectra from the SSPP grid with FERRE, and spectra from the FERRE grid with the n-SSPP. When fitting the SSPP spectra with FERRE, we keep \teff, \logg and \feh fixed to the true values from the SSPP grid, and only fit for \cfe. We fit the spectra between $3900 - 5500$~\AA. When running the n-SSPP carbon routine on the FERRE spectra, we keep \teff and \logg fixed to the true values from the FERRE grid, deriving [C/H]. Both grids were smoothed to $\mathrm{R}=1000$, which is the resolution used in the n-SSPP.

\begin{figure*}
\centering
\includegraphics[width=0.49\hsize,trim={0.0cm 0.0cm 0.0cm 0.0cm}]{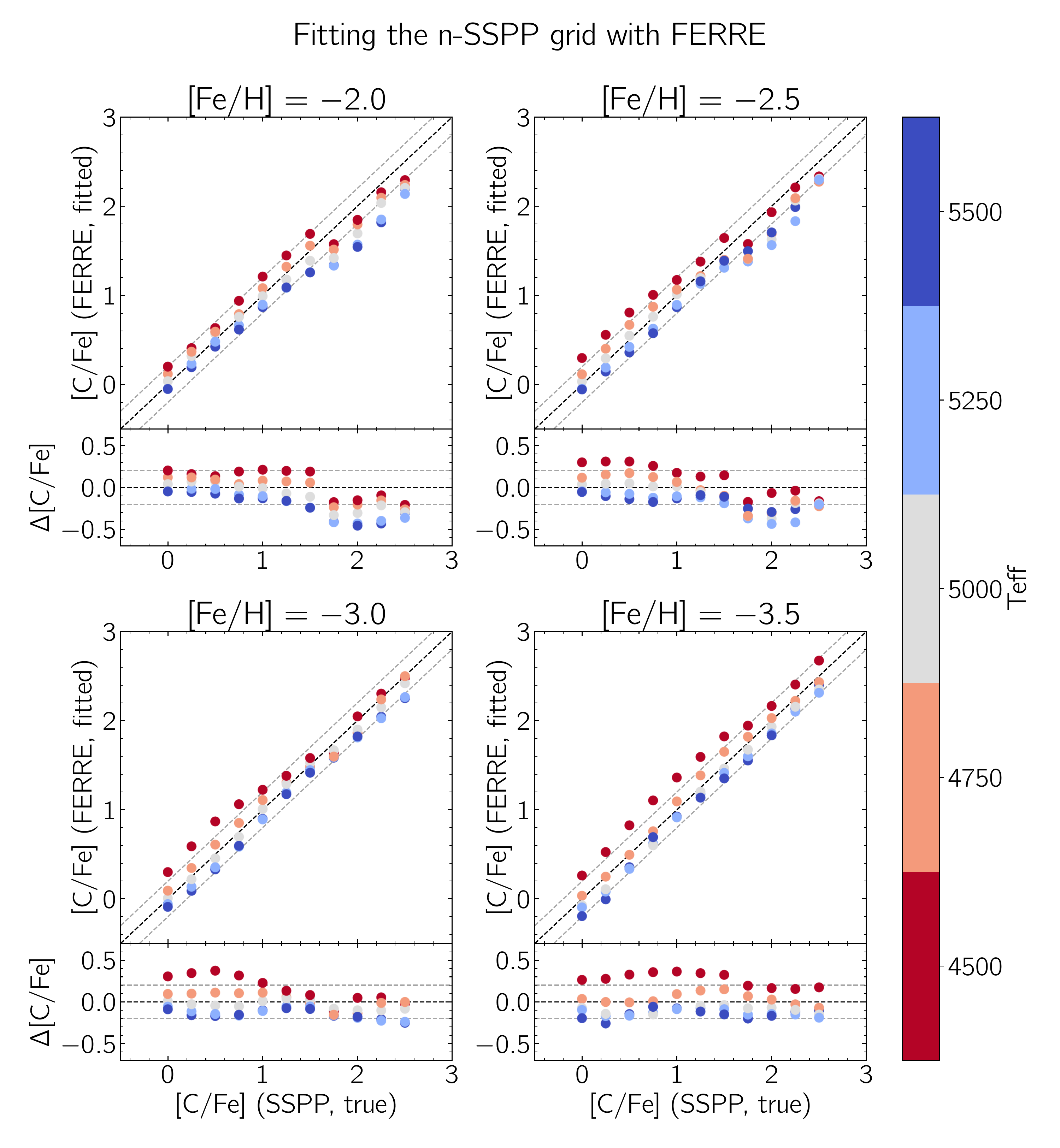}
\includegraphics[width=0.49\hsize,trim={0.0cm 0.0cm 0.0cm 0.0cm}]{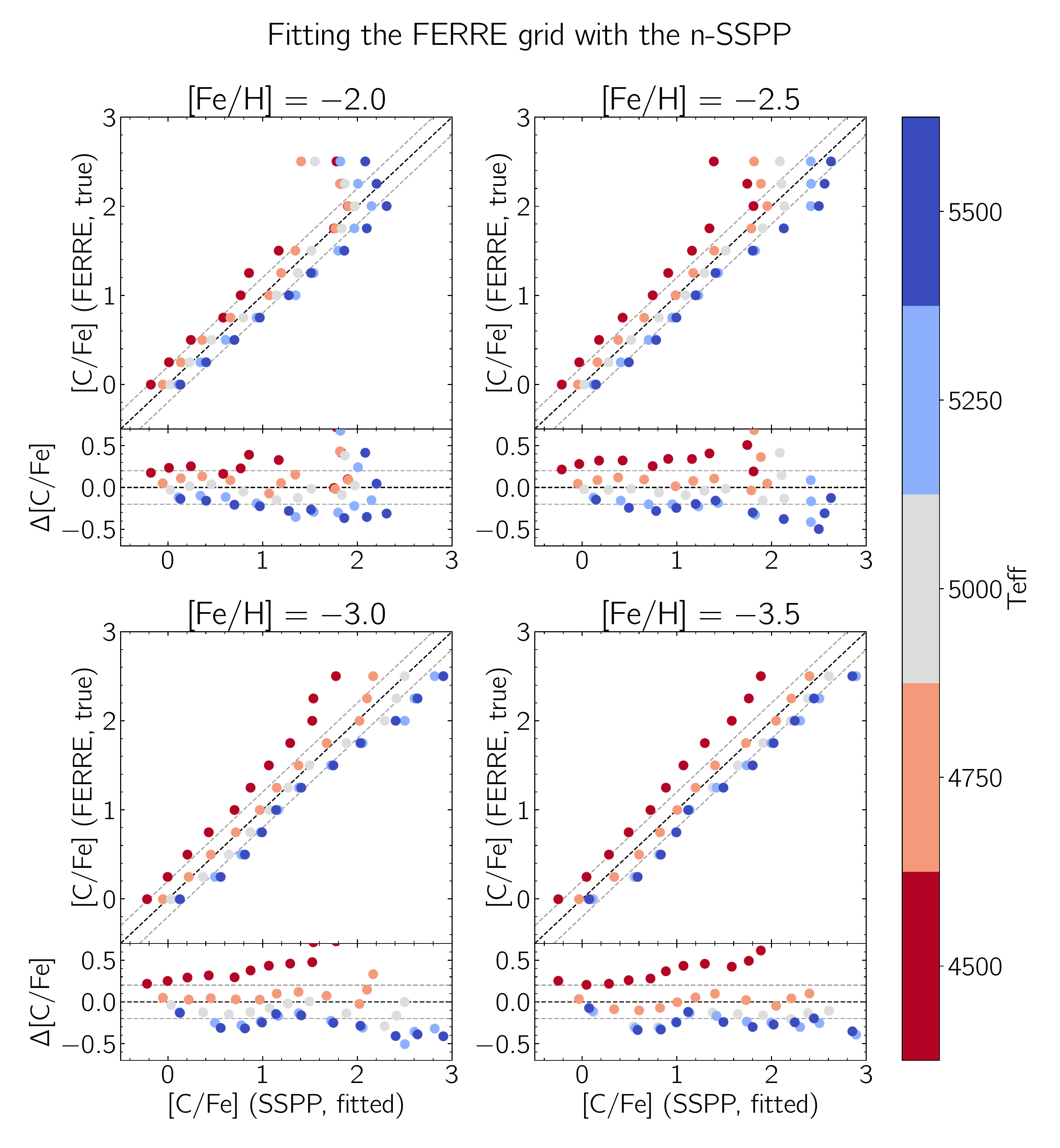}
\caption{Results of the cross-analysis of the FERRE and n-SSPP synthetic grids as function of \cfe, \feh (different panels) and \teff (colour). The left-hand side shows the results of the n-SSPP analysis of the FERRE grid, the right-hand side the FERRE analysis of the n-SSPP grid. The black-dashed line is the one-to-one line, they grey-dashed lines are $\pm 0.2$~dex. The $\Delta \cfe$ in the bottom panels is always FERRE $-$ SSPP. 
}  
    \label{fig:syntcomp}
\end{figure*}

The results are shown in Figure~\ref{fig:syntcomp}, with on the left-hand side the FERRE analysis of the SSPP grid and on the right-hand side the n-SSPP analysis of the FERRE grid, separated by \feh in different panels and colour-coded by \teff. For $\cfe < +1.5$, the average difference (combining all \teff) is close to zero for all metallicities, but there are systematics that vary with the effective temperature, of the order of $\sim 0.1-0.2$~dex. The effect is similar in both analyses: the cool stars typically have higher FERRE \cfe and the warmer stars typically have lower FERRE \cfe. This appears to be a characteristic of the synthetic grids. However, the magnitude of this effect is larger in the n-SSPP analysis, possibly because it is fitting a much smaller wavelength range. 

For the very carbon-rich stars with $\cfe > +1.5$, other systematics start to appear, especially for the more metal-rich cases ($\feh = -2.0$ and $-2.5$). In the n-SSPP grid, the very carbon-rich stars are also strongly enhanced in nitrogen ($\cfe = \nfe$), whereas the in the FERRE grid nitrogen is fixed with respect to iron ($\nfe = 0$). This results in large differences in the spectra, which the different approaches of the n-SSPP and FERRE appear to respond to differently. For very carbon-rich stars, the different normalisation schemes may also introduce differences in the derived \cfe. 

\section{Photometric temperatures}\label{sec:ap_phot}

\begin{figure*}
\centering
\includegraphics[width=0.7\hsize,trim={0.0cm 1.0cm 0.0cm 0.0cm}]{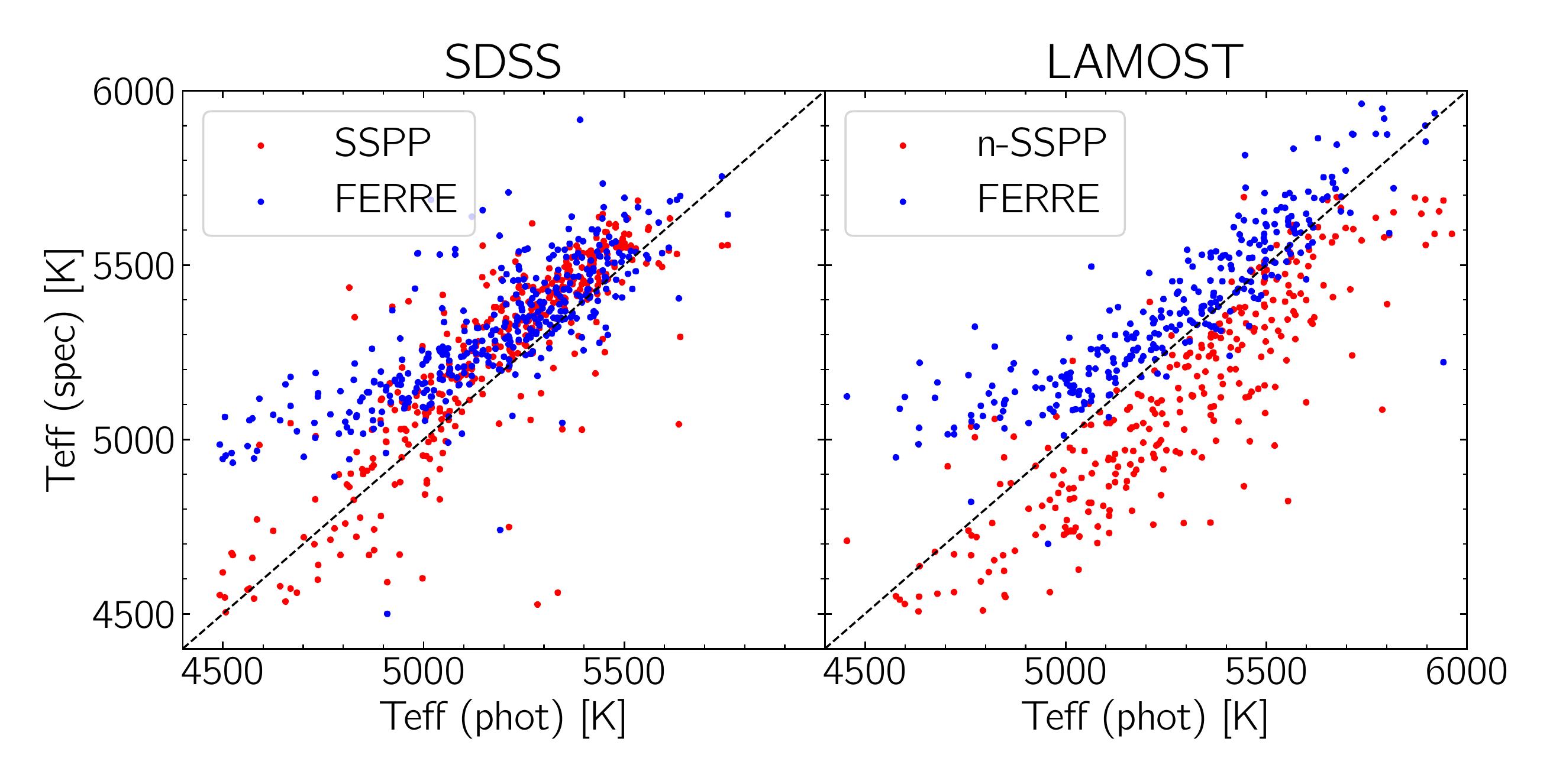}
\caption{Comparison of spectroscopic and photometric effective temperatures for the (n-)SSPP and FERRE analyses of the SDSS and LAMOST samples.}  
    \label{fig:teffphot}
\end{figure*}

The FERRE \teff showed a systematic offset compared to the SSPP \teff values for the coolest stars in the SDSS and LAMOST samples. Here we compute photometric temperatures and compare them to the spectroscopic values to identify which of the two analyses is at the root of this offset. To compute photometric \teff estimates, we use the photometric temperature relation from \citet{casagrande21} applied to \Gaia EDR3 (BP$-$RP) photometry \citep{gaia16,gaiaedr3}, which includes terms that depend on \logg and \feh. The \logg dependence is small, and we adopt $\logg=2.0$ for all stars. For \feh we adopt the SSPP values, but changing to the FERRE values does not affect the results. We deredden the photometry adopting E(B$-$V) values from the Schlegel extinction map \citep{schlegel98, schlafly11}, using the colour-dependent extinction coefficients for the BP and RP photometry from \citet{casagrande21}. The resulting comparison between spectroscopic and photometric effective temperatures is shown in Figure~\ref{fig:teffphot}, for the SDSS sample on the left and the LAMOST sample on the right, with the SSPP results in red and the FERRE results in blue. The issue with the cool stars appears to be an issue in the FERRE analysis, as the same systematic offset is visible. 

This bias was very puzzling at first, because it is not seen for the RAVE/NTT spectra where the \teff is well-determined even for the coolest stars. We then checked the metallicities for the $\teff < 5000$~K stars in the different samples, and found that in this temperature range, the SDSS and LAMOST stars are clearly more metal-poor than those in the NTT sample. The more metal-poor a star is, the fewer spectral lines are present, and the harder it is to estimate \teff and \logg from spectra alone. This is particularly the case for the coolest stars (e.g. $\teff < 5000$~K), where the \teff/\logg-sensitive Balmer lines become very weak. Many of the spectral lines in VMP stars are in the bluest part of the spectrum, and an additional disadvantage of the SDSS spectra is that they do not go bluer than 3820~\AA. 

The FERRE analysis is purely spectroscopic, whereas the SSPP analyses also include photometry in the stellar atmospheric parameter determination. This is likely the reason that the FERRE \teff determinations for cool, very metal-poor stars are less good than the SSPP determinations. 

We finally note that the systematic behaviour between the spectroscopic and photometric temperatures is the same for both FERRE analyses of the SDSS and LAMOST spectra, but that it is different for the two SSPP analyses. The scatter in the photometric vs. SSPP \teff also increases for the LAMOST sample. We already saw larger discrepancies between the other stellar parameters (\logg, \feh) in the FERRE and SSPP analyses of the LAMOST spectra (see the left-hand column of Figure~5). These may be indications of larger systematic issues in the n-SSPP. 

\section{Cross-comparison of the parameters from FERRE and n-SSPP}\label{sec:ap_crossFS}

\begin{figure*}
\centering

\includegraphics[width=0.3\hsize,trim={0.0cm 0.0cm 0.0cm 0.0cm}]{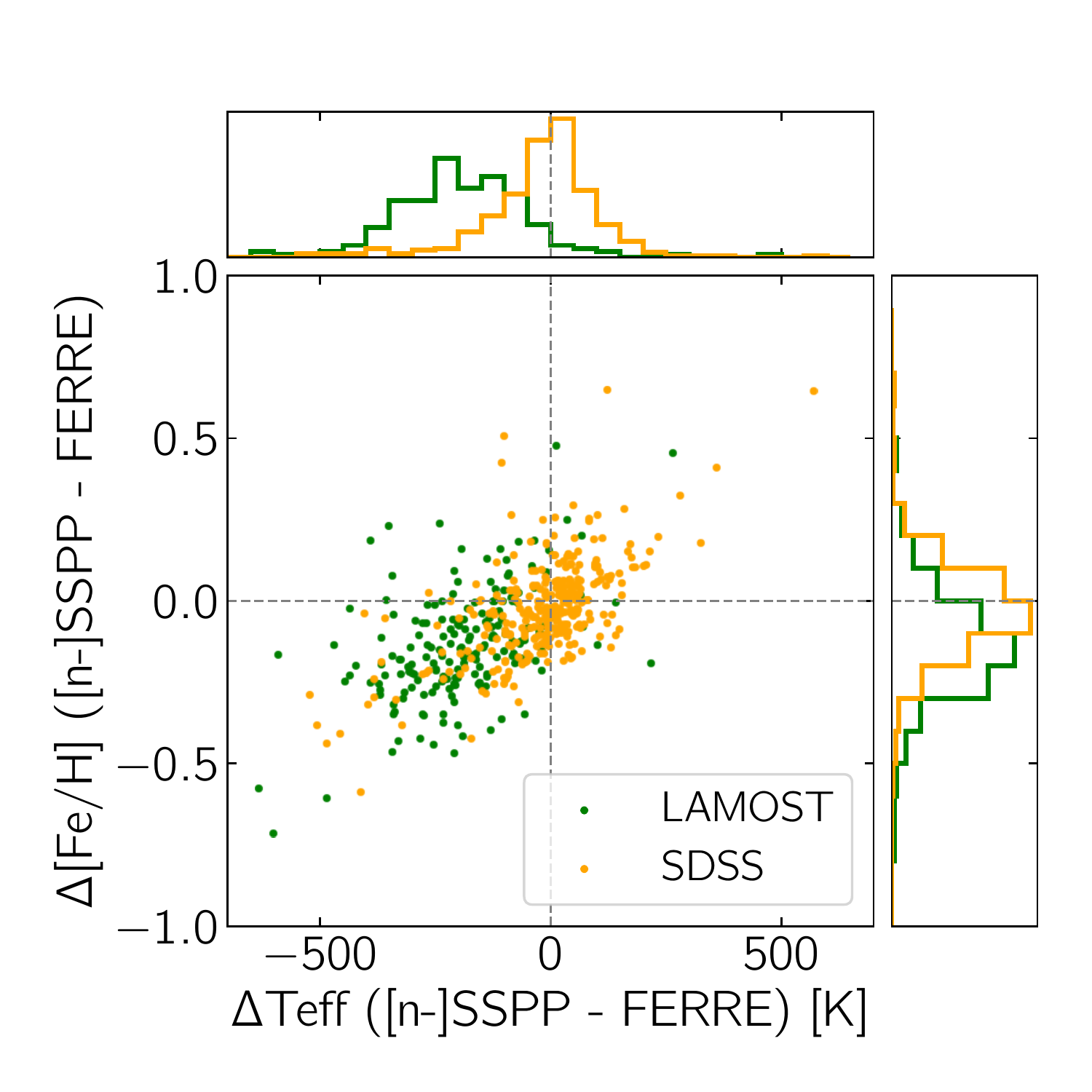}
\includegraphics[width=0.3\hsize,trim={0.0cm 0.0cm 0.0cm 0.0cm}]{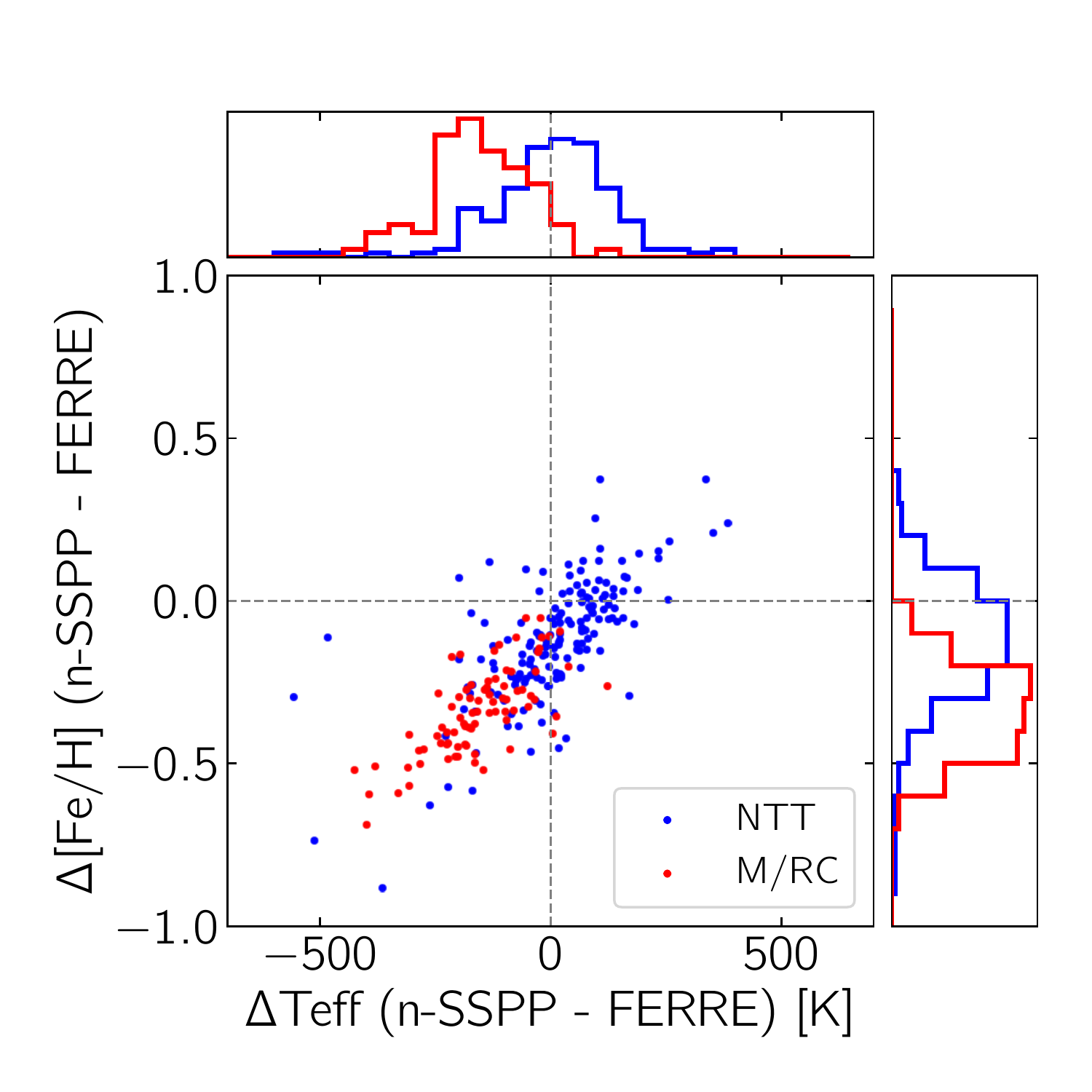}
\includegraphics[width=0.3\hsize,trim={0.0cm 0.0cm 0.0cm 0.0cm}]{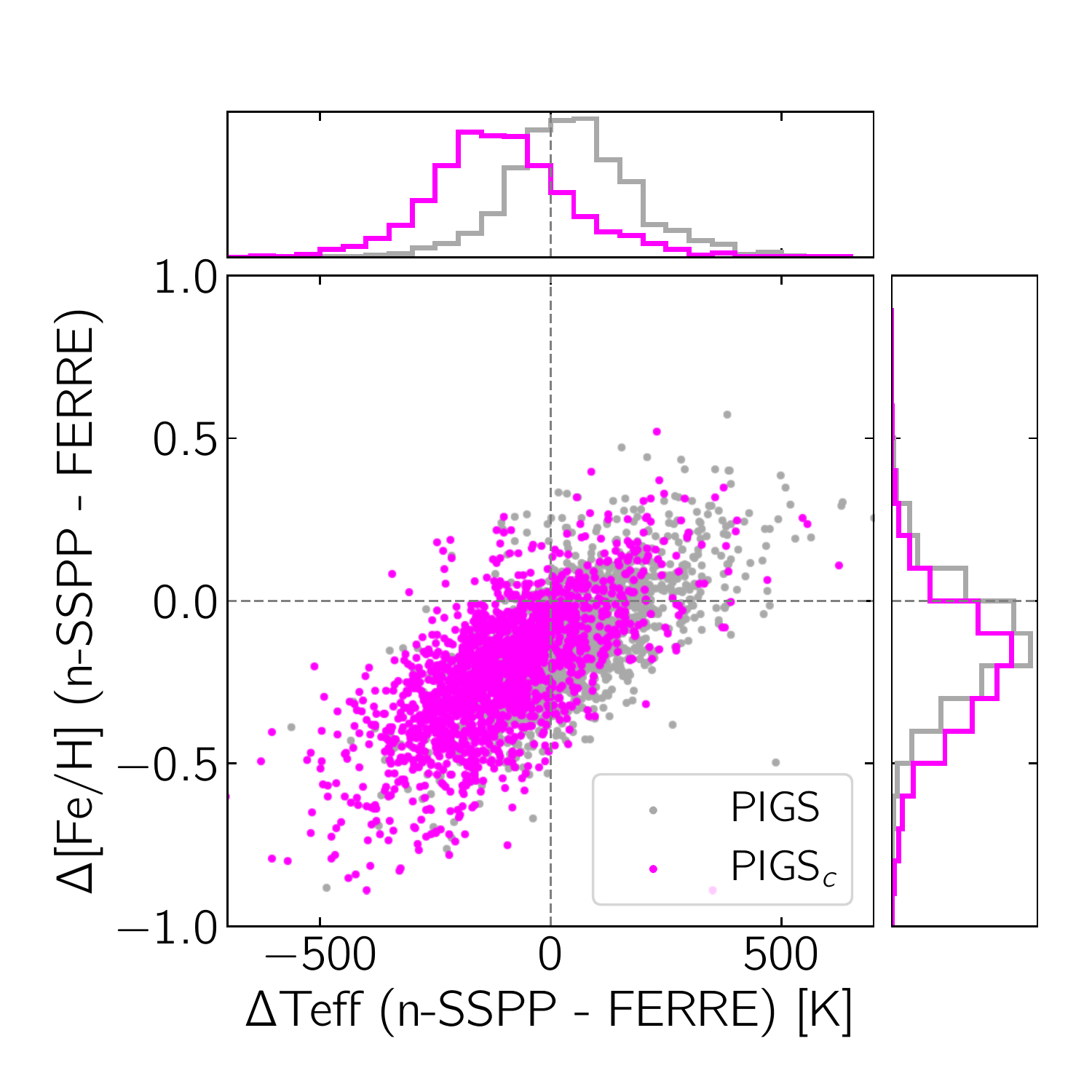}

\includegraphics[width=0.3\hsize,trim={0.0cm 0.0cm 0.0cm 0.0cm}]{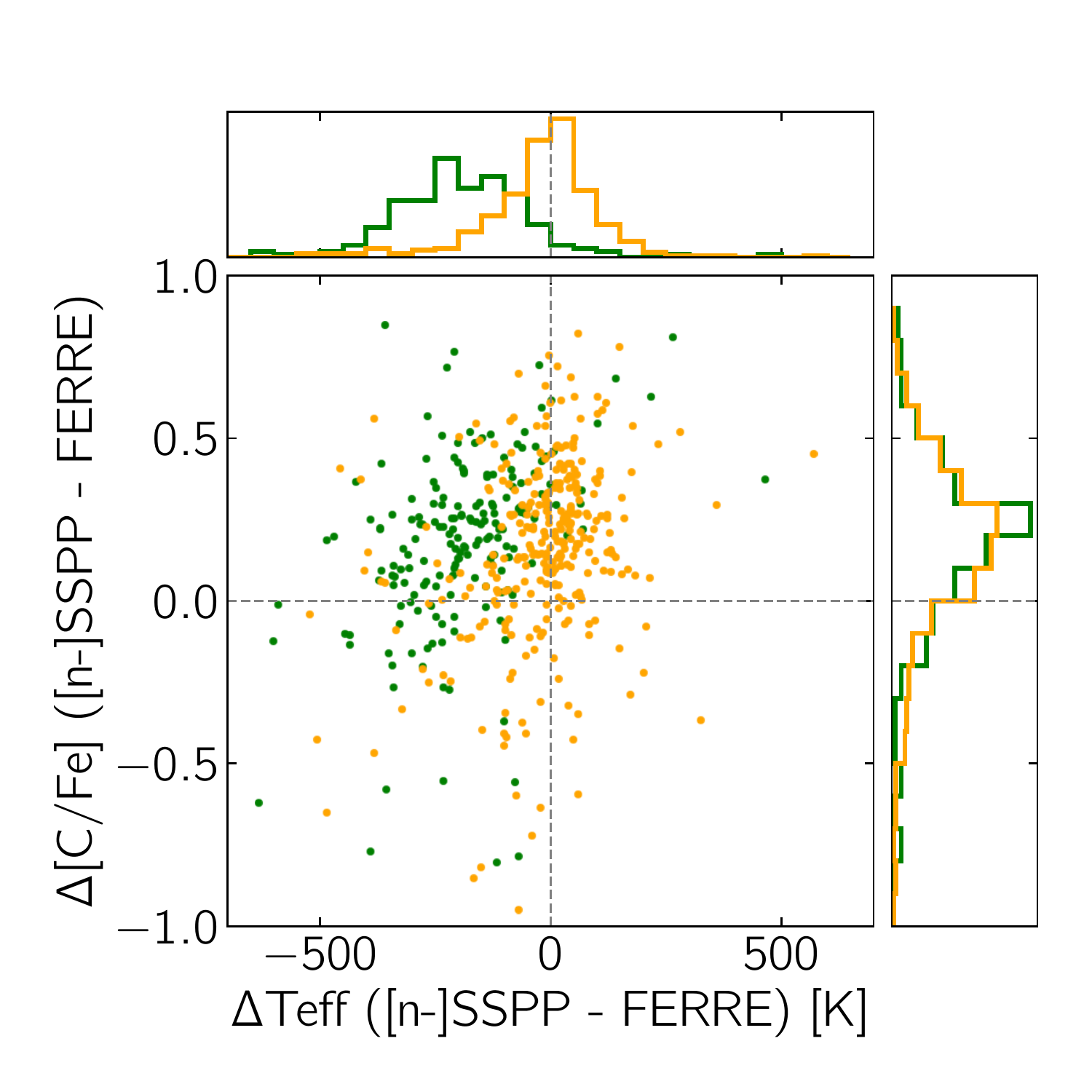}
\includegraphics[width=0.3\hsize,trim={0.0cm 0.0cm 0.0cm 0.0cm}]{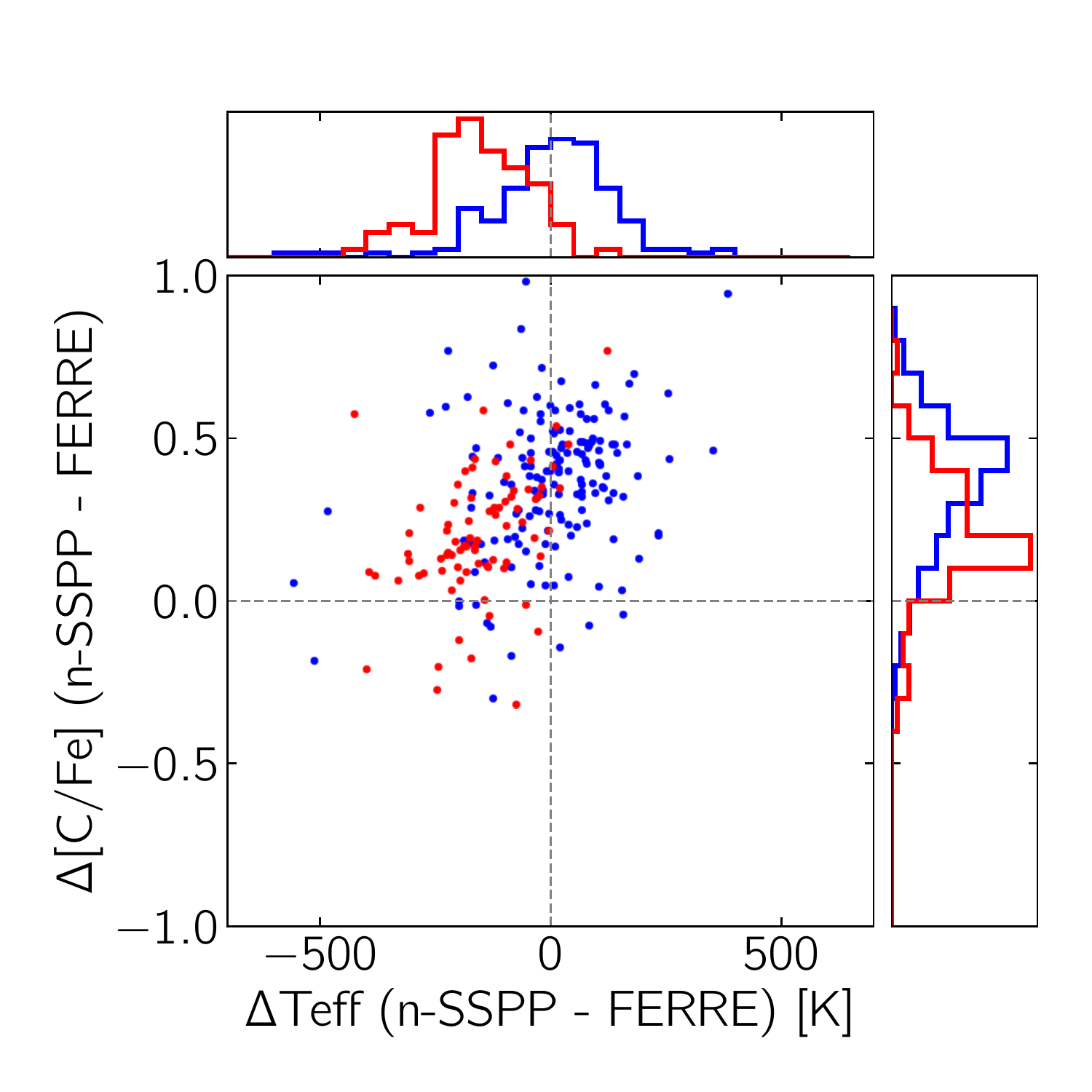}
\includegraphics[width=0.3\hsize,trim={0.0cm 0.0cm 0.0cm 0.0cm}]{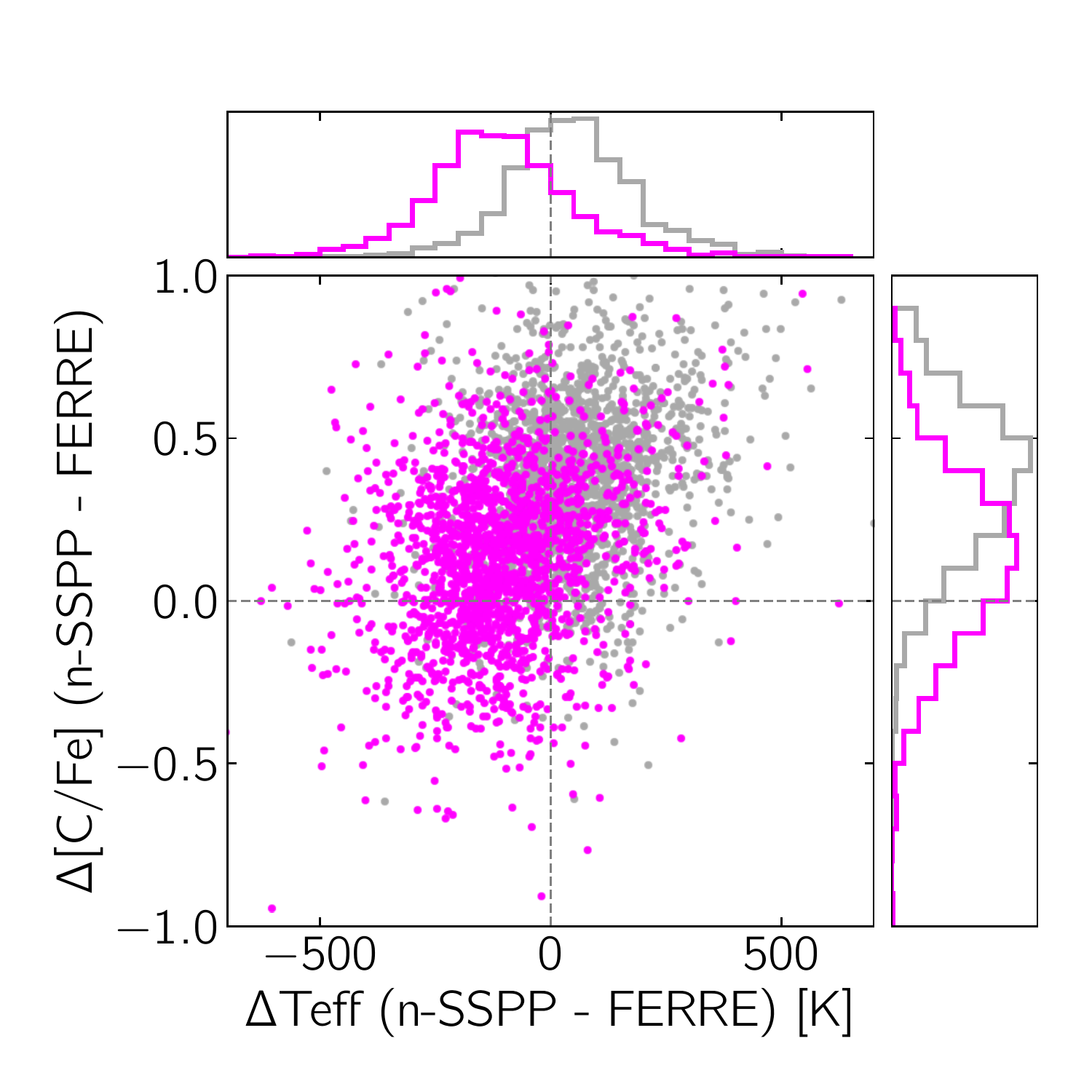}

\includegraphics[width=0.3\hsize,trim={0.0cm 0.0cm 0.0cm 0.0cm}]{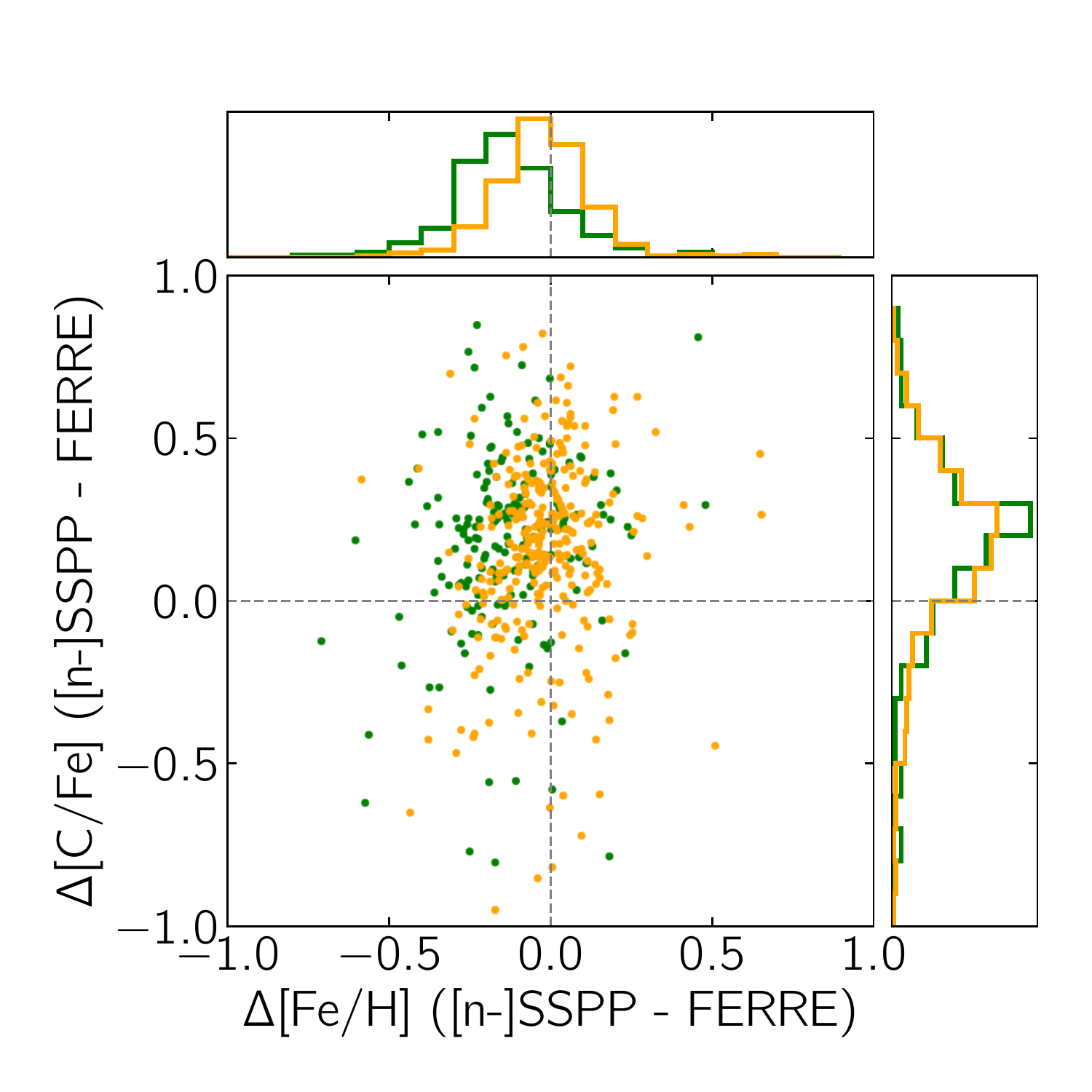}
\includegraphics[width=0.3\hsize,trim={0.0cm 0.0cm 0.0cm 0.0cm}]{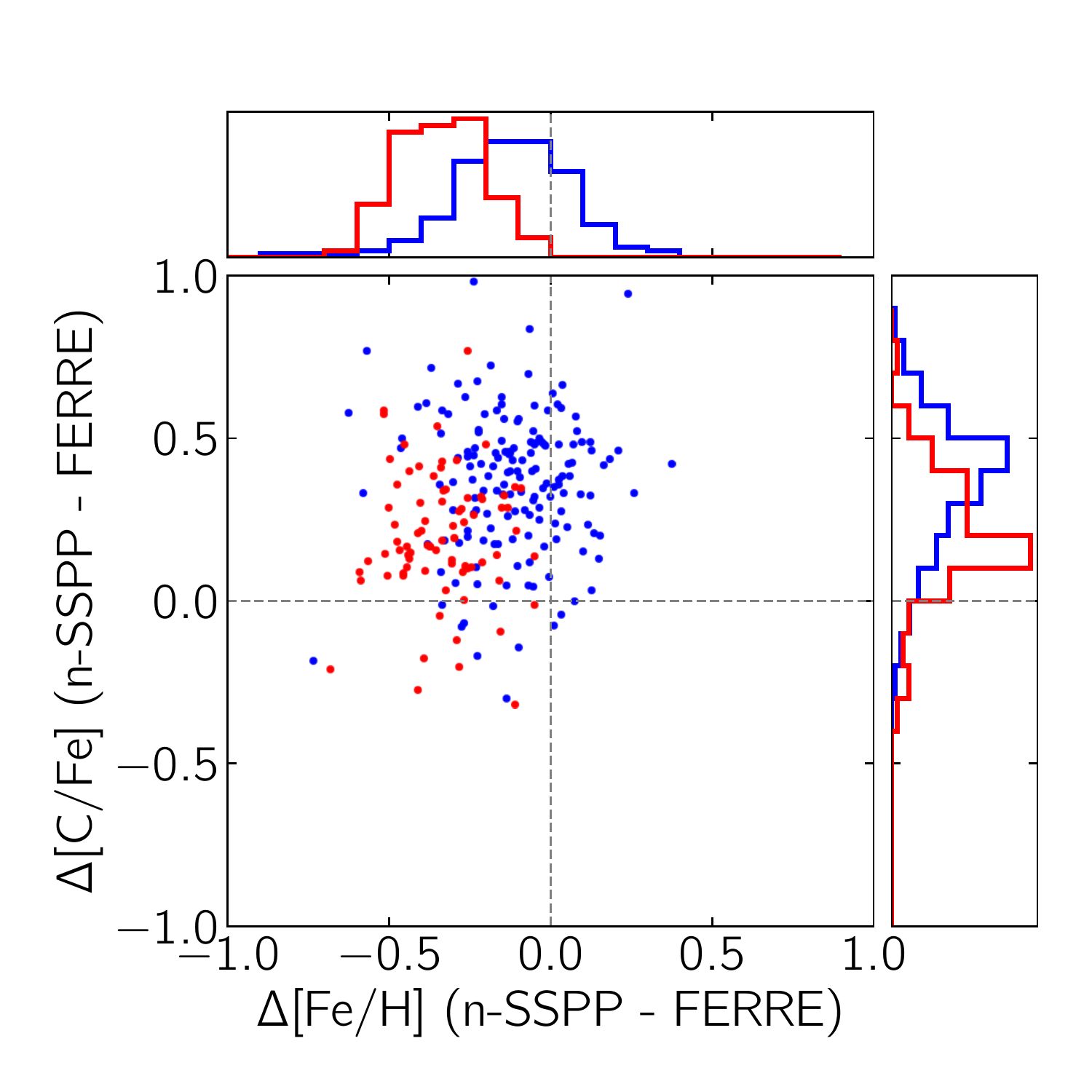}
\includegraphics[width=0.3\hsize,trim={0.0cm 0.0cm 0.0cm 0.0cm}]{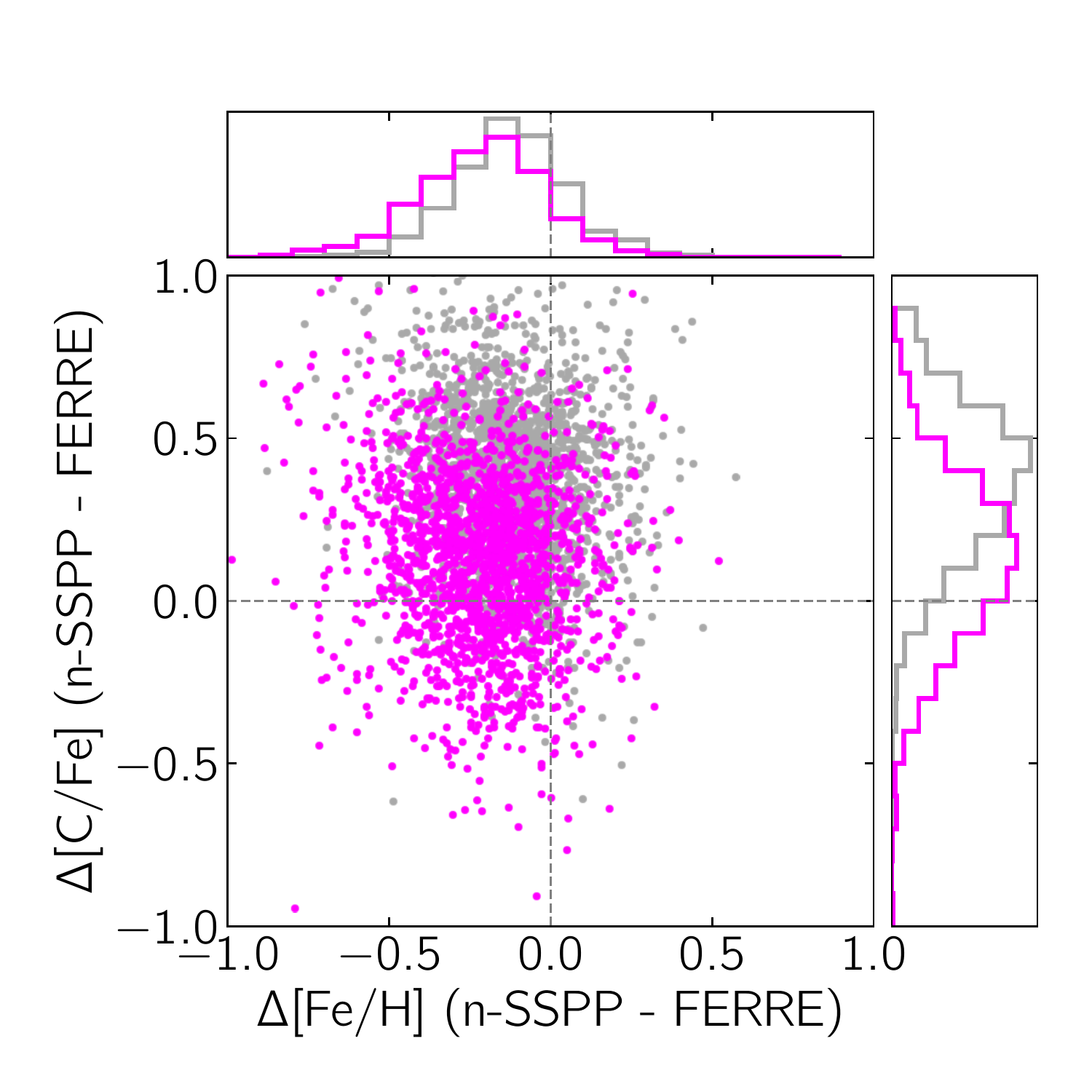}

\includegraphics[width=0.3\hsize,trim={0.0cm 0.0cm 0.0cm 0.0cm}]{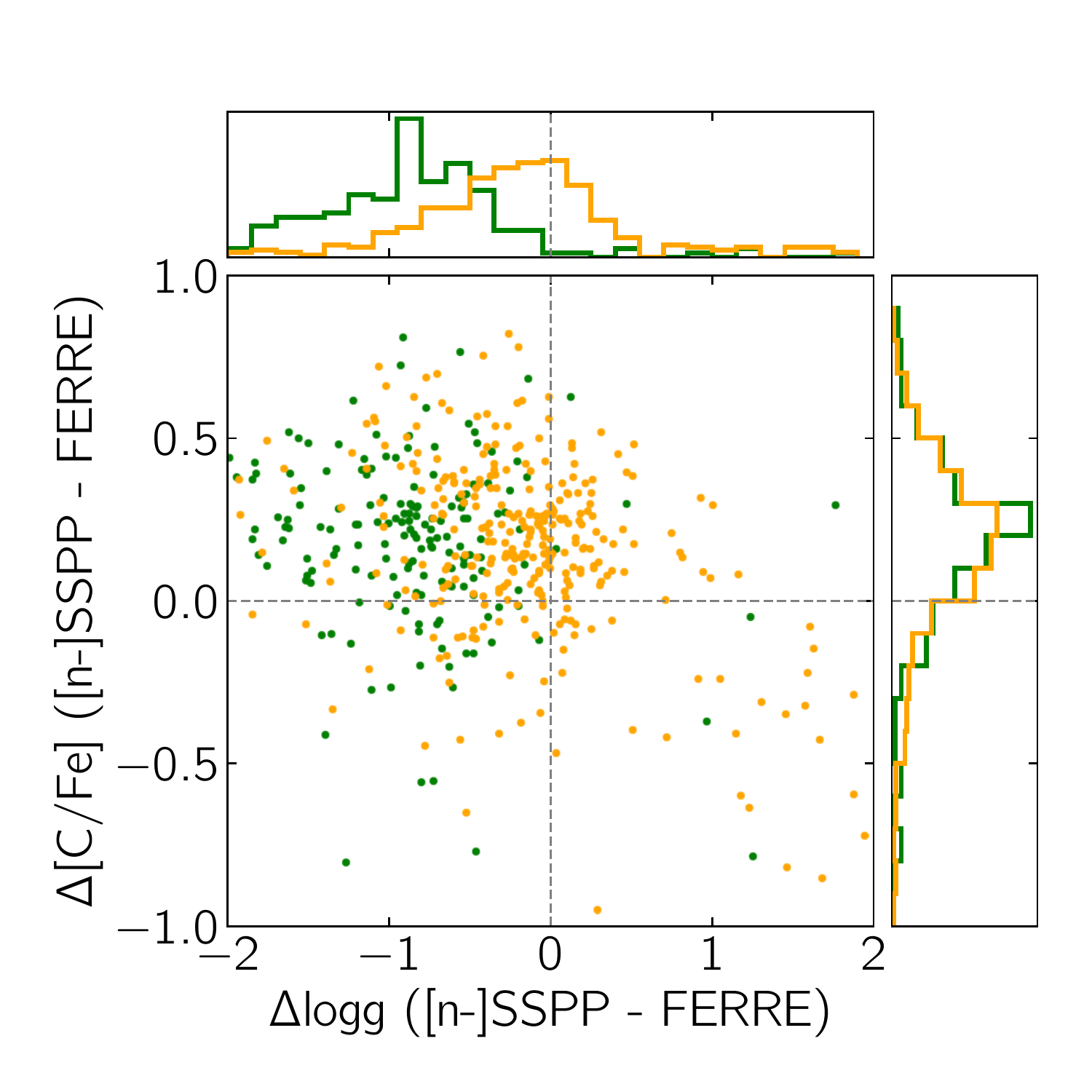}
\includegraphics[width=0.3\hsize,trim={0.0cm 0.0cm 0.0cm 0.0cm}]{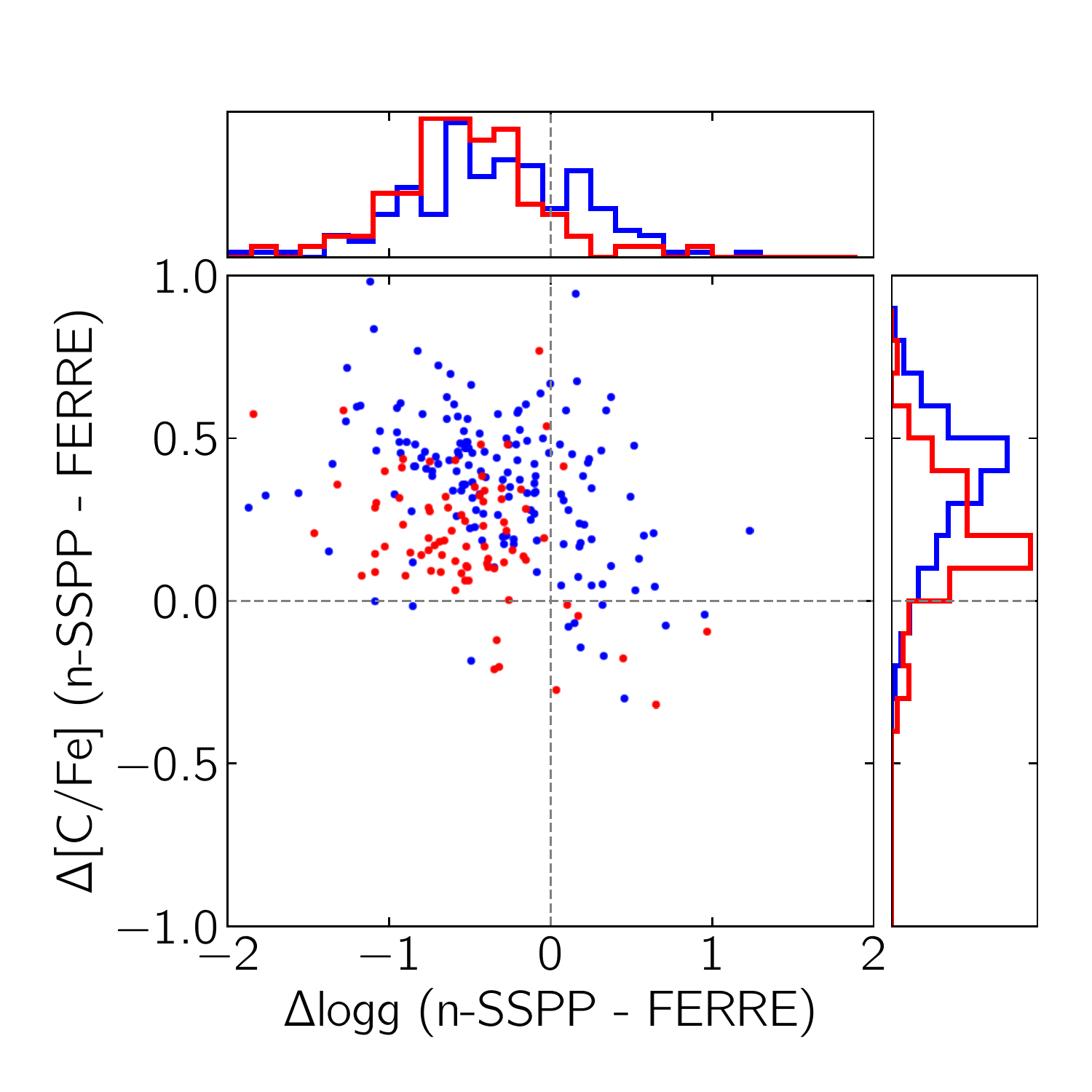}
\includegraphics[width=0.3\hsize,trim={0.0cm 0.0cm 0.0cm 0.0cm}]{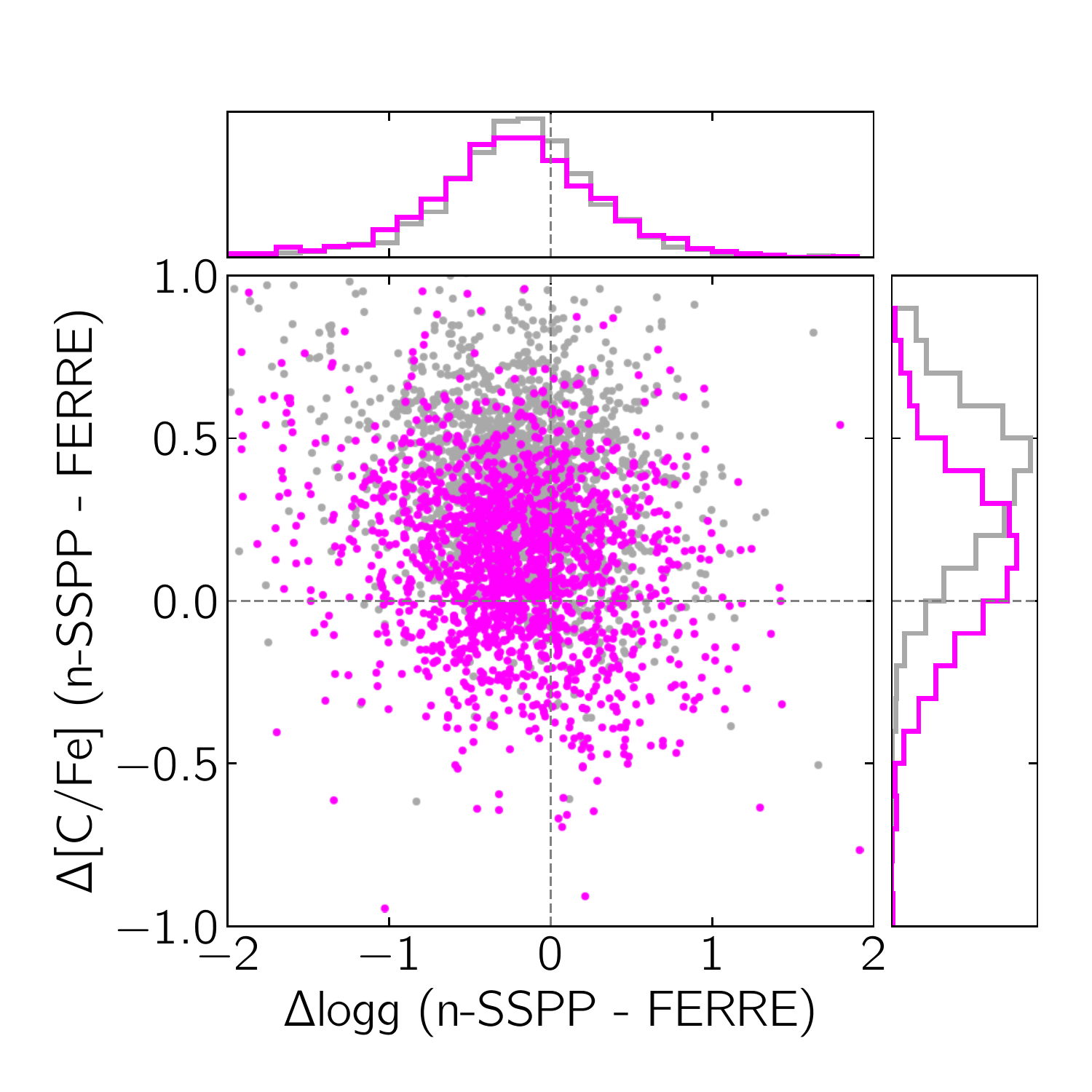}

\caption{As in Figure~5, but comparing the differences between two parameters. For the SDSS/LAMOST panels, only stars with (n-)SSPP $\teff > 5000$~K have been included.}
    \label{fig:ferresspp_delta}
\end{figure*}

We present parameter differences between for two parameters between our FERRE and n-SSPP analyses of the LAMOST, SDSS, RAVE and PIGS samples in Figure~\ref{fig:ferresspp_delta}. This comparison allows to recognise degeneracies between various parameters. The top row shows the well-known correlation between \feh and \teff for low-resolution spectroscopic analyses of metal-poor stars. There also appears to be a slight degeneracy between \teff and \cfe. The other rows, presenting the difference in \cfe versus differences in \feh and \logg, show no strong degeneracies. 

\bsp	
\label{lastpage}
\end{document}